\documentclass[useAMS, usenatbib, a4paper]{mnras}
\usepackage{graphicx}
\usepackage{microtype}
\usepackage{xcolor}
\usepackage{fixltx2e}
\usepackage{booktabs}
\usepackage{siunitx}
\usepackage{color}
\usepackage{enumerate}
\usepackage{pdflscape}
\usepackage{rotating}
\usepackage{hyperref}

\usepackage[T1]{fontenc} 
\usepackage[utf8]{inputenc}

\usepackage{newtxtext}
\usepackage[varvw,smallerops]{newtxmath}

\usepackage{chemgreek}
\activatechemgreekmapping{newtx}

\let\ORIGna\na
\usepackage{musixtex}
\makeatletter
\newcommand\textfermata{
  {\let\extractline\relax
    \setlines10\smallmusicsize \nobarnumbers \nostartrule
    \staffbotmarg0pt \setclefsymbol1\empty \global\clef@skip0pt
    \raisebox{0ex}[0ex][0ex]{
      \startextract\addspace{-\afterruleskip}
      \notes\fermataup{-2}\en
      \zendextract}}}
\makeatother
\let\na\ORIGna

\usepackage{fontawesome}
\usepackage{staves}

\hypersetup{colorlinks=True, linkcolor=blue!50!black, citecolor=black,
  urlcolor=blue!50!black}

\usepackage{etoolbox}
\robustify\bfseries
\robustify\itshape

\makeatletter
\patchcmd\@combinedblfloats{\box\@outputbox}{\unvbox\@outputbox}{}{
  \errmessage{\noexpand\@combinedblfloats could not be patched}
}
\makeatother

\usepackage{bm}

\usepackage{aastex-compat}

\defcitealias{Tarango-Yong:2018a}{Paper~0}

\title[Bow shocks, bow waves, and dust waves. IV.] 
{Bow shocks, bow waves, and dust waves. IV. Shell shape statistics}

\newcommand\AddressCRyA{Instituto de Radioastronom\'{\i}a y Astrof\'{\i}sica,
  Universidad Nacional Aut\'onoma de M\'exico, Apartado Postal 3-72,
  58090 Morelia, Michoac\'an, M\'exico}
\author[Henney et al.]{
  William J. Henney,\thanks{Email: w.henney@irya.unam.mx}
  Jorge A. Tarango-Yong,
  Luis \'Angel Guti\'errez-Soto,\thanks{
    Current address: Observatório do Valongo,
    Universidade Federal do Rio de Janeiro,
    Ladeira Pedro Antonio 43, 20080-090 Rio de Janeiro, Brazil}
  \& S. J. Arthur
  \\
  \AddressCRyA
}

\date{Accepted XXX. Received YYY; in original form ZZZ}

\pubyear{2019}

\providecommand{\abs}[1]{\lvert#1\rvert}

\newcommand{\wind}{\ensuremath{_{\text{w}}}}

\newcommand{\thC}{\(\theta^1\)\,Ori~C}

\newcommand\M{\ensuremath{\mathcal{M}}}
\newcommand\hii{\ion{H}{ii}}

\begin{document}
\label{firstpage}
\pagerange{\pageref{firstpage}--\pageref{lastpage}}
\maketitle
\begin{abstract}
  Stellar bow shocks result from relative motions between stars and
  their environment. The interaction of the stellar wind and radiation
  with gas and dust in the interstellar medium produces curved arcs of
  emission at optical, infrared, and radio wavelengths.  We recently
  proposed a new two-dimensional classification scheme for the shape
  of such bow shocks, which we here apply to three very different
  observational datasets: mid-infrared arcs around hot OB stars;
  far-infrared arcs around luminous cool stars; and H\(\alpha\)
  emission-line arcs around proplyds and other young stars in the
  Orion Nebula.  For OB stars, the average shape is consistent with
  simple thin-shell models for the interaction of a spherical wind
  with a parallel stream, but the diversity of observed shapes is many
  times larger than such models predict.  We propose that this may be
  caused by time-dependent oscillations in the bow shocks, due to
  either instabilities or wind variability.  Cool star bow shocks have
  markedly more closed wings than hot star bow shocks, which may be
  due to the dust emission arising in the shocked stellar wind instead
  of the shocked interstellar medium.  The Orion Nebula arcs, on the
  other hand, have both significantly more open wings and
  significantly flatter apexes than the hot star bow shocks.  We test
  several possible explanations for this difference (divergent ambient
  stream, low Mach number, observational biases, and influence of
  collimated jets), but the evidence for each is inconclusive.
\end{abstract}

\begin{keywords}
  circumstellar matter -- methods: statistical -- stars: winds, outflows
\end{keywords}

% start input ./sec-obs-intro.tex

\section{Introduction}
\label{sec:introduction}

Stellar bow shocks are produced by the relative motion between a star
and its surrounding medium, and are commonly detected as curved arcs
of emission at optical \citep{Gull:1979a, Brown:2005a}, infrared
\citep{van-Buren:1988a, Kobulnicky:2016a}, or radio
\citep{van-Buren:1990a, Benaglia:2010a} wavelengths.  The canonical
theory for these objects is that they are formed by a two-shock
interaction between the stellar wind and the interstellar medium
\citep{Pikelner:1968a, Dyson:1972a}, which is distorted due to the
supersonic motion of the star \citep{Baranov:1970a, Wilkin:1996a}.  In
some instances, however, the absorbed stellar radiation pressure may
be more important than the stellar wind in providing the inner support
for the bow shell \citep[Paper~I]{Henney:2019a} and this may even be
sufficient to break the collisional coupling between gas and dust
grains \citep[Paper~II]{Henney:2019b}.  We have proposed a diagnostic
method to distinguish between these cases, based on the bow shock size
and the luminosity ratio between the bow shock and the star
\citep[Paper~III]{Henney:2019c}. In other cases, the appearance of an
infrared emission arc may be due to the illumination of the inner wall
of an asymmetrical cavity \citep{Mackey:2016a}, rather than the
formation of a dense shell, in which case the relative velocity of the
star may be subsonic with respect to its surroundings
\citep{Mackey:2015a}.

The largest number of bow shocks have been detected around high-mass
OB stars, via their mid-infrared dust emission \citep{van-Buren:1995a,
  Noriega-Crespo:1997b, Povich:2008a, Kobulnicky:2010a, Peri:2012a,
  Peri:2015a, Sexton:2015b, Kobulnicky:2016a, Bodensteiner:2018a}, and
these have typical sizes ranging from \SIrange{0.01}{1}{pc}. In
particularly dense environments, such as the inner Orion Nebula
\citep{Smith:2005a} and the Galactic center region
\citep{Geballe:2004a} they may be as small as \SI{0.003}{pc} and emit
at near-infrared wavelengths \citep{Tanner:2005a,
  Sanchez-Bermudez:2014a}.  Cometary ultracompact \hii{} regions
detected at radio wavelengths \citep{Reid:1985a, Wood:1989a,
  Klaassen:2018a} have also been interpreted as bow shocks
\citep{van-Buren:1990a, Mac-Low:1991a}, although alternative models,
such as a champagne flow caused by steep density gradients
\citep{Cyganowski:2003a, Arthur:2006a, Immer:2014a, Steggles:2017a},
are favored in many cases.  This illustrates a broader point: that it
can be difficult to identify bow shocks from morphology alone, since
other processes can give rise to emission arcs.  In particular, curved
ionization fronts, as seen in evaporating globules \citep{Sahai:2012b}
and proplyds \citep{ODell:1993a} can be mistaken for bow shocks.
Kinematic observations can potentially resolve such ambiguities but
are not always available.

Stellar bow shocks are also observed around other types of stars. Bow
shocks around cool red supergiant and asymptotic giant branch stars
are detected at mid-infrared and far-infrared wavelengths
\citep{Ueta:2006a, Ueta:2008a, Sahai:2010a, Cox:2012a}.  Pulsar bow
shock nebulae are detected principally by their H\(\alpha\) emission
\citep{Kulkarni:1988a, Brownsberger:2014a}.  In the Orion Nebula (M42,
NGC~1976), at least three different classes of stellar bow shock have
been identified. As well as a small number of OB bow shocks
\citep{Smith:2005a, ODell:2001c}, bow shocks are also seen around the
closest proplyds to the dominant O~star \thC{} \citep{Hayward:1994a,
  Bally:1998a, Robberto:2005a}.  The proplyds \citep{ODell:2008b} are
photoevaporating protoplanetary disks around low-mass young stars
\citep{Johnstone:1998a} and the bow shocks have been modeled as the
interaction between the disk's ionized photoevaporation flow and the
supersonic stellar wind from \thC{} \citep{Garcia-Arredondo:2001a}.
The third class of Orion Nebula bow shock is the LL~Ori-type objects
\citetext{\citealp{Gull:1979a}; \S~5 of \citealp{Bally:2000a}; \S~3.2
  of \citealp{Bally:2001a}; \citealp{Henney:2013a}}, which tend to be
found in the outer regions of the nebula.  These are probably due to
interactions between the Orion Nebula's champagne flow
\citep{Zuckerman:1973a} and outflows from T~Tauri stars, which may or
may not be proplyds \citep{Bally:2000a, Gutierrez-Soto:2015a}.

\begin{figure}
  \centering
  \includegraphics[width=0.6\linewidth]{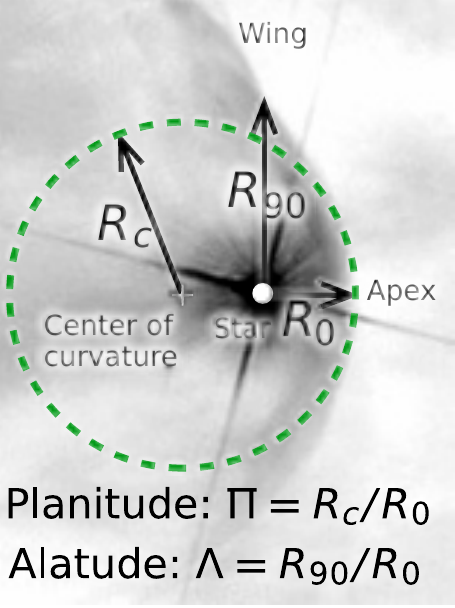}
  \caption[Terminology]{Terminology employed in this paper to describe
    bow shock shapes, following \citet[Paper~0]{Tarango-Yong:2018a}.}
  \label{fig:obs-shape-terminology}
\end{figure}

The shapes of stellar bow shocks are frequently compared with what has
become known as the \textit{wilkinoid} surface \citep{Cox:2012a},
which is the result of the idealized interaction between a spherical
wind and a plane-parallel stream in the hydrodynamic thin-shell
approximation.  Numerical approximations to this shape were used by
various authors \citep{Baranov:1971a, Mac-Low:1991a} before an elegant
analytic solution was found by \citet{Wilkin:1996a} and extended to
the case of interaction between two spherical winds
\citep{Canto:1996}.  In \citet[hereafter, Paper~0]{Tarango-Yong:2018a}
the study of bow shock shapes and their projection on the plane of the
sky was formalized. The term \textit{cantoid} was introduced for the
\citet{Canto:1996} family of shapes, together with \textit{ancantoid}
for a generalization to the case where one of the winds is
anisotropic, as is appropriate for the proplyds.  In addition, Paper~0
proposed the use of two dimensionless parameters, \textit{planitude}
and \textit{alatude}, to describe a general bow shock shape, which we
illustrate in Figure~\ref{fig:obs-shape-terminology}.  The planitude,
\(\Pi = R_c / R_0\), measures the flatness of the bow shock apex, where
\(R_0\) is the star--apex distance and \(R_c\) is the radius of
curvature measured at the apex.  The alatude,
\(\Lambda = R_{90}/R_0\), measures the openness of the bow shock wings,
where \(R_{90}\) is the lateral size of the bow, measured from the
star in the direction perpendicular to the star--apex direction.

In this paper, we investigate the shapes of stellar bow shocks by
calculating the distributions of planitude and alatude for different
classes of bow shock source.  The remainder of the paper is organized
as follows. In \S~\ref{sec:mid-infrared-arcs} we present an analysis
of the shapes of several hundred bow shock candidates associated with
OB stars from the \SI{24}{\um} survey of \citet{Kobulnicky:2016a}.
Our algorithm for automatically fitting and tracing the shapes is
described in \S~\ref{sec:autom-trac-fitt}, together with our ``star
rating'' system for evaluating the fit quality, while in
\S~\ref{sec:ob-shapes} we locate the sources on the planitude--alatude
plane.  In \S~\ref{sec:corr-size} we study the correlations amongst
non-shape parameters of the bow shock sources, such as angular size
and stellar magnitude, while in \S~\ref{sec:corr-shape} we explore the
correlations between these parameters and the planitude and alatude.  In
\S~\ref{sec:far-infrared-arcs} we compare with results for bow shocks
around cool luminous stars and in \S~\ref{sec:stat-emiss-line} we
compare with results for stationary emission-line arcs in the Orion
Nebula.  In \S~\ref{sec:discussion} we discuss the implications of our
findings for physical models of bow shock formation in the different
classes of sources, and in \S~\ref{sec:conclusion} we summarise our
results.  Further details of the statistical tests that we have
applied are provided in Appendix~\ref{sec:distr-p-values} and a simple
model for time-dependent oscillations of the bow shock surface is
presented in Appendix~\ref{sec:perturbed-bows}.

 % end input ./sec-obs-intro.tex
 % start input ./sec-quadrics-observations.tex

\section{Mid-infrared arcs around early-type stars}
\label{sec:mid-infrared-arcs}
\label{sec:comp-with-observ}

\begin{figure*}
  \includegraphics[width=\linewidth]{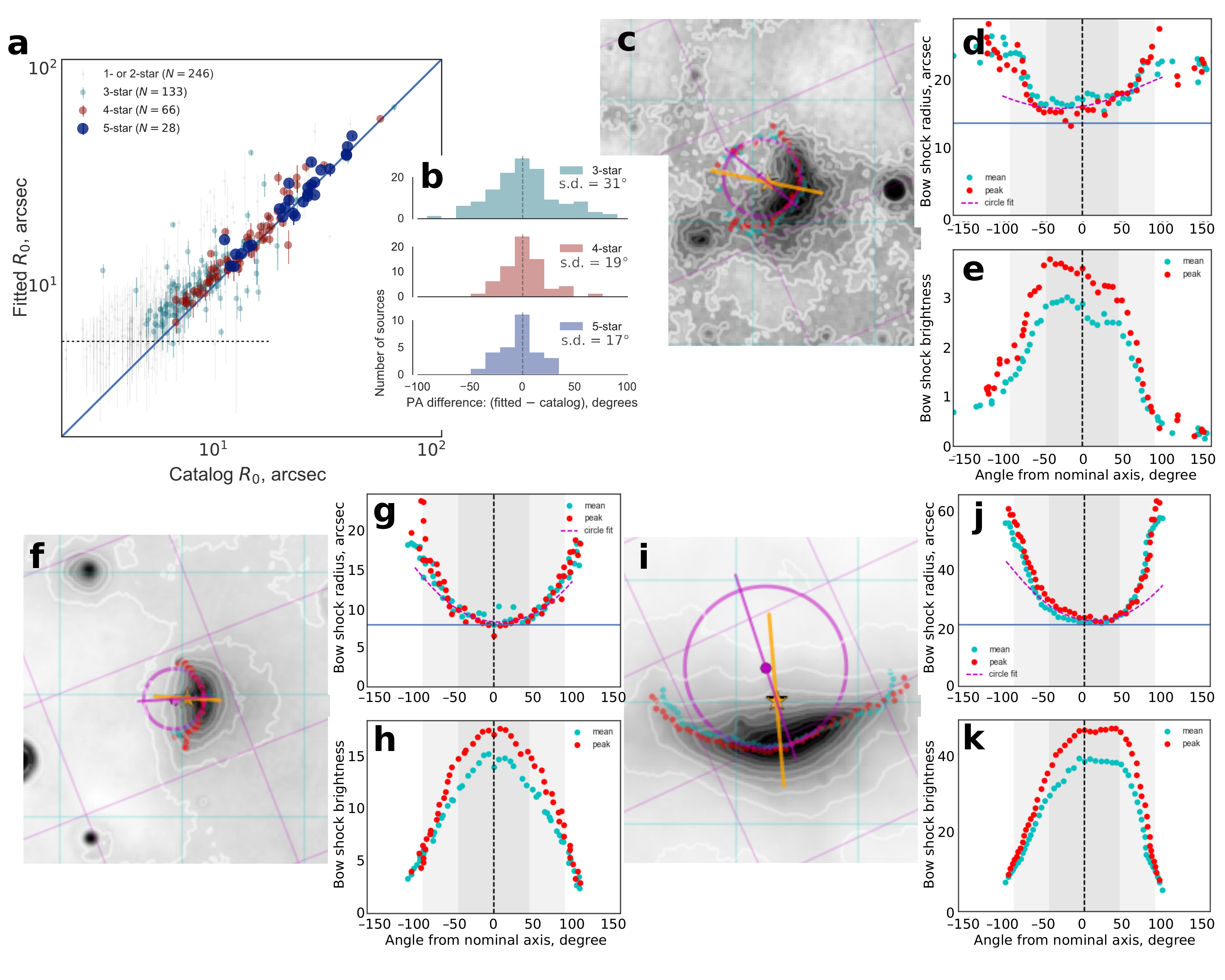}
  \caption[]{Examples of typical fits to the bow shock shapes of
    MIPSGAL sources.  
    (a)~Scatter plot of the bow shock sizes determined from our fits,
    as compared with those tabulated by \citet[K16]{Kobulnicky:2016a}
    for the MIPSGAL sources.  Different symbol sizes and colors
    correspond to different star ratings, which indicate our
    subjective judgement of the quality of the fit.  The horizontal
    dotted line shows the MIPS \SI{24}{\um} point spread function
    FWHM of \(5.5\arcsec\).  
    (b)~Histograms of the difference in bow shock position angles
    between our fits and the K16 values for sources with 3- to 5-star
    ratings.  
    (c--e)~Source K510, with a 3-star rating. 
    (f--h)~Source K506, with a 4-star rating. 
    (i--k)~Source K517, with a 5-star rating. 
    Panels~(c, f, i) show a \SI{24}{\um} gray-scale image of a
    \(160'' \times 160''\) field of view, centered on each source.
    Contours are ten linearly spaced levels between the median
    brightness of the entire image and the maximum brightness of the
    bow shock arc. Grids of galactic coordinates (light blue lines,
    parallel to the box sides) and equatorial coordinates (tilted
    magenta lines) are shown.  The stellar source and the bow shock
    axis, as determined by \citet{Kobulnicky:2016a} are indicated by
    an orange star and an orange line, respectively, where the line
    extends from \(-2 R_0\) to \(+2 R_0\).  The automatically traced
    arc shapes using the ``mean'' and ``peak'' methods (see text) are
    shown by blue and red dots, respectively.  The magenta circle
    shows the fit to the arc points within \(\pm 45^\circ\) of the nominal
    bowshock axis, with the magenta dot showing the center of
    curvature and the magenta line showing the fitted bow shock axis,
    which is the line passing through the source and the center of
    curvature. 
    Panels~(d, g, j) show plots of the radius measured from the source
    of the arc points, plotted as a function of angle \(\theta\) from the
    nominal bow shock axis, and with the same color coding as used on
    the image. Angular ranges of \(\theta = \pm 45\degr\) and
    \(\pm 90\degr\) are shown by gray shaded boxes.  The \(R_0\) value
    tabulated by \citet{Kobulnicky:2016a} is shown by a horizontal
    blue line. 
    Panels~(e, h, k) show equivalent plots for the surface brightness.
  }
  \label{fig:mipsgal-examples}
\end{figure*}

The most extensive observational sample of stellar bow shock nebulae
to date is a catalog of 709 arcs \citep{Kobulnicky:2016a} detected in
mid-infrared surveys of the Galactic Plane by the \textit{Spitzer
  Space Telescope} (\textit{SST}, \citealp{Werner:2004a}) and
\textit{Wide-field Infrared Survey Explorer} (\textit{WISE},
\citealp{Wright:2010a}).  These sources are believed to be powered by
the winds of early-type stars, which are either moving supersonically
through the interstellar medium (runaway stars,
\citealp{Gvaramadze:2008a}), or are interacting with a local bulk
flow, such as the champagne flow from a nearby \hii{} region (weather
vanes, \citealp{Povich:2008a}).

\subsection{Automatic tracing and fitting of bow shocks}
\label{sec:autom-trac-fitt}

In order to study the shapes of these bow shocks, we downloaded data
from the NASA/IPAC Infrared Science Archive archive\footnote{
  \url{http://irsa.ipac.caltech.edu/docs/program_interface/api_images.html}}
and extracted 4\arcmin{} square images in the \SI{24}{\um} bandpass of
the Multiband Imaging Photometer for \textit{Spitzer} (MIPS) centered
on each of the 471 \citet{Kobulnicky:2016a} sources that are covered
by the MIPSGAL \citep{Carey:2009a} survey, which includes most of the
sources with Galactic longitude within \(\pm 60\degr\) of the Galactic
center.

We have developed a method for automatically tracing the arcs and
determining their planitude and alatude, which is an extension of the
method described in Appendix~E of Paper~0.  The steps of the method
are as follows:
\begin{enumerate}[1.]
\item Calculate arrays of celestial coordinates, \(C\), for each pixel
  of the image. In our implementation of the method we use functions
  from the Python library \texttt{astropy.coordinates}
  \citep{Astropy-Collaboration:2018a}.
\item Using the central source coordinates, \(C_0\) and nominal
  bowshock radius, \(R_0\) from \citet{Kobulnicky:2016a}, construct a
  pixel mask that includes only those pixels with separations from the
  source that satisfy \(\frac12 R_0 \le |C - C_0| \le 3 R_0\).  This mask
  will be used for all subsequent operations, which serves to help
  avoid confusion from the star itself and other bright sources in the
  field of view.
\item Define a ``step-back'' point, \(C_1\), which is located at a
  separation \(2 R_0\) from the source, but in the opposite direction
  from the apex of the bow shock. That is, along a position angle
  180\degr{} from the nominal position angle, \(\text{PA}_0\), of the
  bow shock axis.  This point is at one end of the orange line shown
  superimposed on the bow shock images in
  Figure~\ref{fig:mipsgal-examples}.
\item Looping over a grid of 50 position angles, \(\text{PA}_k\),
  within \(\pm 60\degr\) of \(\text{PA}_0\), estimate the location of
  the arc along rays cast from the step-back point, using two
  different methods:
  \begin{enumerate}[(a)]
  \item The pixel with the peak brightness, with coordinates
    \(C_{k,\text{peak}}\) (red dots in
    Fig.~\ref{fig:mipsgal-examples}).
  \item The mean brightness-weighted separation from \(C_1\), with
    coordinates \(C_{k,\text{mean}}\) (light blue dots in
    Fig.~\ref{fig:mipsgal-examples}).
  \end{enumerate}
  For each \(\text{PA}_k\) in the grid, the calculation is performed
  over only those pixels that satisfy
  \(|\text{PA}(C, C_1) - \text{PA}_k| < \frac12 \delta\theta\), where
  \(\delta\theta = 120/50 = 2.4\degr\), which defines a thin radial wedge from
  \(C_1\).  The results are shown as red and blue dots superimposed on
  the images in Figure~\ref{fig:mipsgal-examples}. Each of the two
  methods, ``peak'' and ``mean'', works better in some objects and
  worse in others (according to the subjective judgment of
  ``correctly'' tracing the bow shock shape).  We therefore take the
  average by amalgamating all the \(C_{k,\text{peak}}\) and
  \(C_{k,\text{mean}}\) points into a single set, \(C_{k}\), for the
  following steps.  We have found that the use of the step-back point
  greatly improves the reliability of the method, as compared with the
  simpler option of tracing rays from the stellar source.
\item For each of the points \(C_{k}\), determine the radial
  separation from the central source, \(R_k = |C_k - C_0|\) and the
  angle from the bow shock axis about the central source
  \(\theta_k = \text{PA}(C_k, C_0) - \text{PA}_0\).  These are plotted in
  the upper left panels of Figure~\ref{fig:mipsgal-examples}.  Note
  that, even though the rays are cast from the step-back point \(C_1\)
  within \(\pm 60\degr\) of \(\text{PA}_0\), the angles \(\theta_k\) are
  measured from the source, \(C_0\), which is closer to the bow shock
  than \(C_1\) and therefore \(|\theta_k|\) can be much larger than
  \(60\degr\).
\item Make our own estimate of the axial size, \(R_0\), of the bow
  shock by calculating the mean of \(R_k\) over all points \(C_k\)
  with \(|\theta_k| \le 10\degr\).  Note that this is distinct from the
  nominal value of \(R_0\) given in the \citet{Kobulnicky:2016a}
  catalog, which was ``measured by eye''.  We denote by
  \(\epsilon(R_0)\) the standard deviation of the \(R_k\) that go into
  calculating \(R_0\).\label{step:R0}
\item Estimate the radius of curvature, \(R_c\), by fitting a circle
  to all those points within \(\pm 45\degr\) of the nominal axis
  (\(|\theta_k| < 45\degr\)), but after excluding any point with
  \(R_k < \frac12 R_m\) or \(R_k > 2 R_m\), where \(R_m\) is the median
  \(R_k\) for \(|\theta_k| < 45\degr\).\label{step:Rc}
\item Determine two separate estimates, \(R_{90+}\) and \(R_{90-}\),
  of the perpendicular radius, \(R_{90}\), by taking the mean of
  \(R_k\) over all points \(C_k\) with
  \(|\theta_k - 90\degr| \le 10\degr\) for \(R_{90+}\), and with
  \(|\theta_k + 90\degr| \le 10\degr\) for \(R_{90-}\).  The average of the
  two standard deviations of the \(R_k\) that contribute to
  \(R_{90+}\) and \(R_{90-}\) is denoted by \(\epsilon(R_{90})\).\label{step:R90}
\item The planitude is determined as \(\Pi = R_c / R_0\) and the mean
  alatude as \(\Lambda = \frac12 (R_{90+} + R_{90-}) / R_0\).  We also
  calculate an alatude asymmetry as
  \(\Delta\Lambda = \frac12 (R_{90+} - R_{90-}) / R_0\). Note that in this paper
  we simplify the notation of Paper~0 by no longer using the prime
  symbol (\('\)) to distinguish projected from intrinsic quantities.
  All planitudes and alatudes should be understood as specifying the
  projected bow shock shape on the plane of the sky unless noted
  otherwise.\label{step:Pi-Lambda}
\end{enumerate}
\label{sec:subj-eval-fit}
After these automatic steps, we subjectively evaluate the resulting
fit quality by giving a star rating to each source:

\paragraph*{0 stars} The fitting algorithm failed for some reason. 

\paragraph*{1 star} The fit was formally successful, but the results
for \(R_c\) or \(R_{90}\) are far removed from what a human would
predict by looking at the image.  For example, in the smallest
bowshocks, which are only marginally resolved by Spitzer's 6\arcsec{}
beam, the dispersion in \(R_k\) can be a significant fraction of
\(R_0\), in which case our algorithm tends to erroneously favor
\(R_c < R_0\).

\paragraph*{2 stars} The fit results are not totally outlandish, but
nonetheless some problem is apparent that casts doubt on their
reliability.  For example, a double-shell structure to the bow shock
that leads to large differences between the ``peak'' and ``mean''
methods, or point sources near to the bow shock that interfere with
the tracing procedure.
  
\paragraph*{3 stars} A good fit, but where the dispersion in \(R_k\)
and/or the asymmetry in the bow shock reduces the precision in the
determination of \(R_c\) and \(R_{90}\), giving subjectively estimated
uncertainties around the 20\% level.  An example of a 3-star fit is
shown in Figure~\ref{fig:mipsgal-examples}a.

\paragraph*{4 stars} A high quality fit, with subjectively estimated
uncertainties in \(R_c\) and \(R_{90}\) around the 10\% level. An
example of a 4-star fit is shown in
Figure~\ref{fig:mipsgal-examples}b.

\paragraph*{5 stars} The highest-quality fit, usually corresponding to
large, sharply defined bow shocks, whose shape is determined with high
precision. An example of a 5-star fit is shown in
Figure~\ref{fig:mipsgal-examples}c.
\bigskip

Figure~\ref{fig:mipsgal-examples}d compares the bow shock size,
\(R_0\), determined by our fits (vertical axis) with the corresponding
value given in the \citet{Kobulnicky:2016a} catalog (horizontal axis).
For most sources with 3-star or higher rating, the two estimates agree
to within \(\pm 20\%\), but there are a small number of sources with a
discrepancy of more than a factor of two.  In all cases that we
checked, we believe that our estimates of \(R_0\) are more accurate
than those in the catalog.  It is apparent that the star ratings are
correlated with the bow shock size, with larger bow shocks tending to
receive higher ratings, although there is considerable overlap.  In
particular, most of the 1- and 2-star sources are close to the
resolution limit of the MIPSGAL \SI{24}{\um} images (\(6\arcsec\),
indicated by the dotted horizontal line in the figure).

\begin{figure*}
  \centering
  \setlength\tabcolsep{0pt}
  \begin{tabular}{ll}
    (a) & (b) \\
    \includegraphics[width=0.5\textwidth]
    {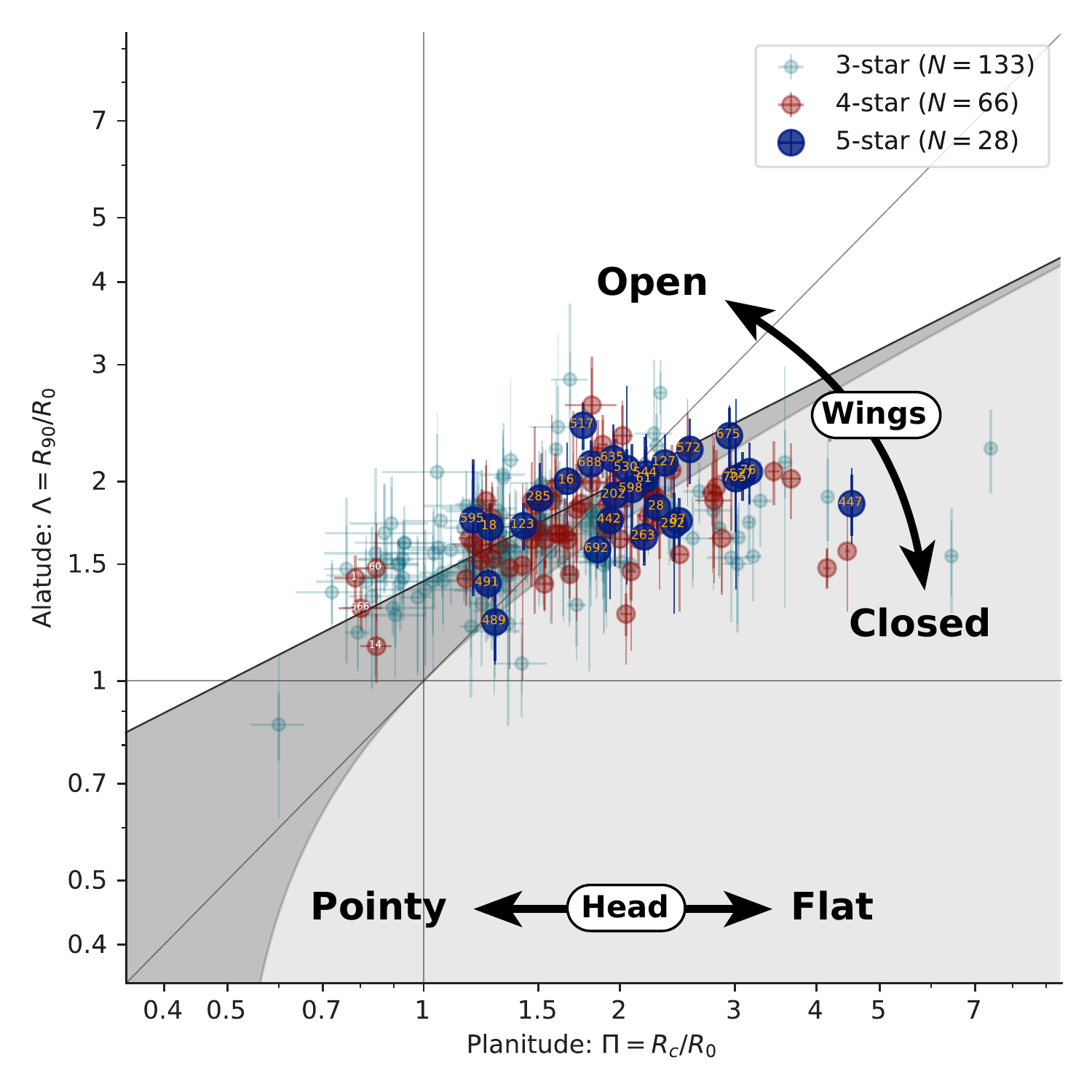}
        & \includegraphics[width=0.5\textwidth]
          {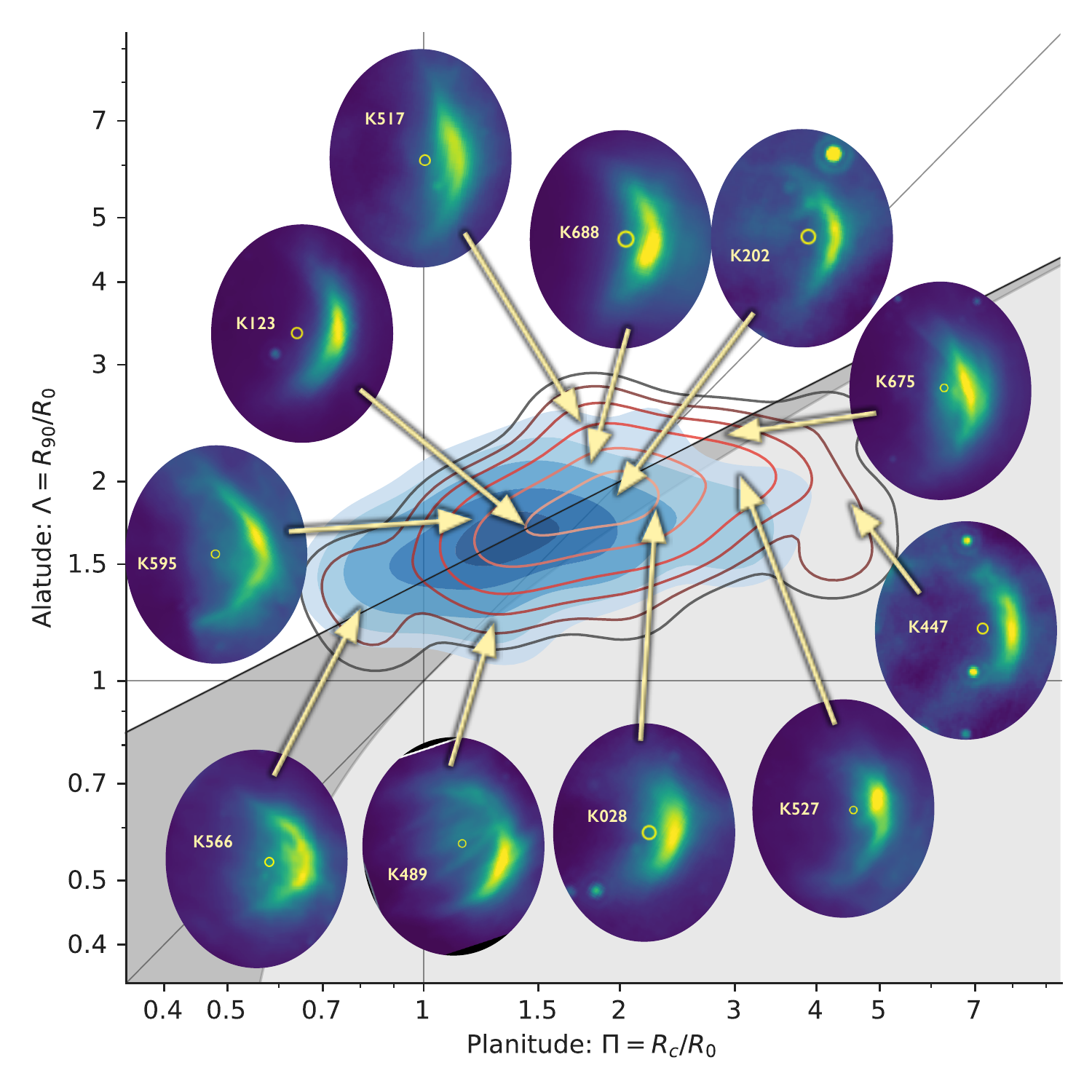} 
  \end{tabular}
  \vspace*{-\baselineskip}
  \caption[]{MIPSGAL sources on the bow shock shape diagnostic diagram
    of dimensionless radius of curvature versus perpendicular radius.
    The regions corresponding to different classes of quadrics are
    shown by shading (see \S~4 of Paper~0): oblate spheroids (light
    gray background); prolate spheroids (darker gray background);
    paraboloids (black line); hyperboloids (white background).
    (a) Individual sources with bow shock fit quality rating of 3-star
    or above.  All 5-star sources plus those 4-star sources with
    \(\Pi < 1\) are labelled with their \cite{Kobulnicky:2016a} catalog
    number.  Horizontal error bars do not directly reflect the
    uncertainty in \(\Pi\) but are instead simply the standard deviation
    from the circle fit of bowshock points \(R_k\) within
    \(\pm 45\degr\) of the axis.  Values on the vertical axis
    represent the average of \(\Lambda_{+}\) and \(\Lambda_{-}\), with thin
    vertical error bars showing the difference between \(\Lambda_{+}\) and
    \(\Lambda_{-}\), and thick vertical error bars showing the rms
    dispersion of \(R_k\) about these values for bow shock points
    within \(\pm 10\degr\) of the \(+90\degr\) and \(-90\degr\)
    directions.  (b) Kernel density estimator (KDE) of the
    distribution for 3-star sources (blue, filled contours) and 4-
    plus 5-star sources (orange/brown, unfilled contours).  The KDE
    uses an anisotropic gaussian kernel with bandwidths of
    \(0.06 \times 0.04\) in \(\log_{10}\) units. Thumbnail images of
    representative 4- and 5-star sources at different points on the
    \(\Pi\)--\(\Lambda\) plane are also shown. The angular scale of each image
    is indicated by a yellow circle of diameter \(7.5''\), centered on
    the stellar source.}
  \label{fig:mipsgal-shapes}
\end{figure*}

In the following analysis, only those sources with a 3-star or higher
rating are used.  These comprise approximately half (227 out of 471)
of all the MIPSGAL arc sources.  In some cases of poor and failed
fits, there is nothing apparently ``wrong'' with the source itself,
and it is likely that minor tweaks to the methodology would improve
matters, but we have elected not to do so, in order to maintain a
uniform methodology across all sources.

The inset of Figure~\ref{fig:mipsgal-examples}d shows histograms of
the difference between the position angle, \(\text{PA}_0\) determined
by our fits and that listed in \citet{Kobulnicky:2016a}.  Although
observational uncertainties undoubtedly contribute in part, the
differences are mainly due to real asymmetries in the bow shocks,
especially for the 4- and 5-star sources.  The
\citeauthor{Kobulnicky:2016a} catalog \(\text{PA}_0\) values are
mostly sensitive to the orientation of the bow shock wings, whereas
our fitted \(\text{PA}_0\) values are determined by the point in the
bow shock head that is closest to the stellar source.  For this
reason, we use the catalog \(\text{PA}_0\) values for defining the
axis when measuring \(R_{90+}\) and \(R_{90-}\). On the other hand,
the fitted values of \(\text{PA}_0\) are better correlated with the
position of the bow shock's brightness peak, as is apparent in the
lower left panels of Figure~\ref{fig:mipsgal-examples}a and c.

\subsection{OB bow shock shapes on the diagnostic plane}
\label{sec:ob-shapes}

The derived bowshock shapes of all the 3-, 4-, and 5-star sources are
shown in Figure~\ref{fig:mipsgal-shapes} on the \(\Pi\)--\(\Lambda\) plane,
which was discussed extensively in Paper~0.
Figure~\ref{fig:mipsgal-shapes}a shows each individual source with
horizontal error bars that represent the dispersion of traced points
from the circle fit (step~\ref{step:Rc} of the previous section) and
vertical error bars that show the wing asymmetry \(\Delta\Lambda\)
(step~\ref{step:Pi-Lambda}).  The horizontal error bars, which are a
proxy for observational uncertainties, become very small for the 4-
and 5-star sources, whereas the vertical error bars remain at roughly
the \(10\%\) level since they reflect real asymmetries in the shapes
of the bow shocks.

The horizontal axis corresponds to the shape of the head of the bow
shock near its apex, ranging from sharper, pointier shapes with
\(\Pi < 1\) to flatter, snubber shapes with \(\Pi \gg 1\), where it must be
understood that all judgments of sharpness/flatness are with respect
to the axial separation, \(R_0\), between the source and the bow shock
apex (see Fig.~\ref{fig:obs-shape-terminology}).  The vertical axis
corresponds to the shape of the bow shock wings, ranging from closed
``C'' shapes for smaller values of \(\Lambda\) to open ``V'' shapes for
larger values of \(\Lambda\).  The background shading shows the regions of
the \(\Pi\)--\(\Lambda\) plane occupied by simple quadric shapes (see \S~4 of
Paper~0): oblate spheroids (light gray), prolate spheroids (dark
gray), and hyperboloids (white). The boundary between closed and open
corresponds to the paraboloids, and is shown by the solid line
that divides the dark gray and white regions of the graph.

In Figure~\ref{fig:mipsgal-shapes}b, we present a smoothed version of
the same data.  The contours show the kernel density estimator (KDE,
see \citealp{Leiva-Murillo:2012a, Scott:2015a}) of the two-dimensional
distribution of points on the \(\Pi\)--\(\Lambda\) plane.  The KDE contours
indicate that the distribution of 3-star sources is very similar to
that of 4- and 5-star sources, although the higher-rated sources are
shifted slightly to the upper right.  Possible reasons for this are
discussed in \S~\ref{sec:corr-shape} below.
The bulk of the sources are concentrated around the paraboloid line,
with \(1 < \Pi < 3\), and \(1.2 < \Lambda < 2\).  But significant minorities
are found in three other regions: (1)~a clump with \(\Pi \la 1\); (2)~a
vertical spur towards higher \(\Lambda\) at \(\Pi \approx 1.8\); and (3)~a broad
horizontal tail towards higher \(\Pi\) at \(\Lambda \la 2\).  Thumbnail images
of the \SI{24}{\um} emission of selected sources from different
regions of the distribution are shown, in order to illustrate the
range of bow shock morphologies encountered.

\subsection{Correlations between non-shape parameters of the OB sources}
\label{sec:corr-size}

\begin{figure*}
  \centering
  \includegraphics[width=0.86\textwidth]{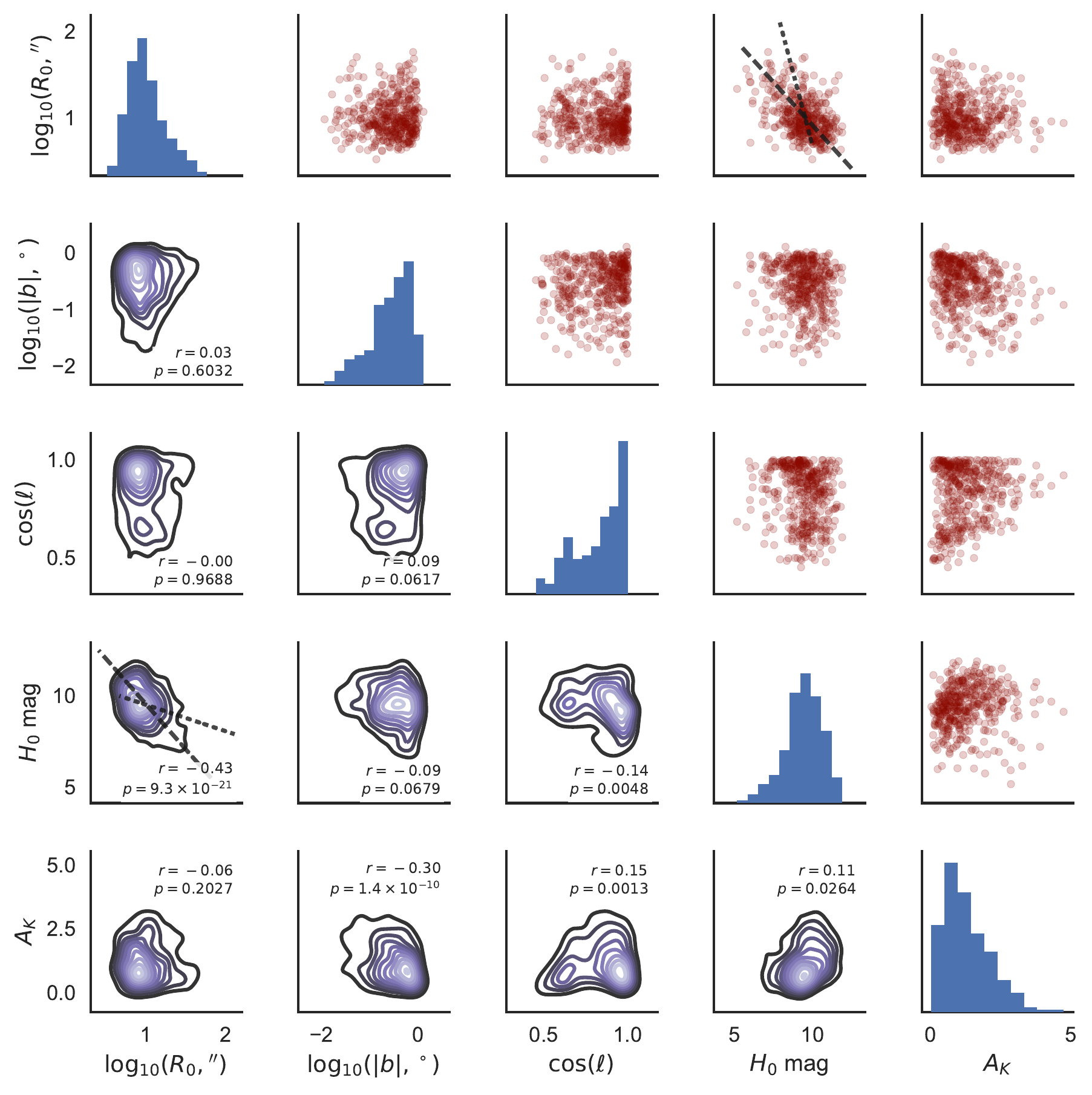}
  \vspace*{-\baselineskip}
  \caption[]{Matrix of pair plots that illustrate distributions of and
    correlations between the non-shape parameters of all MIPSGAL bow
    shock sources from \citet{Kobulnicky:2016a}.  Plots on the leading
    diagonal show histograms of the following parameters: bow shock
    angular size, \(\log_{10} R_0\); Galactic latitude,
    \(\log_{10}|b|\); Galactic longitude, \(\cos \ell\);
    extinction-corrected \(H\)-band magnitude of the stellar source,
    \(H_0\); \(K\)-band extinction, \(A_K\).  Scatter plots in the
    upper triangle show the joint distribution of each pair of
    parameters.  These are repeated in the lower triangle but showing
    the KDEs of the joint distributions, which are annotated with the
    Pearson linear correlation coefficient, \(r\), for each pair. The
    straight lines shown superimposed on the plots of stellar magnitude
    versus bowshock size correspond to toy model results for the same star
    at a sequence of distances (dashed lines) and a sequence of
    stellar luminosities at a fixed distance (dotted lines).  See text
    for details. }
  \label{fig:mipsgal-pairplot}
\end{figure*}

In Figure~\ref{fig:mipsgal-pairplot} we show the distributions over
all MIPSGAL bow shock sources of the bow shock size, Galactic
coordinates, extinction-corrected stellar source magnitude, and dust
extinction.  For the bow shock size, \(R_0\), we use the results from
our model fitting rather than the values given in the
\citet{Kobulnicky:2016a} catalog, but the distribution is very
similar, as can be seen by comparing the top-left plot of
Figure~\ref{fig:mipsgal-pairplot} with \citeauthor{Kobulnicky:2016a}'s
Figure~8.

The catalog gives the \(K\)-band extinction, \(A_K\), derived using
the method of \citet{Majewski:2011a}. However, that assumes an
intrinsic color of \((H - [\SI{4.5}{\um}])_0 = +0.08\) magnitudes,
which is too red if the sources are assumed to be OB stars.  We
therefore re-derive \(A_K\) from the catalog magnitudes combined with
the \citet{Indebetouw:2005a} reddening law, but assuming
\((H - [\SI{4.5}{\um}])_0 = -0.1\) magnitudes, which is more typical
of early type stars.  This does not make very much difference (compare
the top-right plot of our Fig.~\ref{fig:mipsgal-pairplot} with
\citeauthor{Kobulnicky:2016a}'s Fig.~9), but it does eliminate some of
the apparent negative extinctions that are found in the catalog.  The
same reddening law gives \(A_H = 1.55 A_K\), and this is used to
derive extinction-corrected \(H\)-band apparent magnitudes, \(H_0\).

\begin{figure*}
  \setlength\tabcolsep{0pt}
  \begin{tabular}{ll}
    (a) & (b) \\
    \includegraphics[width=0.5\textwidth]{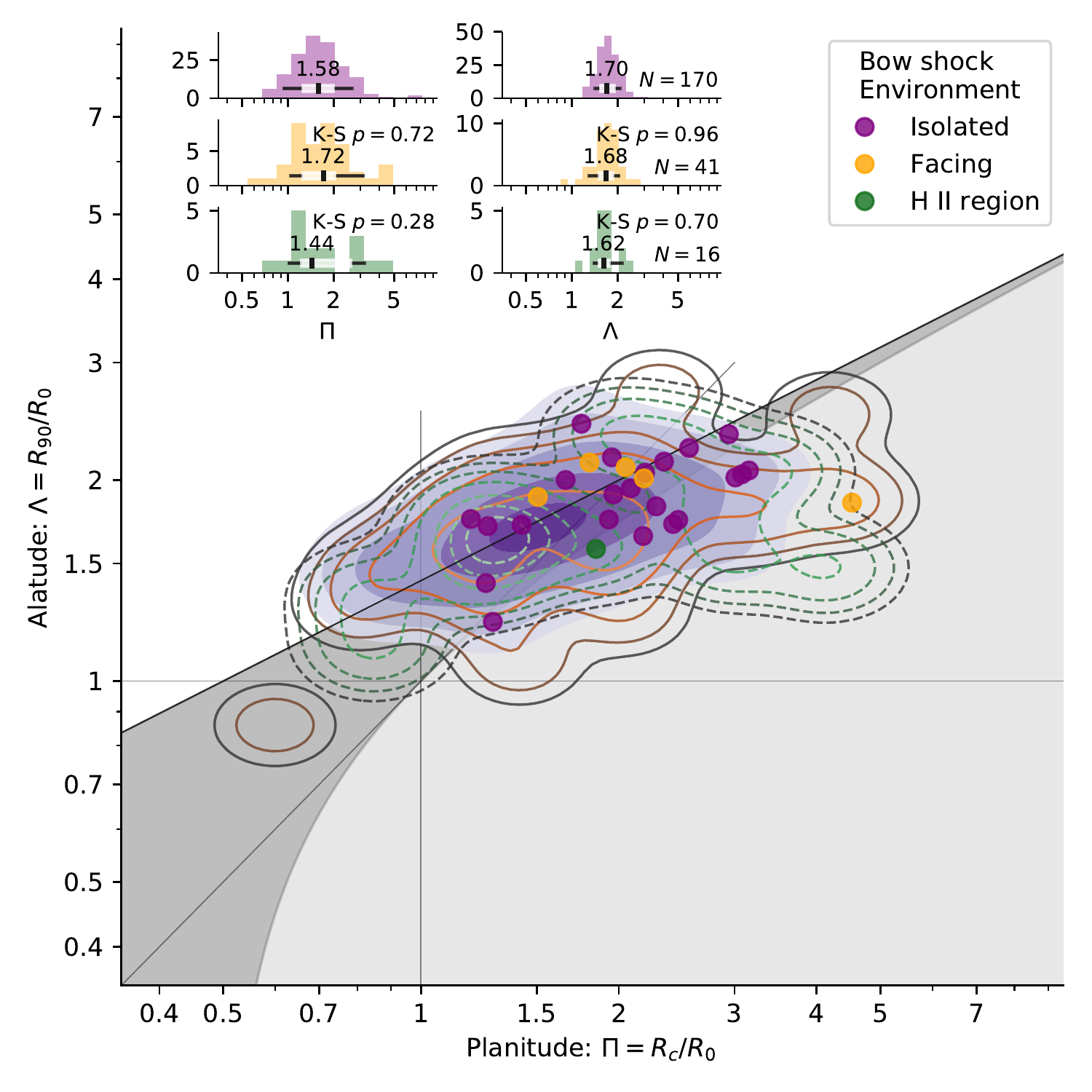} &
    \includegraphics[width=0.5\textwidth]{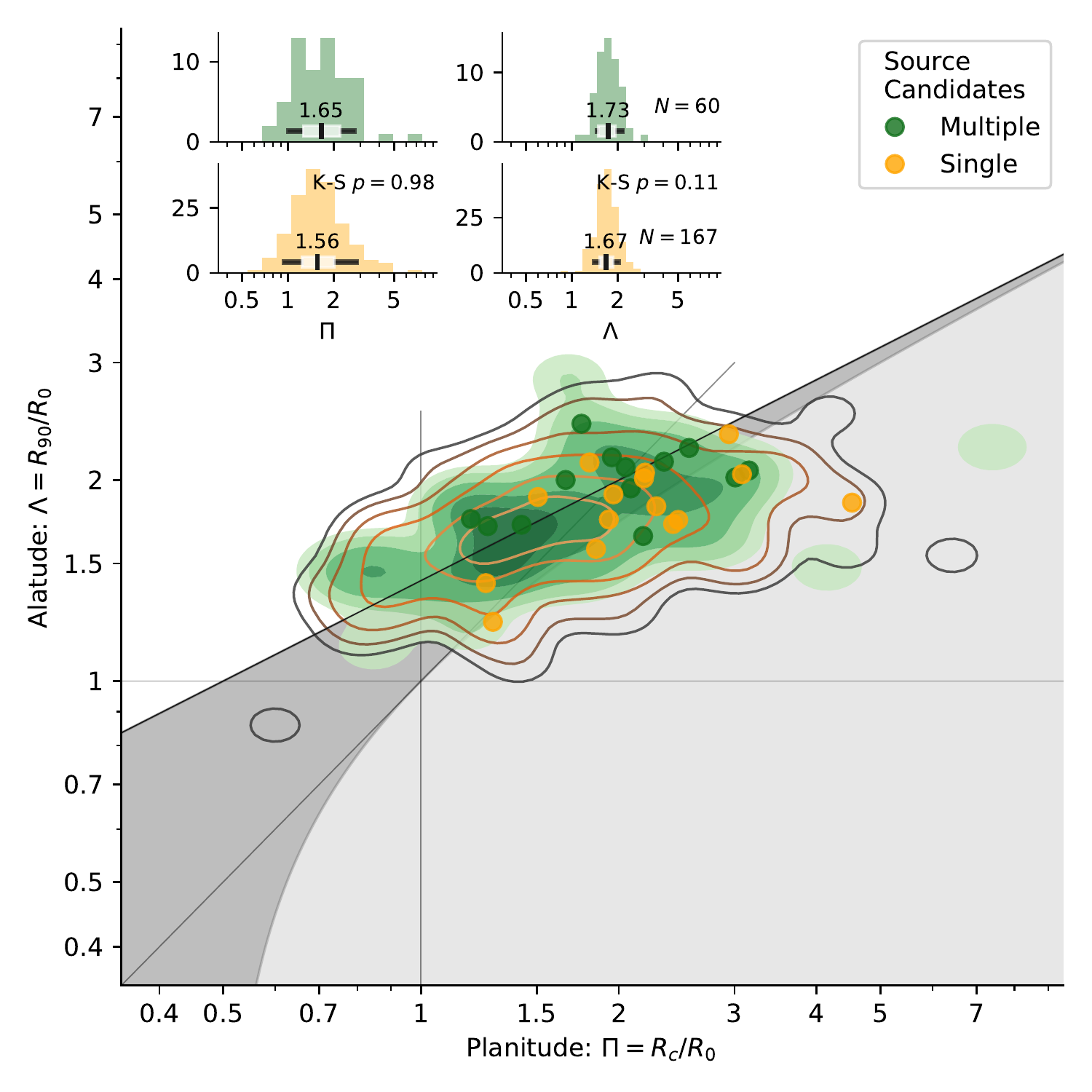} 
  \end{tabular}
  \vspace*{-\baselineskip}
  \caption[]{Example comparisons between the distribution of bowshock
    shapes when the sources are divided into two or more sub-samples
    according to the value of a categorical parameter.  Contours show
    the KDE of the distribution of each sub-sample for all 3-, 4-, and
    5-star sources, while filled circle symbols show 5-star sources
    only. Inset histograms show the marginal distributions on the two
    shape axes. (a)~Source environment, divided into three
    sub-samples: ``Isolated'' (purple symbols and purple filled
    contours), ``Facing \hii{} region or \SI{8}{\um} bright-rimmed
    cloud'' (orange symbols and orange-brown hollow continuous
    contours), and ``Within \hii{} region'' (green symbols and green
    hollow dashed contours). (b)~Uncertainty in stellar source
    identification, divided into two sub-samples: ``Multiple
    candidates for stellar source'' (green symbols and green filled
    contours) and ``Single candidate for stellar source'' (orange
    symbols and orange-brown hollow continuous contours).}
  \label{fig:mipsgal-uncorrelated}
\end{figure*}
\begin{figure*}
  \setlength\tabcolsep{0pt}
  \begin{tabular}{ll}
    (a) & (b) \\
    \includegraphics[width=0.5\textwidth]{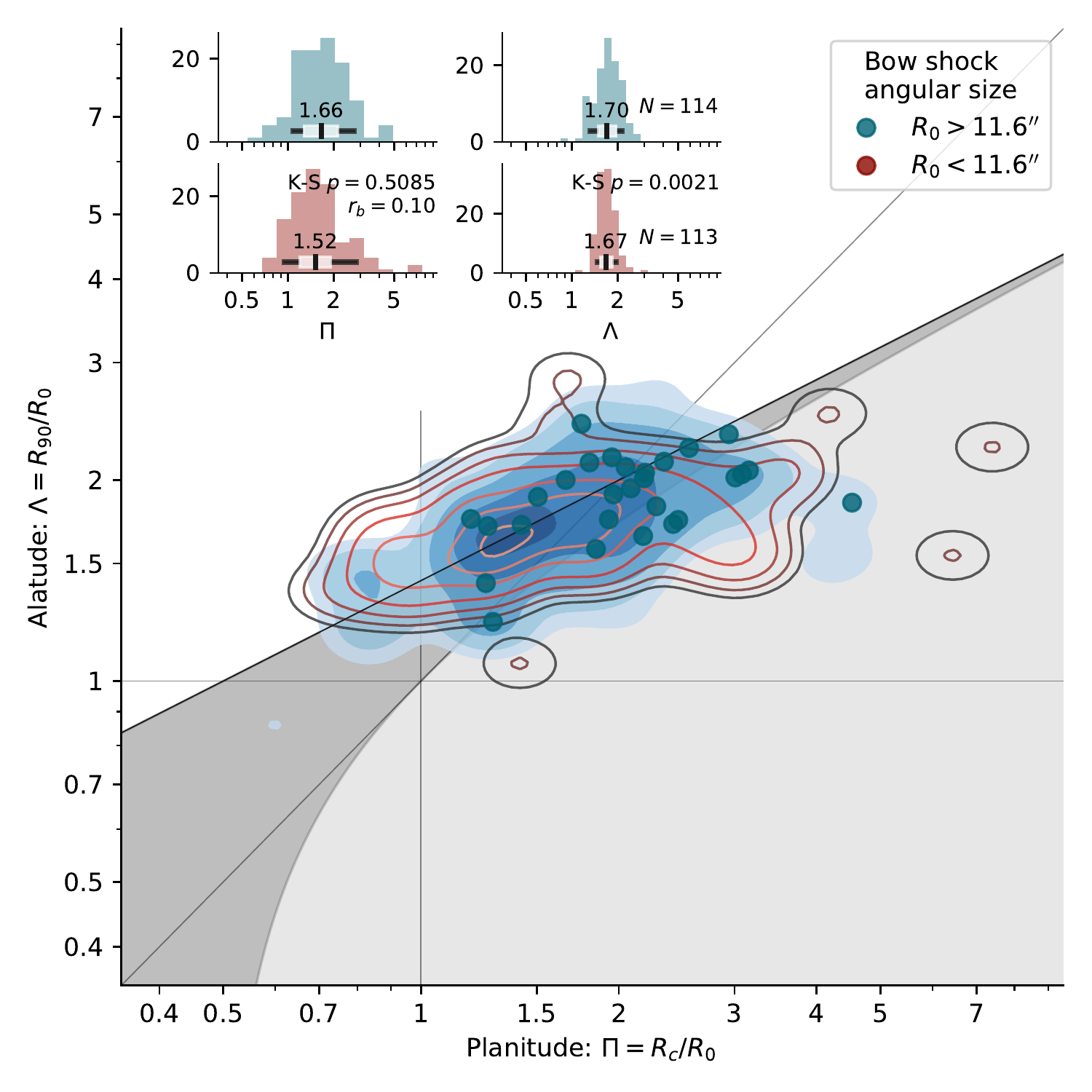} &
    \includegraphics[width=0.5\textwidth]{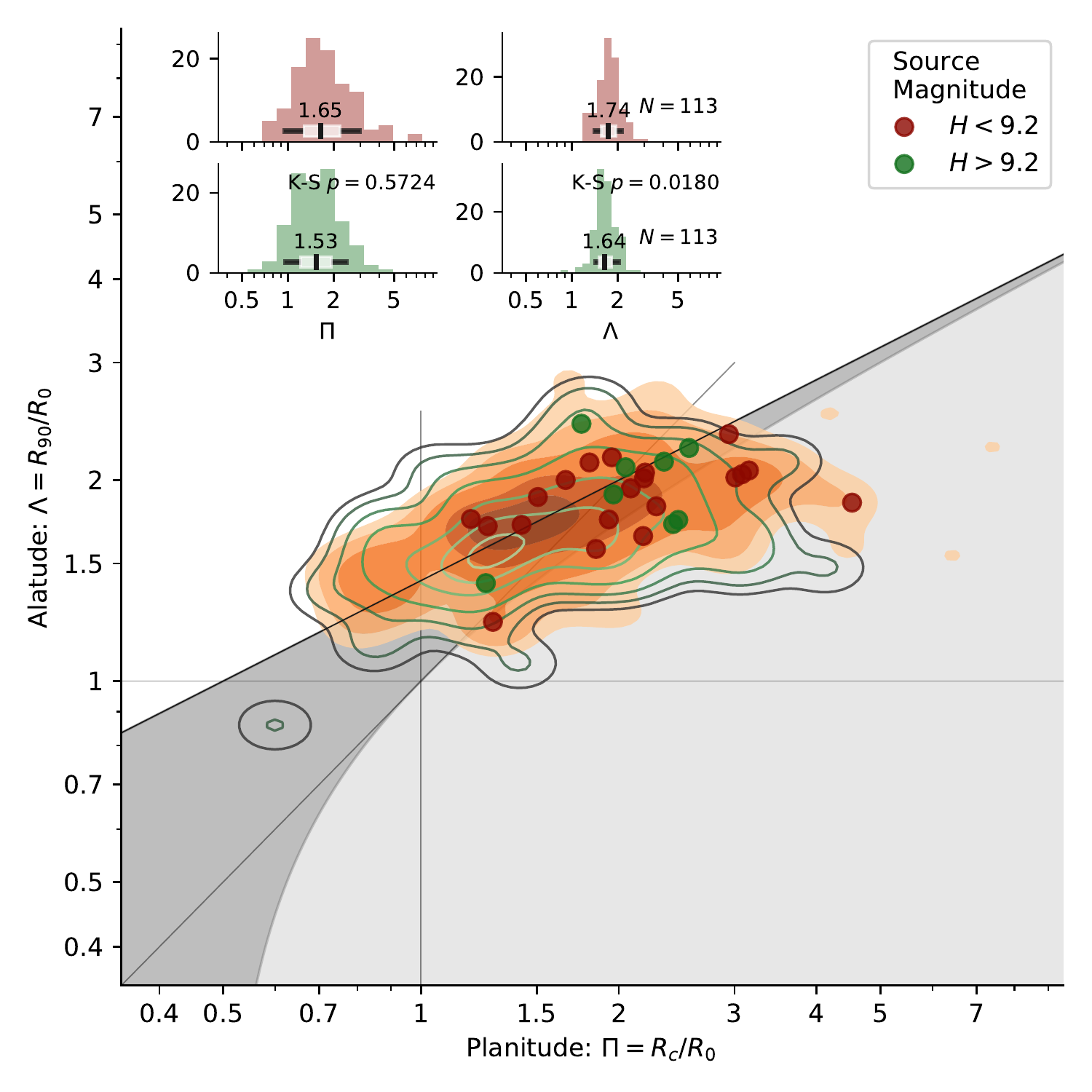} 
  \end{tabular}
  \vspace*{-\baselineskip}
  \caption[]{As Fig.~\ref{fig:mipsgal-uncorrelated} but for median
    splits of continuous parameters. (a)~Bow shock angular size,
    \(R_0\), divided into two equal-sized sub-samples: large (blue
    symbols and blue filled contours) and small (red hollow contours).
    (b)~Extinction-corrected \(H\)-band magnitude of the stellar
    source, divided into two equal-sized sub-samples: bright (red
    symbols and orange filled contours) and faint (green symbols and
    green hollow contours).}
  \label{fig:mipsgal-correlated}
\end{figure*}

The most significant linear correlation between any pair of parameters
in Figure~\ref{fig:mipsgal-pairplot} is that between bow shock size
and stellar source brightness: \(H_0\) versus \(\log_{10} R_0\), with
correlation coefficient \(r = -0.43\).  The distribution of \(H_0\)
depends on the absolute magnitude, \(M_H\), and the distance, \(d\),
to the source.  It is likely that variation in \(d\) is the more
important of the two because \(M_H\) changes relatively little for
main-sequence OB stars, ranging from \(M_H \approx -4\) (early-O) to
\(M_H \approx -1.5\) (mid-B).  This is because part of the increase in
bolometric luminosity, \(L\), as one ascends the main sequence is
offset by an increase in the effective temperature,
\(T_{\text{eff}}\), which shifts the peak of the stellar spectrum
farther away from the \(H\)~band, resulting in
\(L_H \propto L / T_{\text{eff}}^3 \sim L^{0.53}\), where the last step uses
the upper main-sequence mass--luminosity and mass--radius scalings from
\citet{Eker:2015a}. It is true that evolved OB supergiants can be much
brighter, reaching \(M_H \approx -7\), but such stars are expected to be
relatively rare.  Assuming a B2V star (\(M_H = -2\)), then the
observed range \(H_0 = 5\)--\(12\) corresponds to distances
\(d = 100\)--\SI{6300}{pc}, and the histogram peak at
\(H_0 \approx 9.5\) corresponds to \(d \approx \SI{2000}{pc}\), which is all
perfectly reasonable.

Turning now to the distribution of bow shock angular size, \(R_0\),
this will also be affected by distance to at least some degree, since
for a constant physical size the angular size will vary as
\(R_0 \propto d^{-1}\).  For instance, if we assume that the physical size
of all bow shocks is \SI{0.1}{pc} and the absolute magnitude of all
stars is \(M_H = -2\), as above, then we find the relation
\(H_0 = 14.57 - 5 \log_{10} R_0\) if \(R_0\) is measured in
arcseconds. This is shown as a dashed line on the relevant panels of
Figure~\ref{fig:mipsgal-pairplot} for values of \(R_0\) that
correspond to \(d = \SI{300}{pc}\) to \SI{8000}{pc}.  It can be seen
that this relation is in excellent agreement with the linear trend in
the data. On the other hand, the correlation coefficient of
\(r = -0.43\) means in broad terms that only a fraction
\(r^2 \approx 20\%\) of the total variance in \(H_0\) is ``explained'' by
changes in \(R_0\), and vice~versa, implying that one or both of
\(H_0\) and \(R_0\) is only a very imperfect proxy for \(d\).  We have
already seen that the spread in \(H_0\) probably \emph{is} mostly due
to a spread in distance, rather than a spread in \(H\)-band stellar
luminosity.  If this is true, it follows that it is \(R_0\) that
depends only weakly on \(d\) and is more influenced by other factors.

One such factor is the stellar/environmental momentum-loss ratio,
\(\beta\), between the two supersonic flows that form the bow shock.  All
other things being equal, we have \(R_0 \propto \beta^{1/2}\) for
\(\beta \ll 1\), as is typically the case.  If the environment flow is
constant and the OB star wind has mass-loss rate \(\dot{M}\) and
terminal velocity \(V_\infty\), then
\(\beta \propto \dot{M}{V_\infty}\).  Empirical and theoretical studies of hot star
winds (e.g., \citealp{Puls:1996a}) imply
\(\dot{M}{V_\infty} \sim L^{1.88} R^{-1/2} \sim L^{1.80} \sim L_H^{3.40}\), where
the final two steps apply only to main-sequence stars and again use
the relations of \citet{Eker:2015a}.  If we assume as above that a B2V
star with \(M_H = -2\) has a bow shock physical size of \SI{0.1}{pc},
and consider a sequence of stars with varying \(H\)-band luminosities
but all at a fixed distance of \(d = \SI{2000}{pc}\), then we find the
relation \(H_0 = 8.02 - 1.49 \log_{10} R_0\).  This is shown as a
dotted line in the relevant panels of
Figure~\ref{fig:mipsgal-pairplot} for the absolute magnitude range
\(M_H = -3.6\) (O6V) to \(M_H = -1.5\) (B5V).  It can be seen that
this relation does not match the linear trend in the data, and
predicts a much larger spread in \(R_0\) over a narrow range in
\(H_0\) than is observed.  This could mean one of two things: first,
it may be that the range of stellar luminosities is significantly
narrower than we have supposed, implying that B stars vastly outnumber
O stars among the sources.  Alternatively, there may be a positive
correlation between the stellar luminosity and the momentum of the
environmental flow, with the result that \(\beta\) varies less steeply
with \(L_H\) than we have assumed.  That could arise if more luminous
stars were preferentially found in denser environments, or, in the
case of runaways, if more luminous stars tended to be faster moving.

A third factor that may influence \(R_0\) is the inclination, \(i\),
of the bow shock axis with respect to the plane of the sky.
Figure~11b of Paper~0 shows that for \(\Pi > 1\), then the projected
\(R_0'\) becomes larger than the true \(R_0\) as \(|i|\) increases
(see also Figs.~26 and 27 of Paper~0).  It can be seen that the effect
is relatively modest, with an increase in \(R_0\) of no more than a
factor of 2 to 3.

\subsection{Correlation between bow shock shape and other parameters}
\label{sec:corr-shape}

\begin{figure}
  \centering
  \includegraphics[width=\linewidth]
  {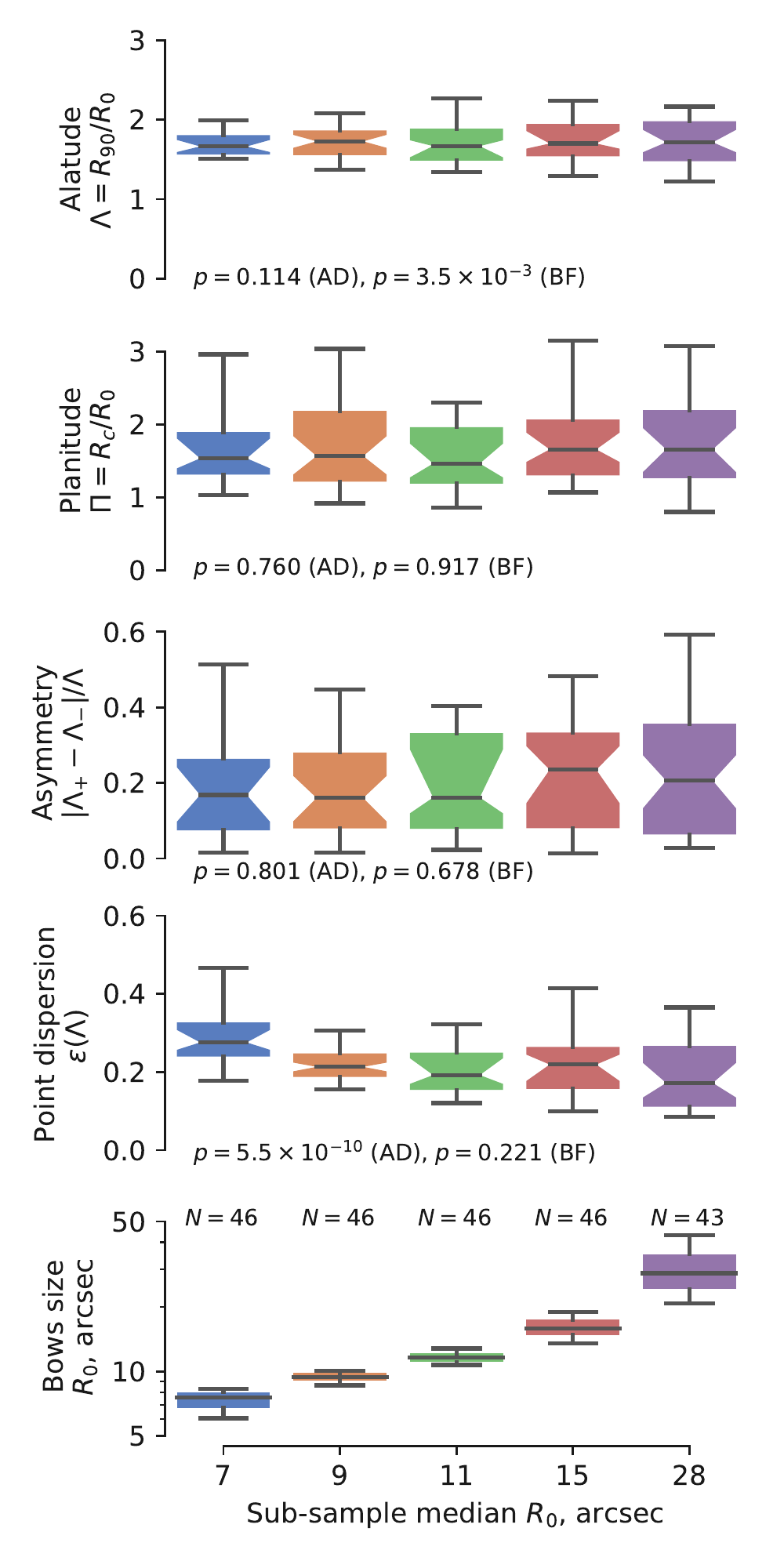}
  \vspace*{-\baselineskip}
  \caption{Box plots of the distributions of bow shock shape
    parameters, \(\Pi\) and \(\Lambda\), together with a measure of the tail
    asymmetry, \(2 |\Delta\Lambda| / \Lambda\), after partitioning on the bow shock
    size, \(R_0\).  All 3-, 4-, and 5-star sources are sorted
    according to \(R_0\) and divided into 5 non-overlapping
    sub-samples of roughly equal size, each labelled by their median
    value of \(R_0\) in arcseconds, as illustrated in the lower panel.
    The colored boxes show the interquartile range of the data, with
    the median indicated by a horizontal line and the notch showing
    the 90\% confidence limits of the median, as determined from 1000
    bootstrap resamplings.  The 5th-to-95th centile range of the data
    is indicated by error bars.  A 5-sample Anderson--Darling test and
    Brown--Forsythe test is performed for each dependent variable, with
    resultant \(p\)-value given at the bottom of each panel.  It is
    apparent that the dispersion in \(\Lambda\) (upper panel) increases
    systematically with \(R_0\), although the central value is roughly
    constant.  No clear systematic changes are apparent in \(\Pi\)
    (second panel).}
  \label{fig:mipsgal-boxplot}
\end{figure}

We now investigate if the bow shock shapes of the MIPSGAL sources are
correlated with any other parameters via the following methodology:
\begin{enumerate}[1.]
\item For each of the parameters in the \citet{Kobulnicky:2016a}
  catalog, we divide the sources into two or more sub-samples,
  according to the value of the parameter.  For quantitative
  parameters, such as those discussed in the previous section, we use
  two sub-samples of equal size, with membership determined by whether
  the parameter is larger or smaller than the median value.  But we
  also use categorical parameters, such as the type of bow shock
  environment, or the presence/absence of \SI{8}{\um} emission, in
  which case the sub-samples are of unequal size.
\item We plot the KDEs of the two sub-samples separately on the
  \(\Pi\)--\(\Lambda\) plane (see Figs.~\ref{fig:mipsgal-uncorrelated} and
  \ref{fig:mipsgal-correlated}) to check for any obvious
  differences.
\item We check if there is any statistically significant difference
  between the bow shock shapes of the sub-samples by applying three
  different non-parametric tests to \(\Pi\) and \(\Lambda\)
  separately.  
\end{enumerate}

The principal test used is Kuiper's variant of the two-sample
Kolmogorov--Smirnov test \citep[e.g.,][]{Stephens:1970a, Paltani:2004a}, as implemented by the Python library function \texttt{astropy.stats.kuiper\_two} \citep{Astropy-Collaboration:2018a}.   This is a
general test of the \textit{null hypothesis} that the two sub-samples
are drawn from the same distribution.  It returns a \(p\)-value, which
is the estimated probability that the observed difference between the
two sub-samples would be as large as it is purely by chance if they
were all were drawn from the same distribution.  We consider two
different thresholds for significance: \(p < 0.05\) (approximately
2-\(\sigma\) for a normal distribution) and \(p < 0.003\) (approximately
3-\(\sigma\)).  We show in Appendix~\ref{sec:distr-p-values} that, given
the large number of potential relationships that we are testing, the
more stringent \(p < 0.003\) condition is required in order to
confidently reject the null hypothesis and declare a ``significant''
difference between the two sub-samples, but we also consider the less
strict threshold of \(p < 0.05\) as an indicator of ``possible''
difference.  We supplement the general-purpose Kuiper test
with two tests that probe specific features of the sample
distributions: the Mann--Whitney--Wilcoxon \(U\) test
\citep{Mann:1947a}, which is sensitive to differences in the central
value (e.g., median) and the Brown--Forsythe test \citep{Brown:1974a},
which is sensitive to differences in the variance, or width, of the
distribution (see Appendix~\ref{sec:distr-p-values} for details).
  
As shown in detail in Table~\ref{tab:big-p}, there is remarkably
little variation in the bow shock shape distributions as a function of
most of the other parameters.  Two examples in which there is
\textit{no} significant shape difference between the sub-samples are
shown in Figure~\ref{fig:mipsgal-uncorrelated}.  This lack of
difference is interesting because in both examples there are a priori
grounds to suspect that differences might exist.  The first example
(Fig.~\ref{fig:mipsgal-uncorrelated}a) is the bow shock environment,
which was categorized in \citet{Kobulnicky:2016a} as ``Isolated'' (I),
``Facing a large \hii{} region'' (FH), ``Facing a \SI{8}{\um}
bright-rimmed cloud'' (FB), and ``Within a giant \hii{} region'' (H),
and where we have merged the FH and FB categories, labelled ``Facing''
in the figure.\footnote{Similar results are also found for the FB and
  FH categories separately.}  The shapes might be expected to vary
with environment because the FB and FH categories should be associated
with ``weather vane'' interactions \citep{Povich:2008a} between the
stellar wind and a divergent photoevaporation flow.  This is expected
to give a more open bow shock than in the ``runaway'' case of
interaction of a moving star with a static environment.  In the
thin-shell approximation, the predicted shapes are a cantoid for
weather vanes and a wilkinoid for runaways, see \S~6 of Paper~0.  The
fact that no such difference is detected could be explained in one of
two ways: (i)~the momentum ratio \(\beta\) for the weather vanes could be
small, since the cantoid becomes indistinguishable from the wilkinoid
as \(\beta \to 0\), or (ii)~many of the bow shocks classified as
``Isolated'' might also be weather vanes rather than runaways.

The second example (Fig.~\ref{fig:mipsgal-uncorrelated}b) divides the
sources according to whether or not \citet{Kobulnicky:2016a} judged
there to be multiple candidates for the identity of the central star
that drives the bow shock.  If the central star were to be
misidentified in a significant number of sources, then the measured
value of \(R_0\) for those sources would be erroneous, which would
increase the scatter in both \(\Pi\) and \(\Lambda\).  The fact that no
significant difference is seen in the shape distributions between
sources with/without multiple candidates implies that such mistakes in
identification of the central star must be rare.

We also tested all the other source parameters listed in
\citeauthor{Kobulnicky:2016a}'s catalog, finding no significant shape
differences for sources with/without \SI{8}{\um} emission, with
low/high extinction, closer/farther from the Galactic plane, or
closer/farther from the Galactic center. Details are given in
Table~\ref{tab:big-p}.  In all these cases, differences in mean or
median \(\Lambda\) less than \(0.06\) and in \(\Pi\) less than
\(0.16\) are found, corresponding to rank biserial correlation
coefficients (a non-parametric dimensionless measure of the difference
between two samples, see Appendix~\ref{sec:distr-p-values}) of
\(r_b < 0.15\), which are not significant even at the 2-\(\sigma\) level.

The only parameters that \emph{do} show a possible correlation with
the bowshock shape are the bow shock angular size, \(R_0\), and the
extinction-corrected magnitude of the central star, \(H_0\), which are
illustrated in Figure~\ref{fig:mipsgal-correlated}.  The general
purpose Kuiper test indicates differences in the sub-sample
distributions of \(\Lambda\) at the 2-\(\sigma\) level for \(H_0\) and at the
3-\(\sigma\) level for \(R_0\).  It is apparent from the histograms shown
in the right-hand inset graphs of Figure~\ref{fig:mipsgal-correlated}a
that in the case of the small/large \(R_0\) sub-samples it is the
width rather than the central tendency of the distributions that is
different, which is confirmed by the more specific Brown--Forsythe
test, which indicates a difference between the sub-sample dispersions
at the 3-\(\sigma\) level.  In the case of the faint/bright \(H_0\)
sub-samples (Fig.~\ref{fig:mipsgal-correlated}b), it is less clear
what feature of the distributions differ.

In order to investigate these effects in more detail and look for
systematic trends, we divide the independent parameter (\(R_0\) or
\(H_0\)) into 5 rather than 2 equal-sized sub-samples, with results in
the case of \(R_0\) shown as box plots in
Figure~\ref{fig:mipsgal-boxplot}.  This time the \(k\)-sample
Anderson--Darling test \citep{Anderson:1952a, Scholz:1987a,
  Makarov:2017a} is used to determine the statistical significance of
any observed differences.  A systematic increase with \(R_0\) in the
dispersion of \(\Lambda\) (upper panel of Fig.~\ref{fig:mipsgal-boxplot}a)
is apparent from both the interquartile range (colored boxes) and
interdecile range (error bars).  As a check on whether observational
uncertainties might be contributing to this trend, the third and
fourth rows of box plots show the statistics for, respectively, the
fractional asymmetry of the bowshock wings and the standard deviation,
\(\epsilon(\Lambda)\), of the individual points on the bow shock that go into
determining \(\Lambda\) (see step~\ref{step:R90} of the tracing/fitting
methodology described in \S~\ref{sec:autom-trac-fitt}).  It can be
seen that neither of these quantities tends to increase with \(R_0\),
and in fact there is a significant \emph{decrease} in
\(\epsilon(\Lambda)\) with \(R_0\).  This implies that the increase with angular
size of the diversity of bow shock wing shapes is real, and not due to
observational uncertainties.  On the other hand, we find that no clear
trends are evident as a function of source magnitude \(H_0\) (not
illustrated), which leads us to suspect that the \(p < 0.05\) result
obtained for the 2-sample Kuiper test was a \emph{false positive}, as
is also supported by the analysis in Appendix~\ref{sec:distr-p-values}
and Figure~\ref{fig:histo-p-values}.

As mentioned in \S~\ref{sec:ob-shapes} there is also a shape
difference between the 3-star sources and the 4/5-star sources (see
Fig.~\ref{fig:mipsgal-shapes}b).  The \(p\)-values of statistical
tests (see Table~\ref{tab:big-p}) indicate that this is much more
significant than any correlation with the other parameters discussed
in this section (\(p < 10^{-4}\) for \(\Lambda\) and \(p < 10^{-5}\) for
\(\Pi\)).  This means that it cannot be simply due to the tendency of
the higher-rated sources to have larger angular sizes.  However, the
subjective nature of the star ratings makes this result hard to
interpret.

\section{Far-infrared arcs around late-type stars}
\label{sec:far-infrared-arcs}

\begin{figure*}
  \setlength\tabcolsep{0pt}
  \setkeys{Gin}{width=0.5\linewidth}
  \begin{tabular}{ll}
    (a) & (b) \\
    \includegraphics[trim=10 0 65 20, clip]{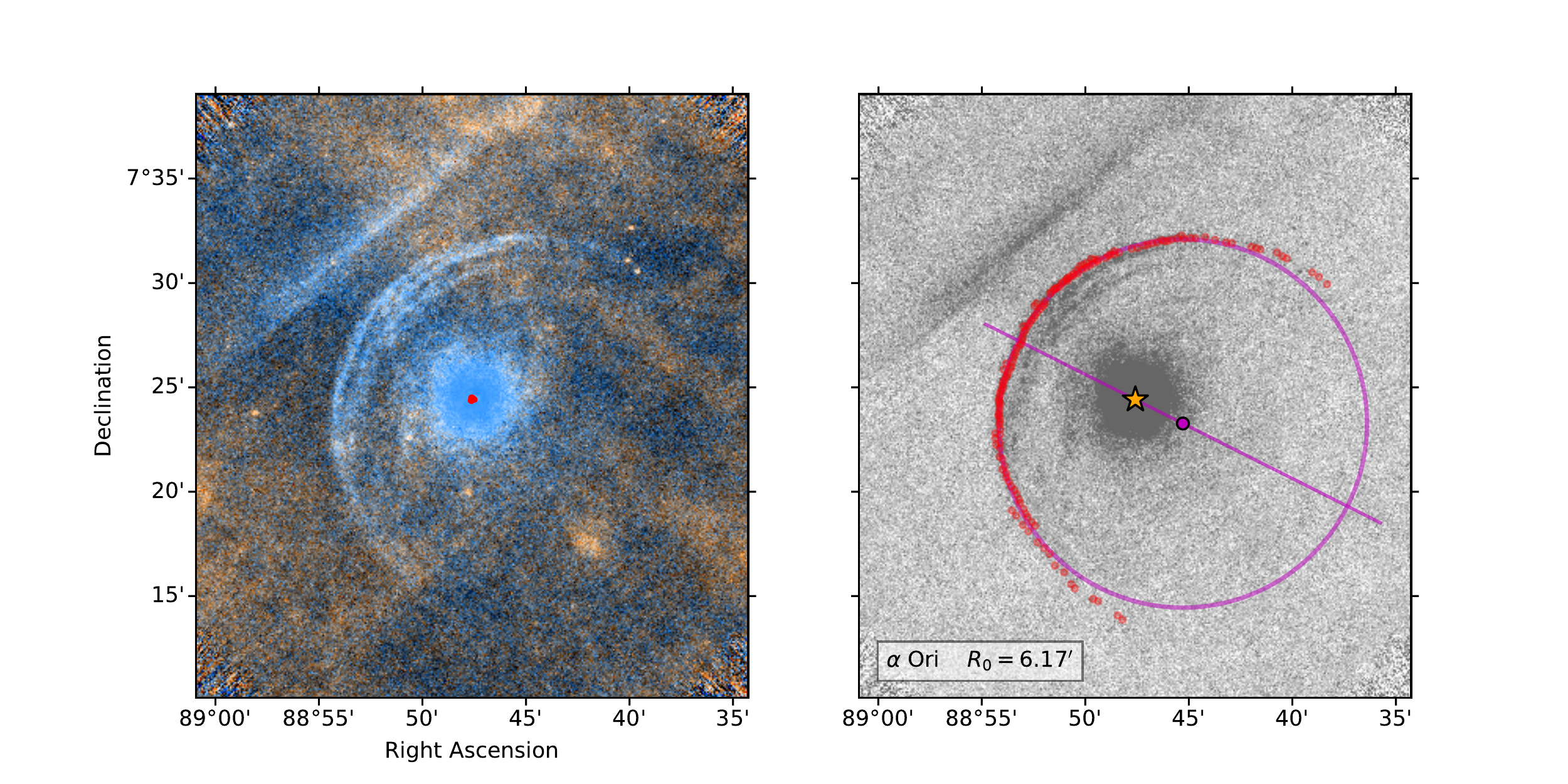}
    & \includegraphics[trim=10 0 65 20, clip]{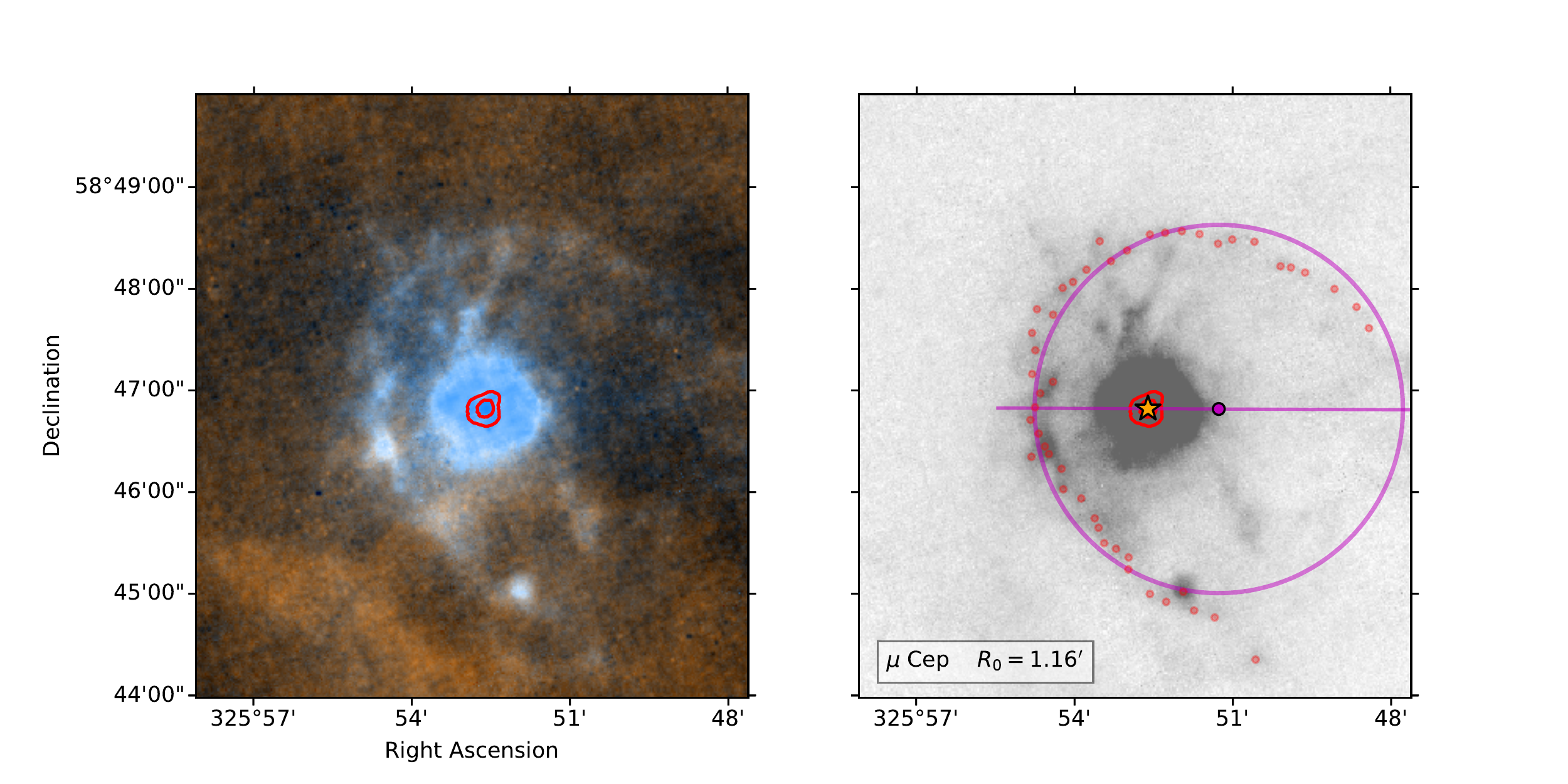}\\
    (c) & (d) \\
    \includegraphics[trim=10 0 65 20, clip]{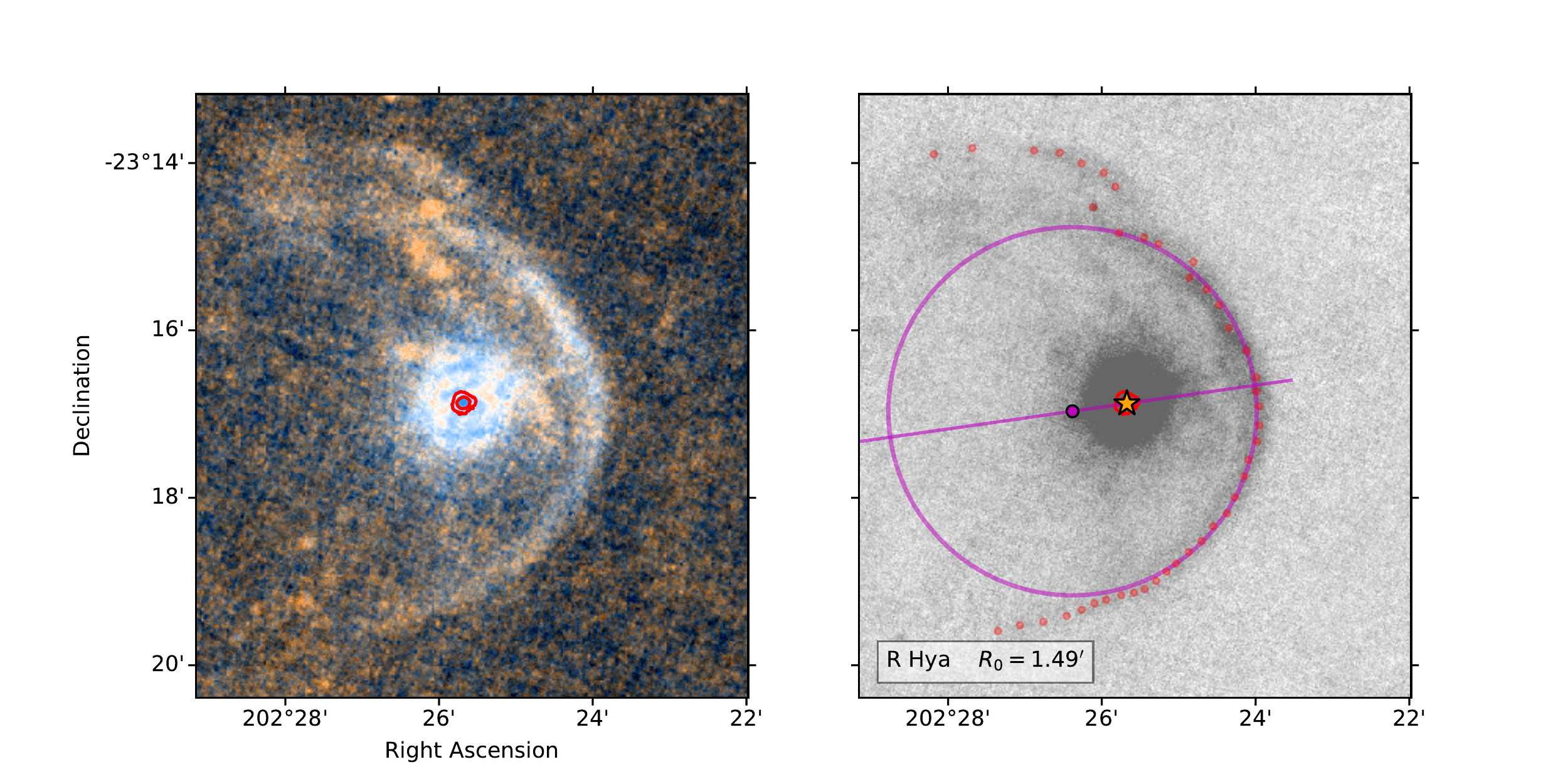}
    & \includegraphics[trim=10 0 65 20, clip]{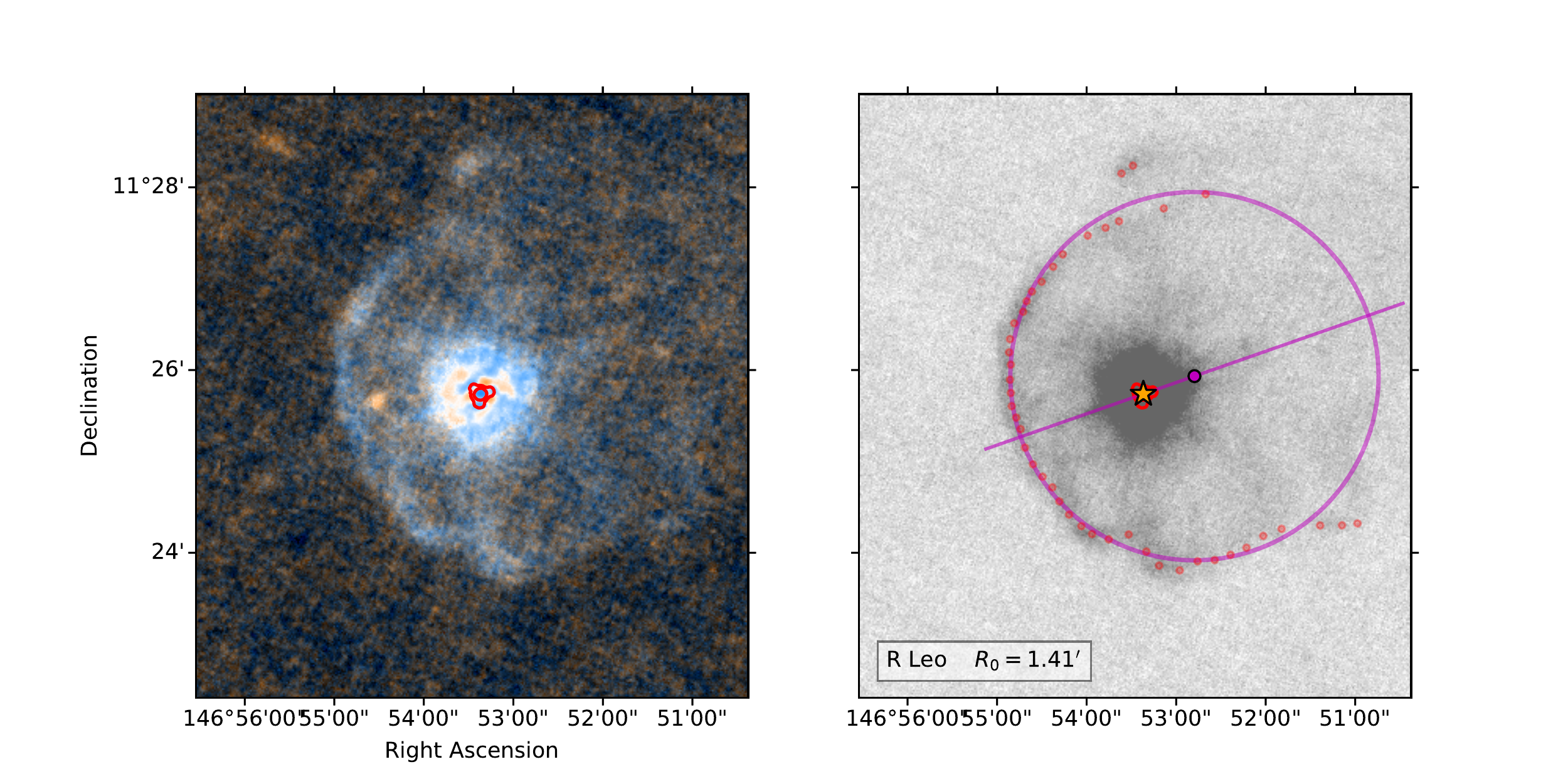}\\
    (e) & (f) \\
    \includegraphics[trim=10 0 65 20, clip]{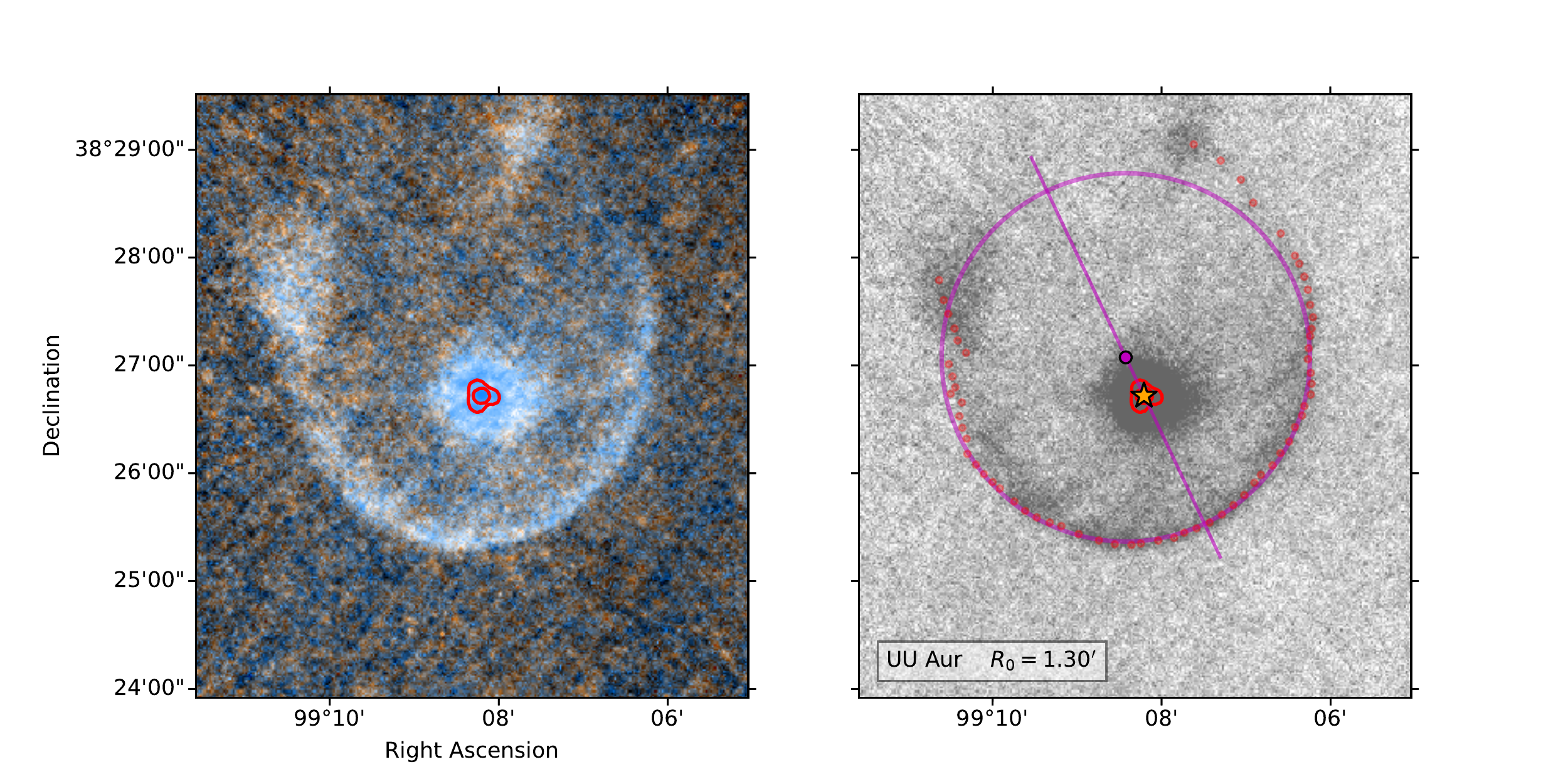}
    & \includegraphics[trim=10 0 65 20, clip]{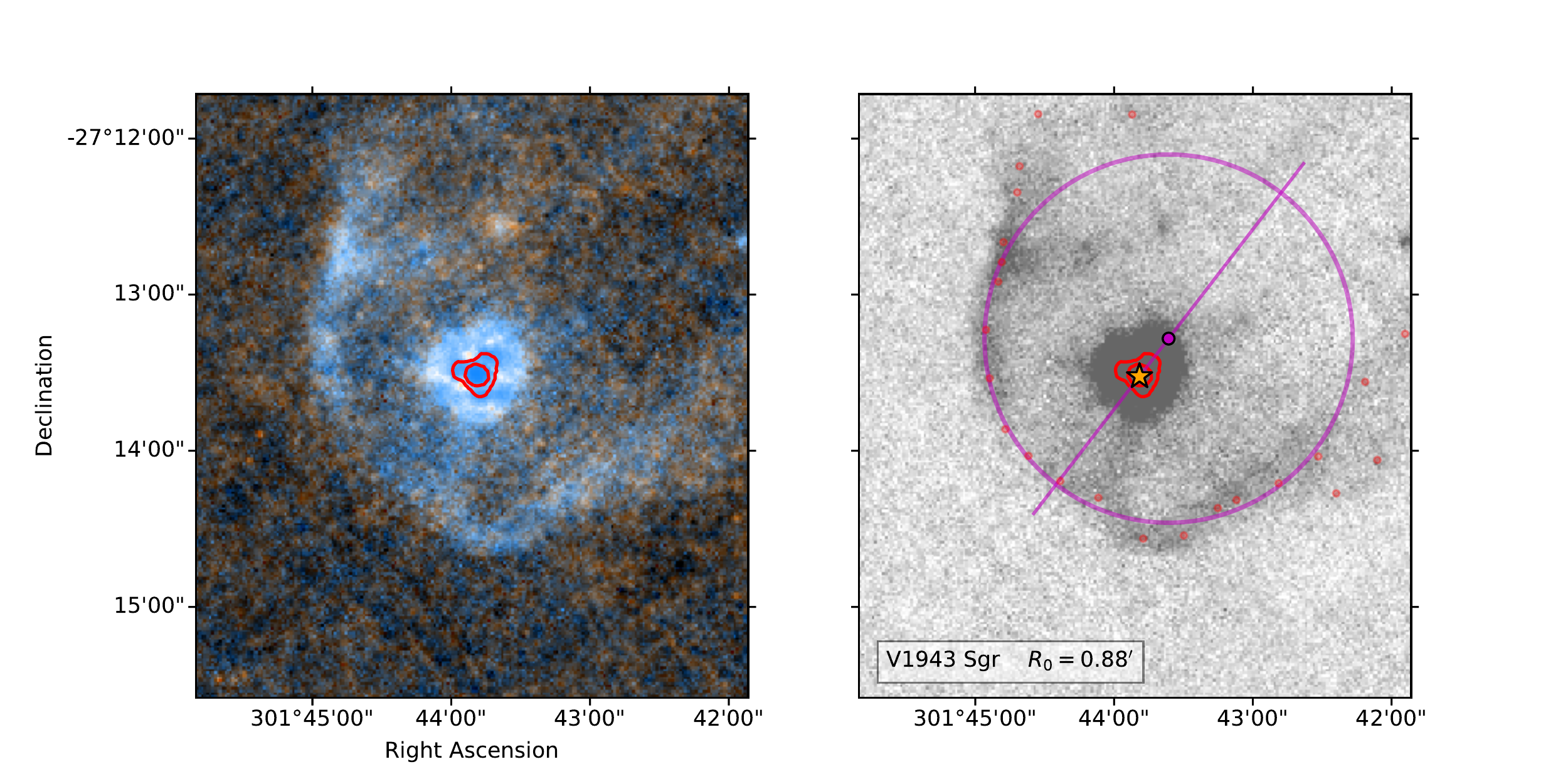}\\
    (g) & (h) \\
    \includegraphics[trim=10 0 65 20, clip]{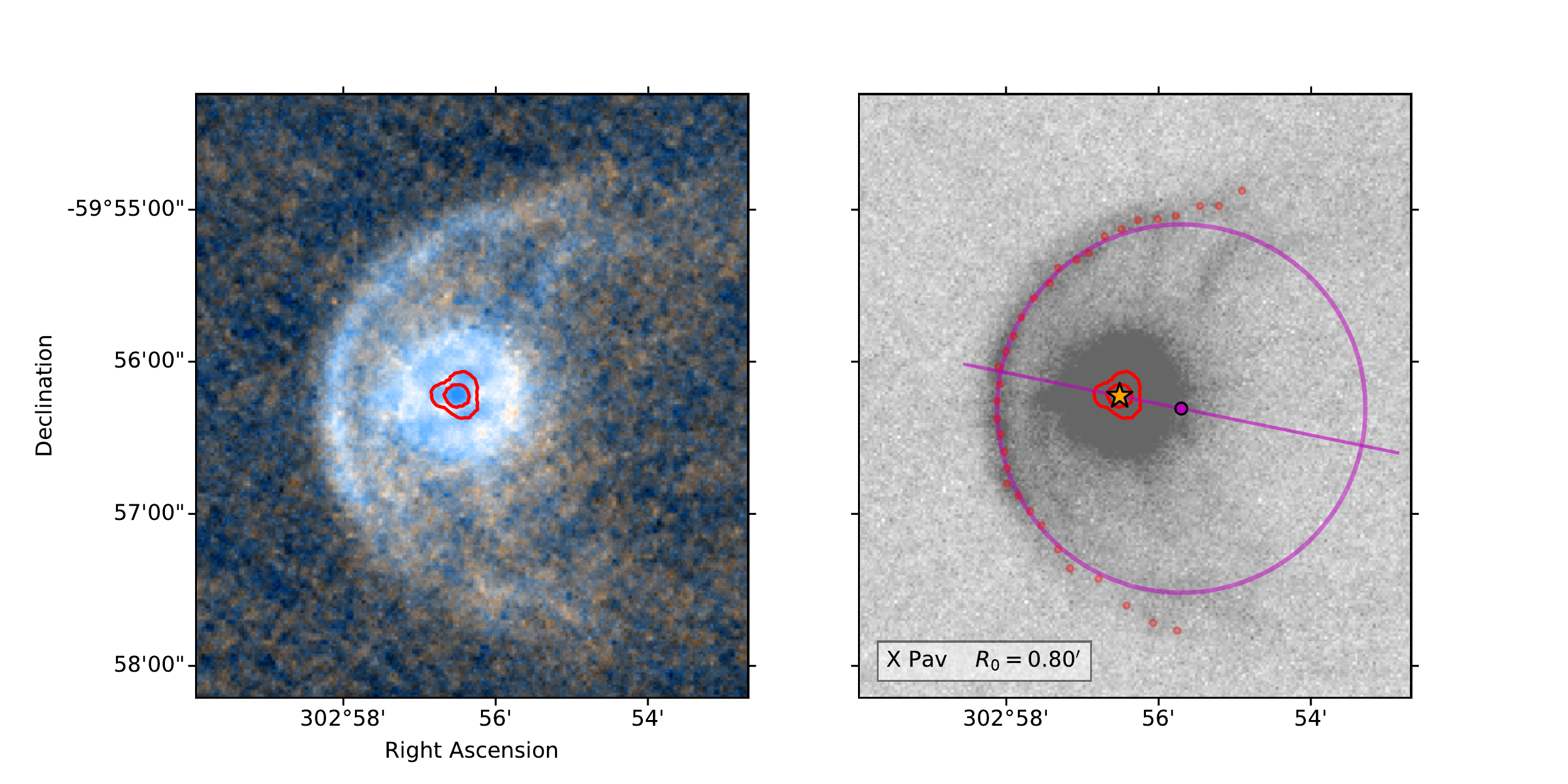}
    & \includegraphics[trim=10 0 65 20, clip]{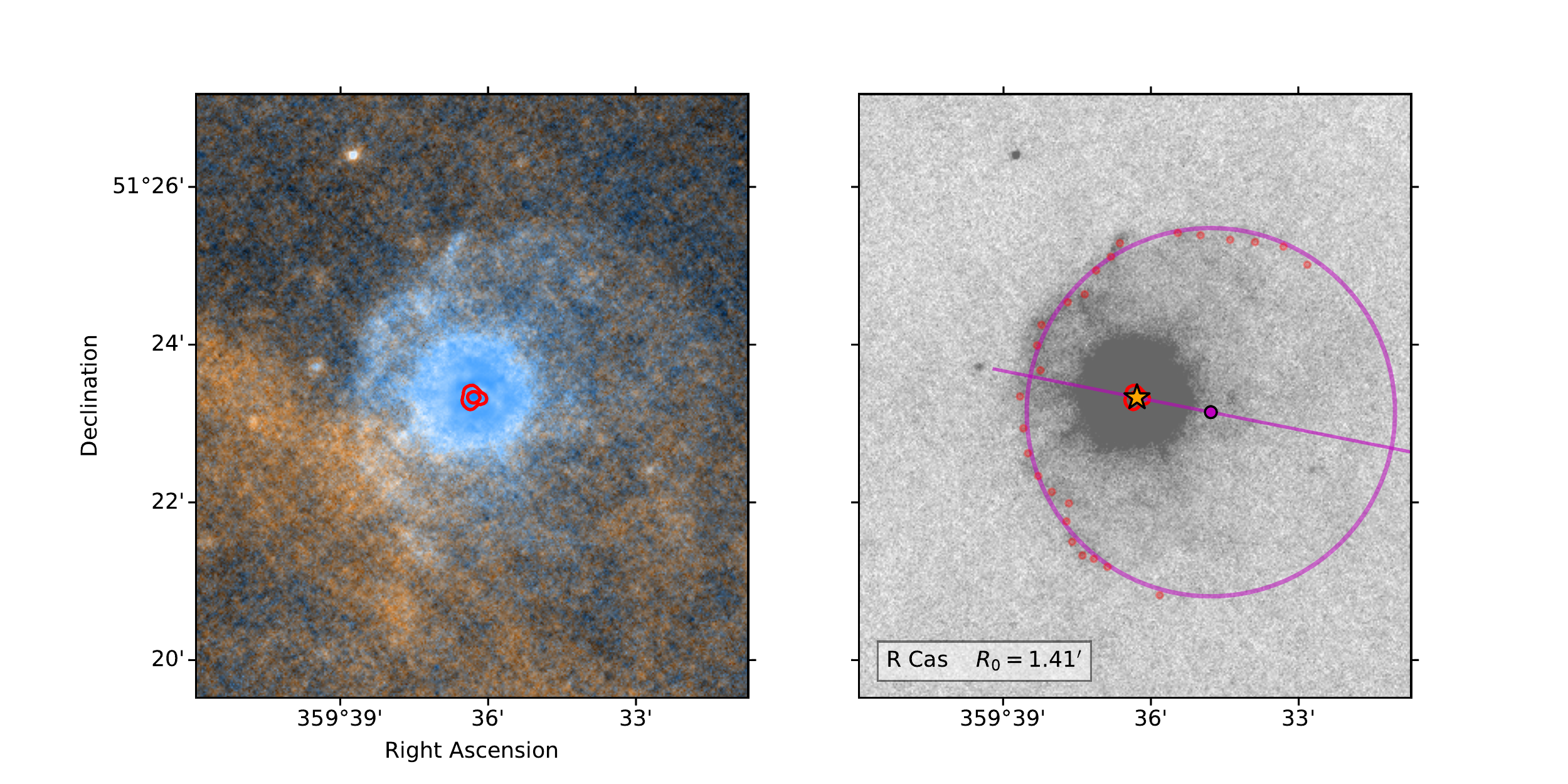}
  \end{tabular}
  \caption{RSG and AGB shells observed with Herschel.  Selected
    Class~I sources (``fermata-like'') from the MESS survey
    \citet{Cox:2012a}, where the arc structure is particularly clear
    and symmetrical.  Left panels show PACS \SI{70}{\um} surface
    brightness (blue) and \SI{160}{\um} (orange).  Right panels show
    tracing of the bow shock arc (red symbols) and circle fit (magenta
    lines and symbols) superimposed on a low-contrast image of the
    \SI{70}{\um} surface brightness. (a)~\(\alpha\)~Ori. (b)~\(\mu\)~Cep.
    (c)~R~Hya. (d)~R~Leo. (e)~UU~Aur. (f)~V1934~Sgr. (g)~X~Pav.
    (h)~R~Cas.}
  \label{fig:herschel-arc-fits}
\end{figure*}

\begin{figure*}
  \setlength\tabcolsep{0pt}
  \setkeys{Gin}{width=0.5\linewidth}
  \begin{tabular}{ll}
    (a) & (b) \\
    \includegraphics[trim=10 0 65 20, clip]{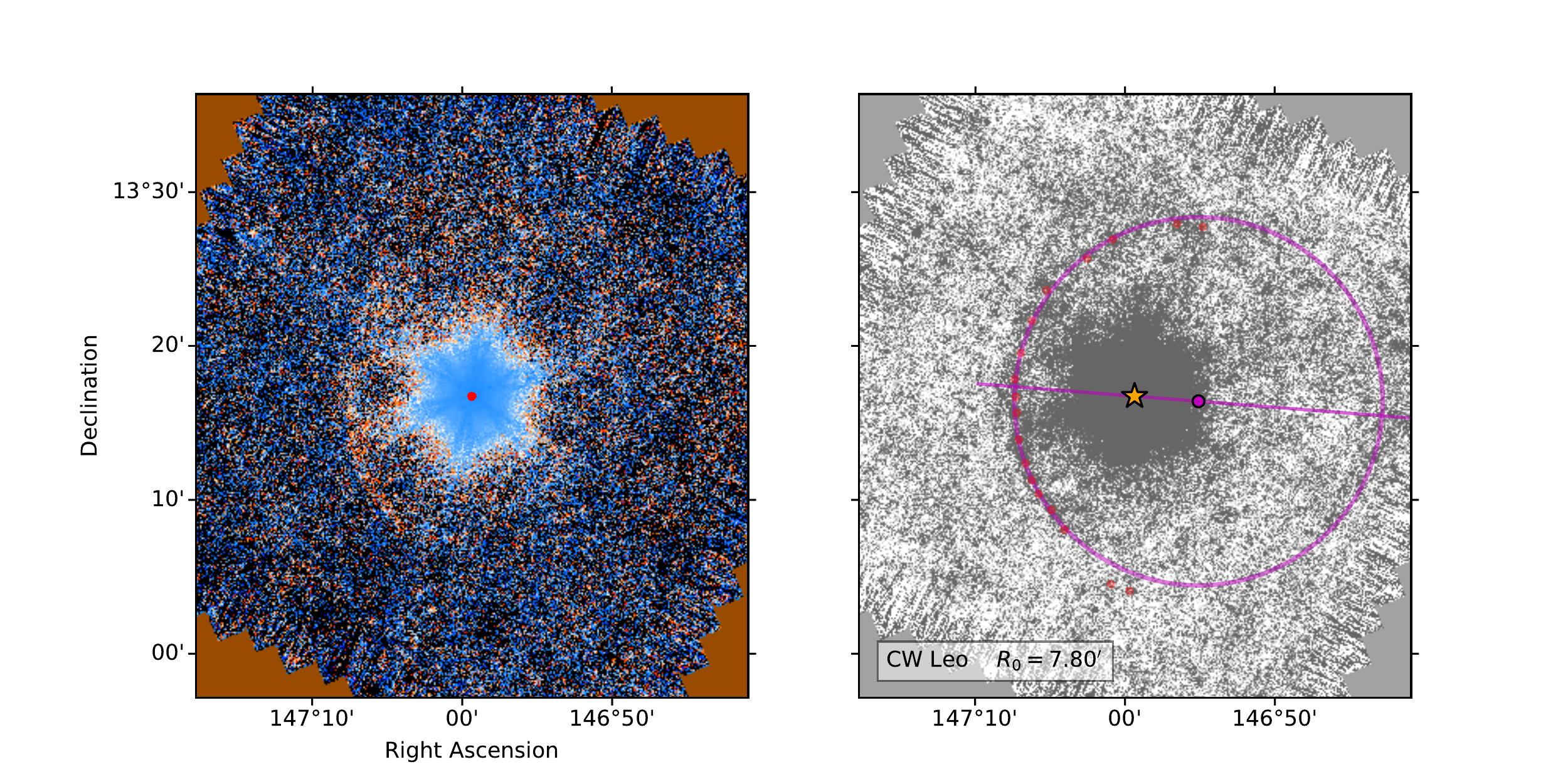}
    & \includegraphics[trim=10 0 65 20, clip]{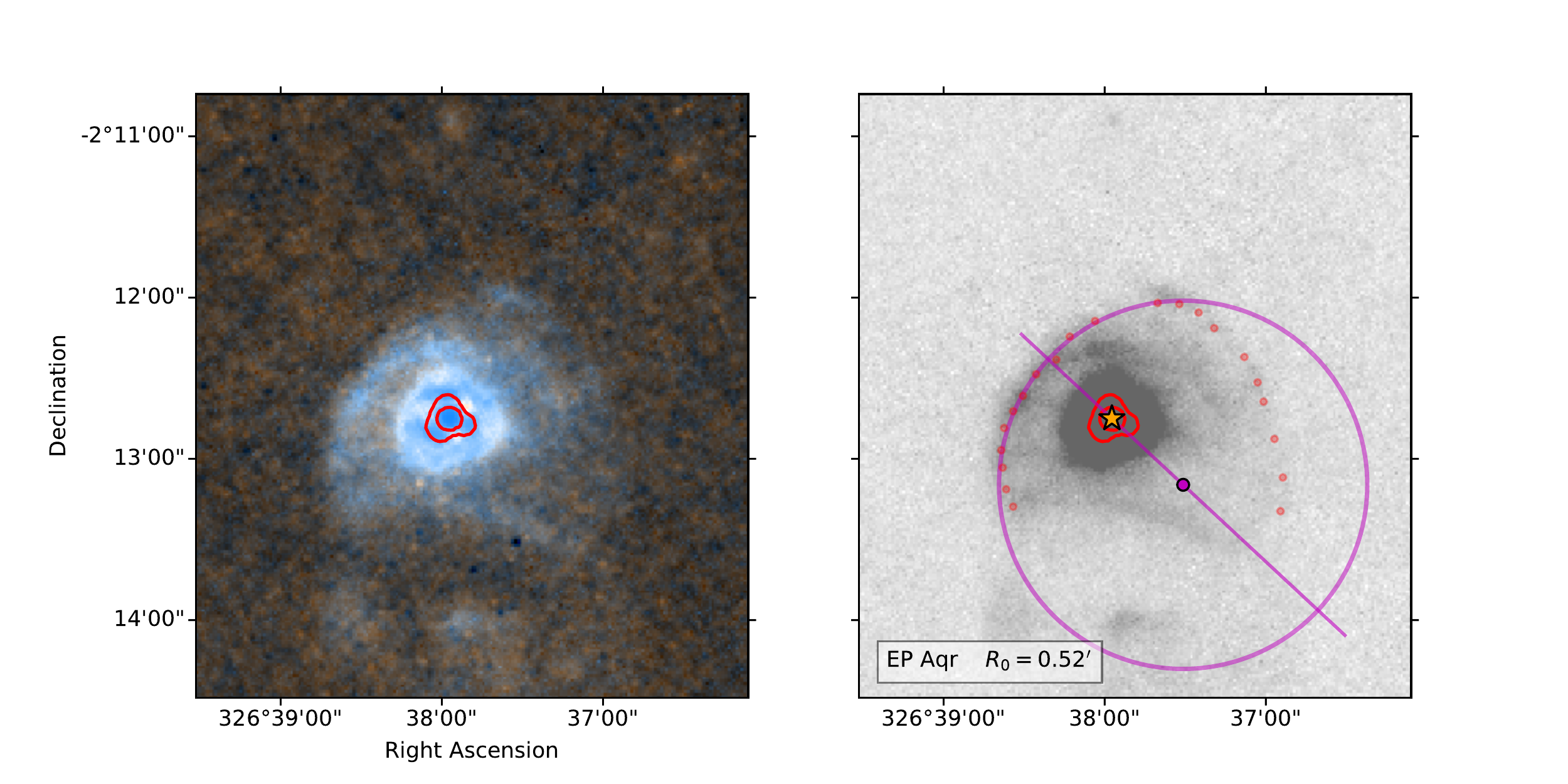} \\
    (c) & (d) \\
    \includegraphics[trim=10 0 65 20, clip]{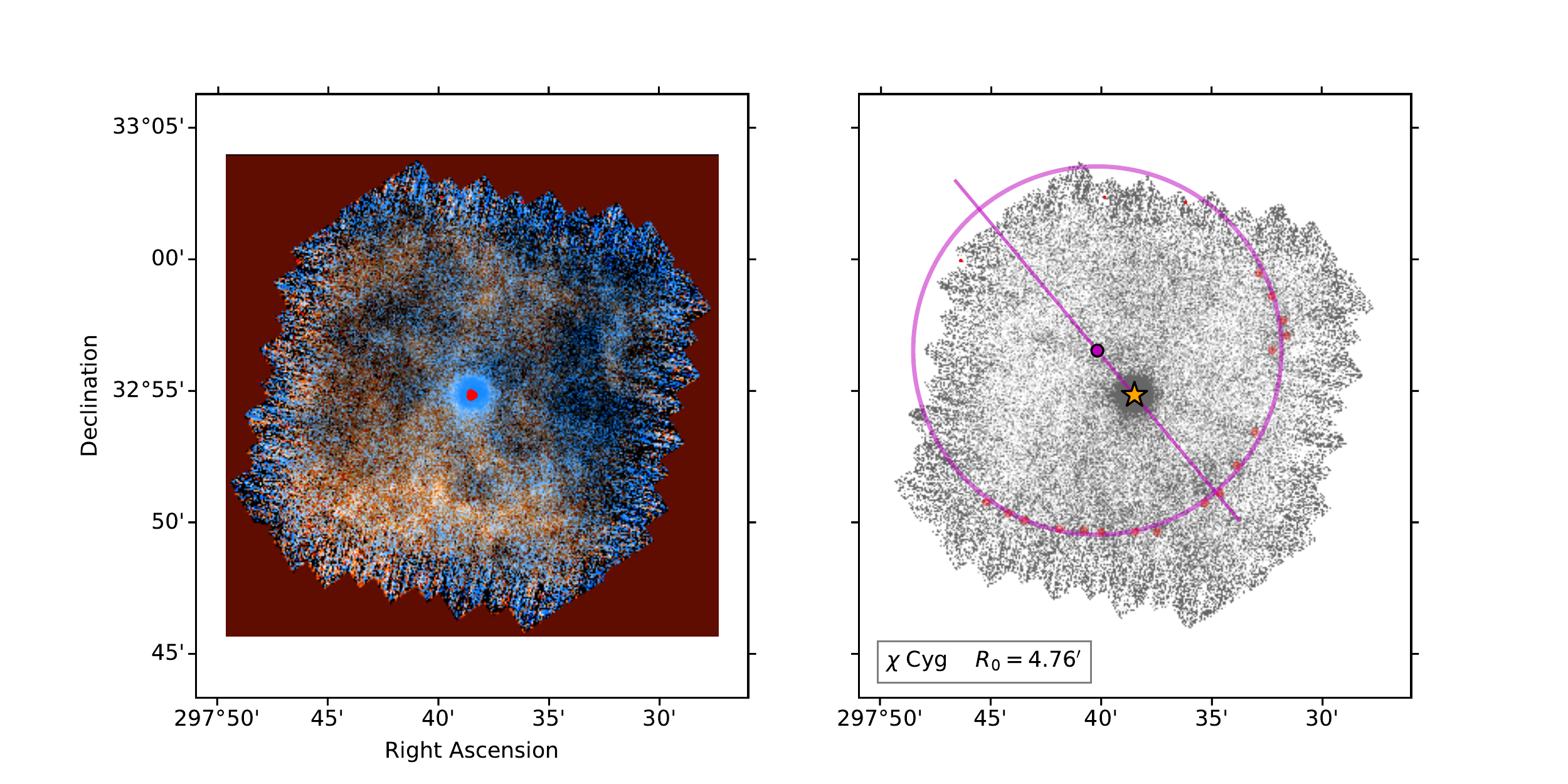}
    & \includegraphics[trim=10 0 65 20, clip]{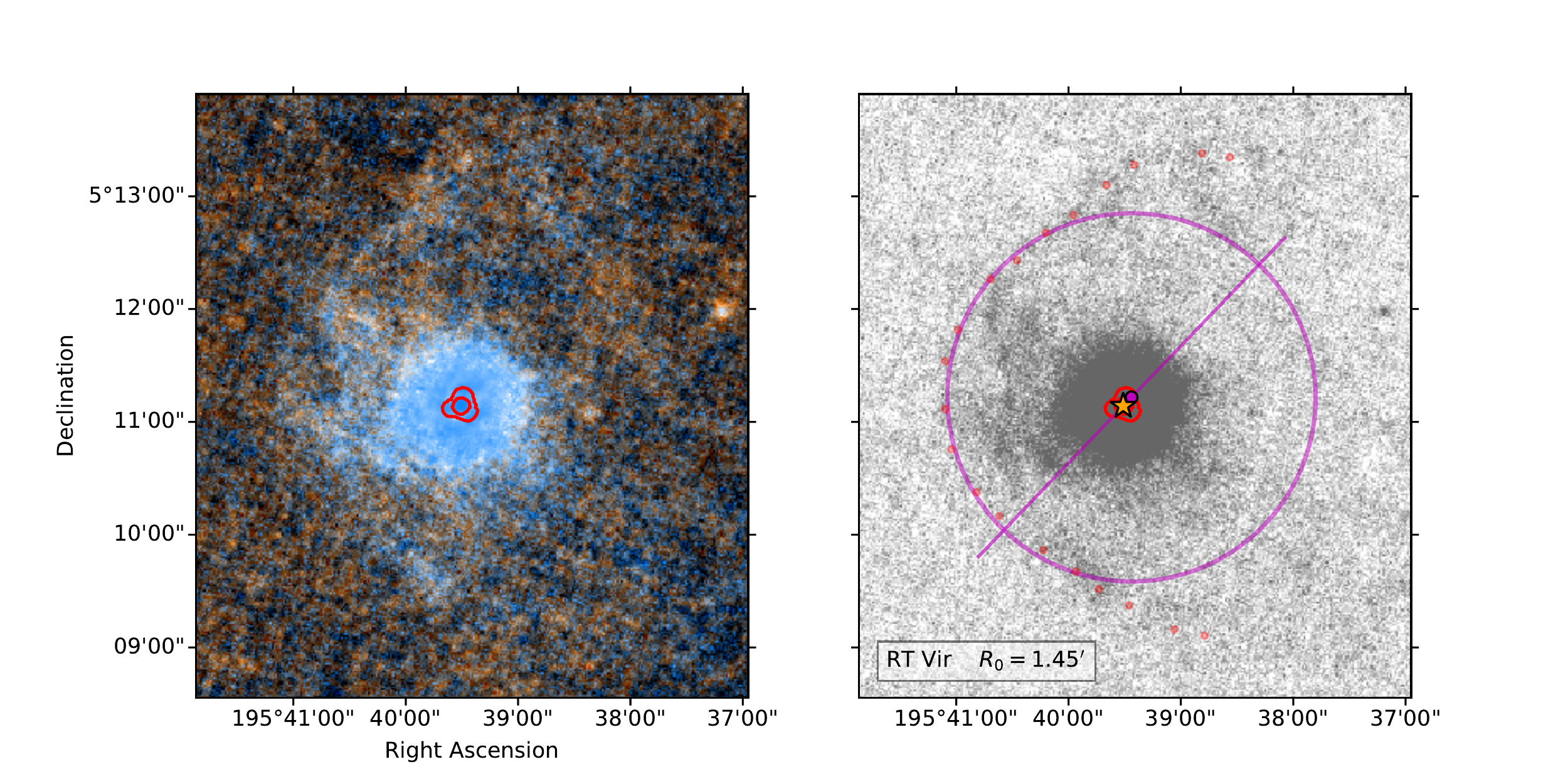} \\
    (e) & (f) \\
    \includegraphics[trim=10 0 65 20, clip]{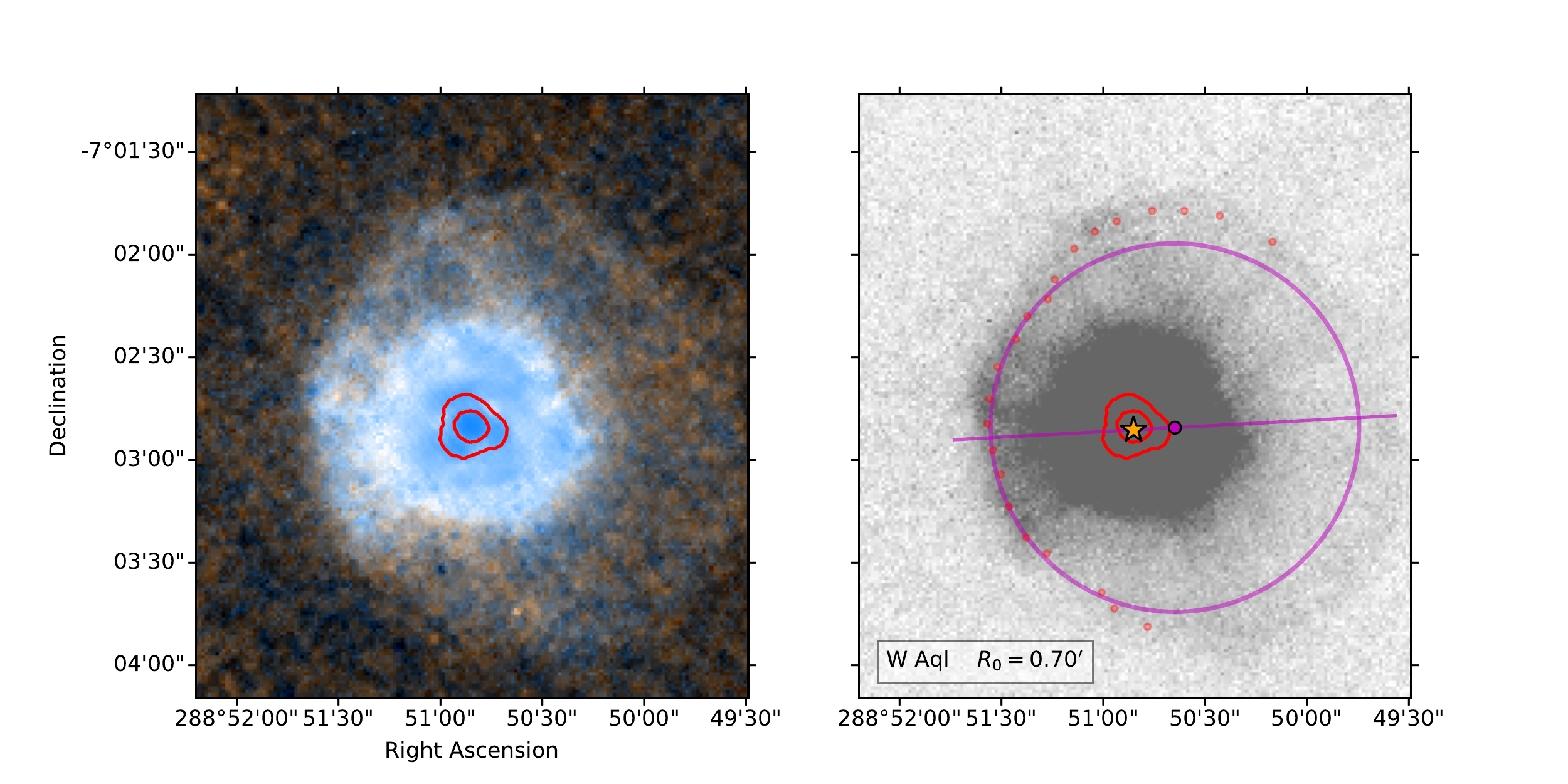}
    & \includegraphics[trim=10 0 65 20, clip]{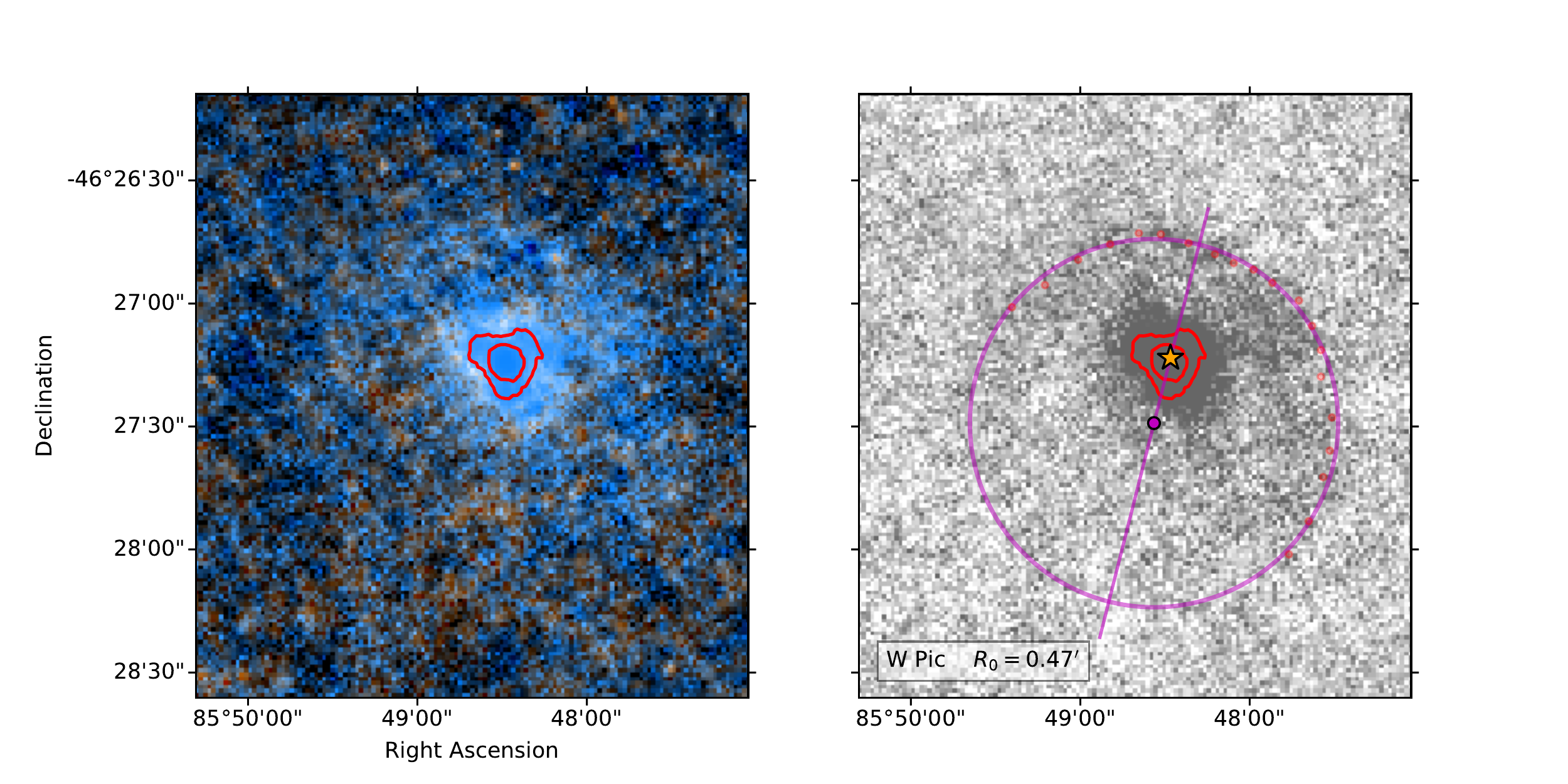} \\
    (g) & (h) \\
    \includegraphics[trim=10 0 65 20, clip]{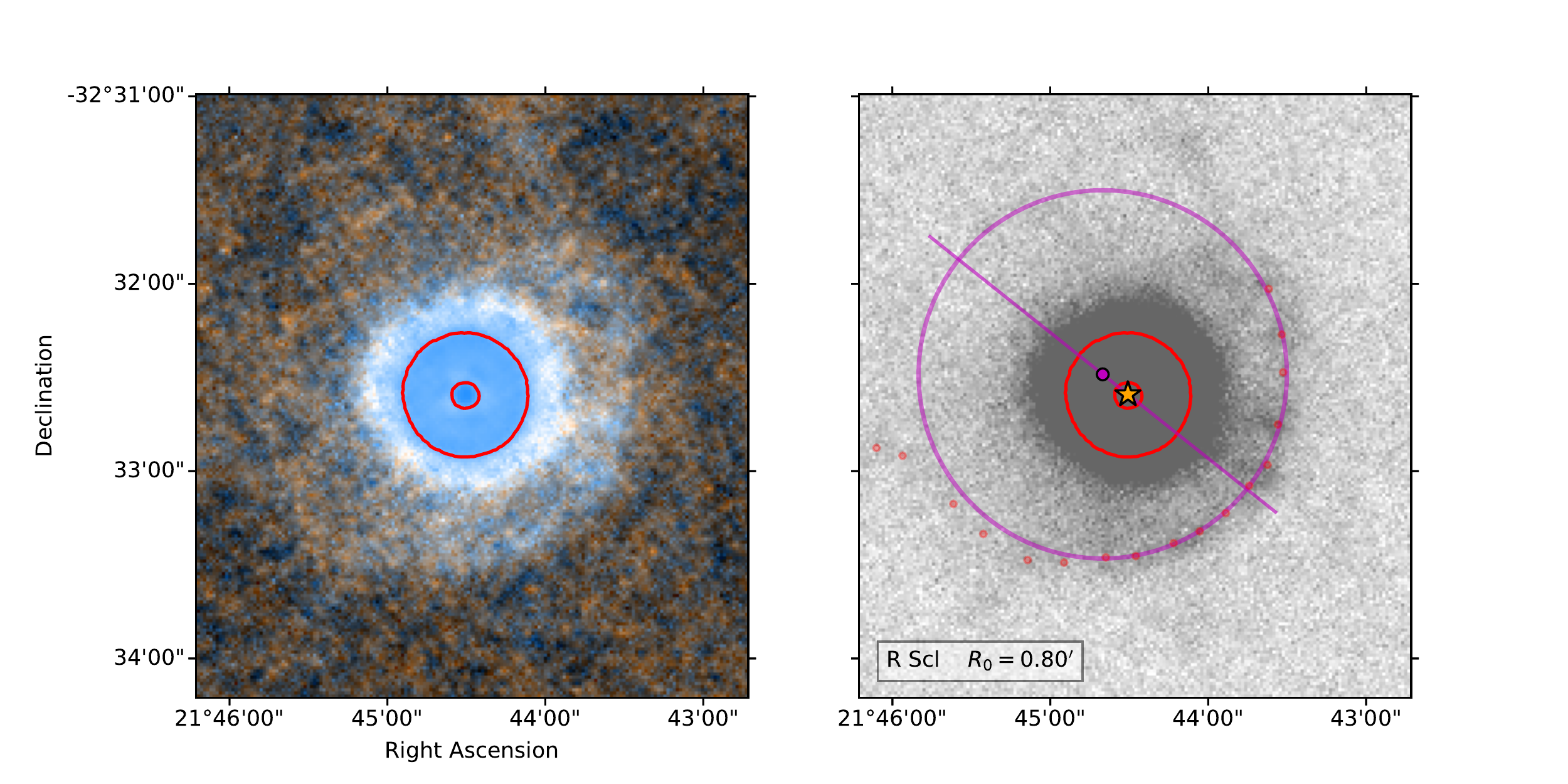}
    & \includegraphics[trim=10 0 65 20, clip]{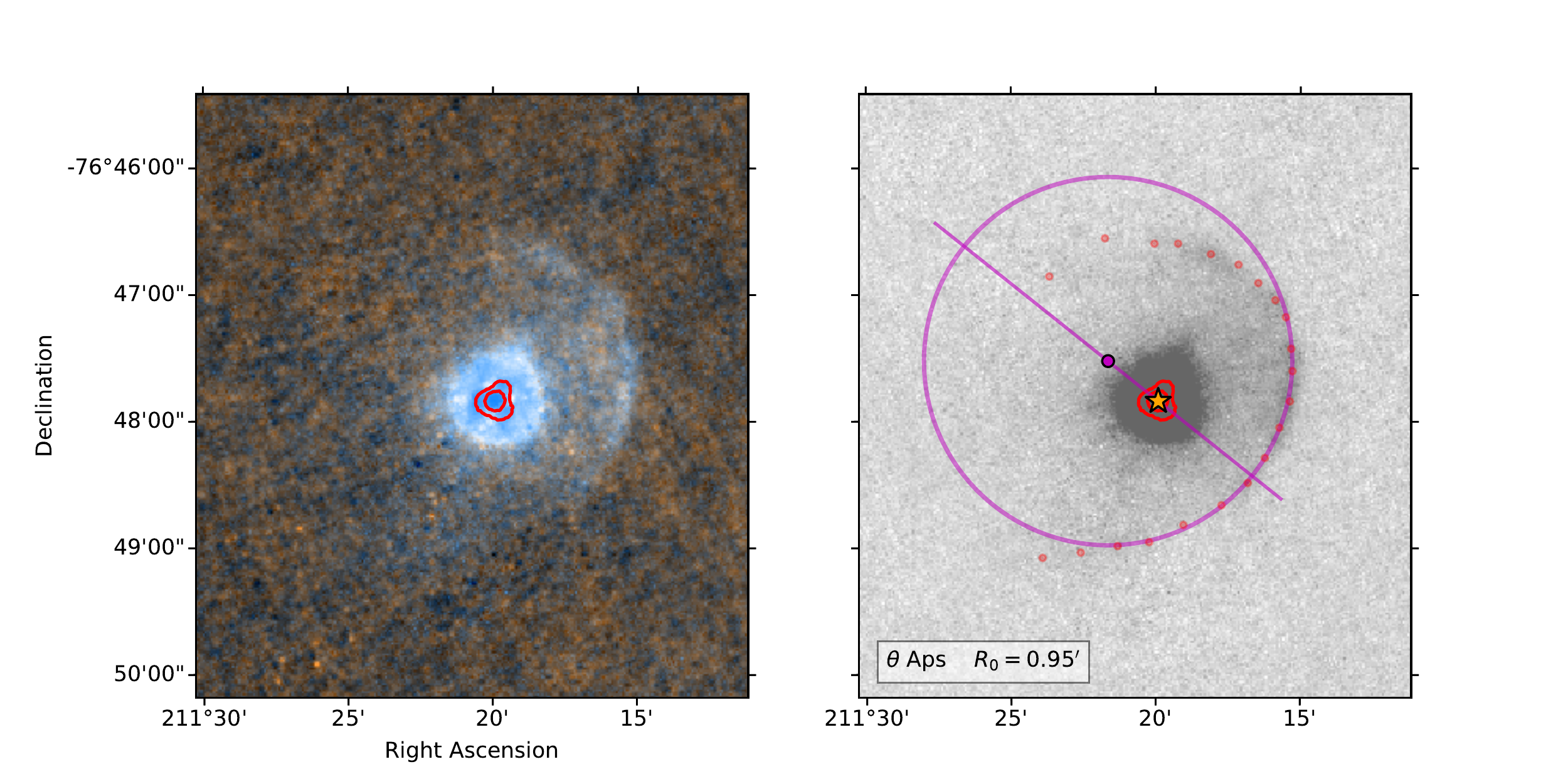}
  \end{tabular}
  \caption{As Fig~\ref{fig:herschel-arc-fits}, but for a further 8
    Class~I sources from \citet{Cox:2012a} where the arc structure is
    more diffuse, weak, and/or asymmetric. (a)~CW~Leo. For this source
    only, the right-hand panel shows a grayscale image of the
    \SI{160}{\um} rather than \SI{70}{\um}
    emission. (b)~EP~Aqr. (c)~\(\chi\)~Cyg. (d)~RT~Vir. (e)~W~Aql. (f)~W~Pic. (g)~R~Scl.
    (h)~\(\theta\)~Aps.}
  \label{fig:herschel-arc-fits-poor}
\end{figure*}

\begin{figure}
  \centering
  \includegraphics[width=\linewidth]{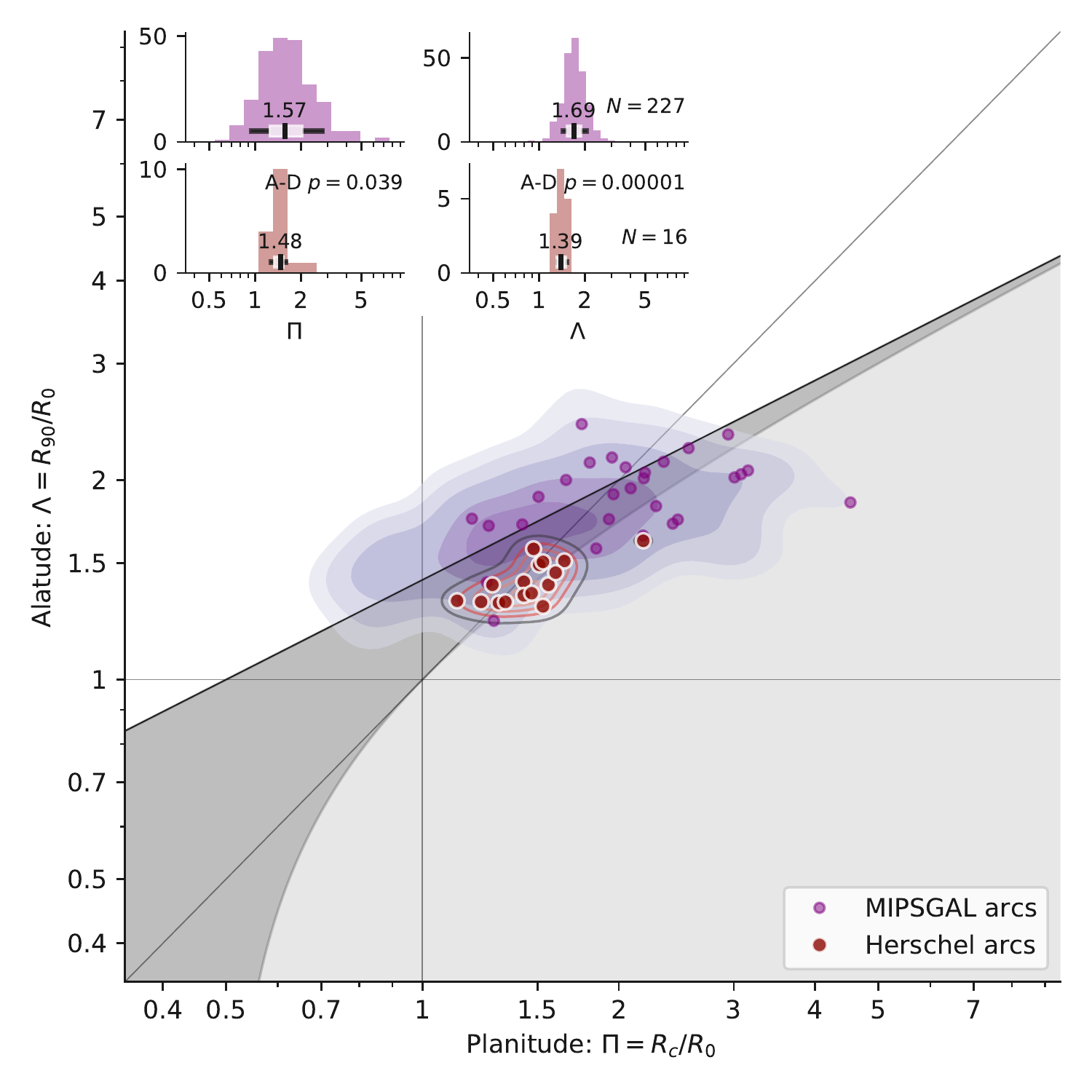}
  \vspace*{-\baselineskip}
  \caption[]{Comparison of bow shock shapes between Herschel RSG/AGB
    arcs (large red symbols and hollow red contours) and MIPSGAL OB
    star arcs (small purple symbols and filled purple contours).
    Other details of the plots are as in
    Figs.~\ref{fig:mipsgal-shapes}, \ref{fig:mipsgal-uncorrelated},
    and \ref{fig:mipsgal-correlated}.}
  \label{fig:herschel-compare-mipsgal}
\end{figure}

We obtain a second sample of bow shocks from a far-infrared survey
\citep{Cox:2012a} of circumstellar shells around known asymptotic
giant branch (AGB) stars and cool supergiants, obtained as part of the
\textit{Herschel} MESS (Mass-loss of Evolved StarS) program
\citep{Groenewegen:2011a}.  The survey sources were divided into four
classes by \citeauthor{Cox:2012a}, according to the overall shape of
the extended emission at \SI{70}{\um} (see their Table~4):
``fermata~\textfermata'', or arcs (Class~I); ``eyes~\faEye''
(Class~II); ``rings~{\Large \(\odot\)}'' (Class~III), and
``irregular~\staveXXV'' (Class~IV).  Of these, only the Class~I
sources clearly correspond to bow shocks, which represent 22 out of 50
total sources detected with extended emission.  For our sample, we
select the 16 Class~I sources where the shape parameters can be
reliably measured, and which are shown in
Figures~\ref{fig:herschel-arc-fits}
and~\ref{fig:herschel-arc-fits-poor}. The remaining six Class~I
sources (Fig.~1 of \citealp{Cox:2012a}) are too asymmetrical or
irregular to reliably determine \(R_{c}\) and \(R_{90}\), and are
therefore excluded from our sample.

In most sources, the bow shock shell is most clearly visible in the
\SI{70}{\um} band, with the exception being CW~Leo where the
\SI{180}{\um} band is used instead.  In several sources, the shell is
split into multiple filaments, with the clearest example being
\(\alpha\)~Ori (Betelgeuse), shown in
Figure~\ref{fig:herschel-arc-fits}(a).  For such sources, we take the
outer envelope of the filaments as the bow shock arc.  Given this
complication, and the fact that the angular resolution relative to the
bow shock size is much better than in the MIPSGAL sources, we judge
that the arc tracing is best performed by eye and this is carried out
using the SAOImage~DS9 FITS viewer \citep{Joye:2003a}, with results
shown as small red open circles in Figures~\ref{fig:herschel-arc-fits}
and~\ref{fig:herschel-arc-fits-poor} (more details of this technique
are given in \S\S~6 and 7 of Paper~0).  Subsequently, the bow shock
parameters \(R_0\), \(\Pi\), and \(\Lambda\) are determined by circle fits, as
in steps 5--9 of \S~\ref{sec:autom-trac-fitt}, with results shown in
magenta on the figures.

Figure~\ref{fig:herschel-compare-mipsgal} compares the distributions
of bow shock shapes between the Herschel (RSG/AGB) and MIGSGAL (OB
star) samples.  The Kuiper test gives a highly significant
difference between the \(\Lambda\) distributions of the two samples, but
only a marginally significant difference between the \(\Pi\)
distributions (detailed results are shown in Table~\ref{tab:big-p}).
The Herschel sources show considerably smaller alatude (median
\(\Lambda \approx 1.4\)), implying bow shock wings that are more closed than in
the MIPSGAL sources (median \(\Lambda \approx 1.7\)).  For the planitude, there is
only a slight difference in average values: median \(\Pi \approx 1.5\) for
Herschel versus \(\approx 1.6\) for MIPSGAL, which is not statistically
significant (rank biserial \(p = 0.29\)).  On the other hand, the
dispersion in \(\Pi\) is four times smaller for the Herschel sample,
which is marginally significant (Brown--Forsythe \(p = 0.015\)).  In
particular, the MIPSGAL sample shows a substantial minority of very
flat-nosed shapes (\(\Pi > 2\)), but these are absent in the Herschel
sample.  The same can be seen directly by comparing the RSG/AGB bow
shock shapes in Figures~\ref{fig:herschel-arc-fits}
and~\ref{fig:herschel-arc-fits-poor} with the representative OB bow
shocks of Figure~\ref{fig:mipsgal-shapes}b.  The MIPSGAL sources K489
and K123 have shapes that are similar to specific Herschel sources
(UU~Aur and R~Hya, respectively), while sources such as K447 (extreme
flat head) or K517 (extreme open wings) have no analog among the
Herschel sources.

\section{Stationary emission line arcs in M42}
\label{sec:stat-emiss-line}

\begin{figure*}
  \centering
  \includegraphics[width=\linewidth]{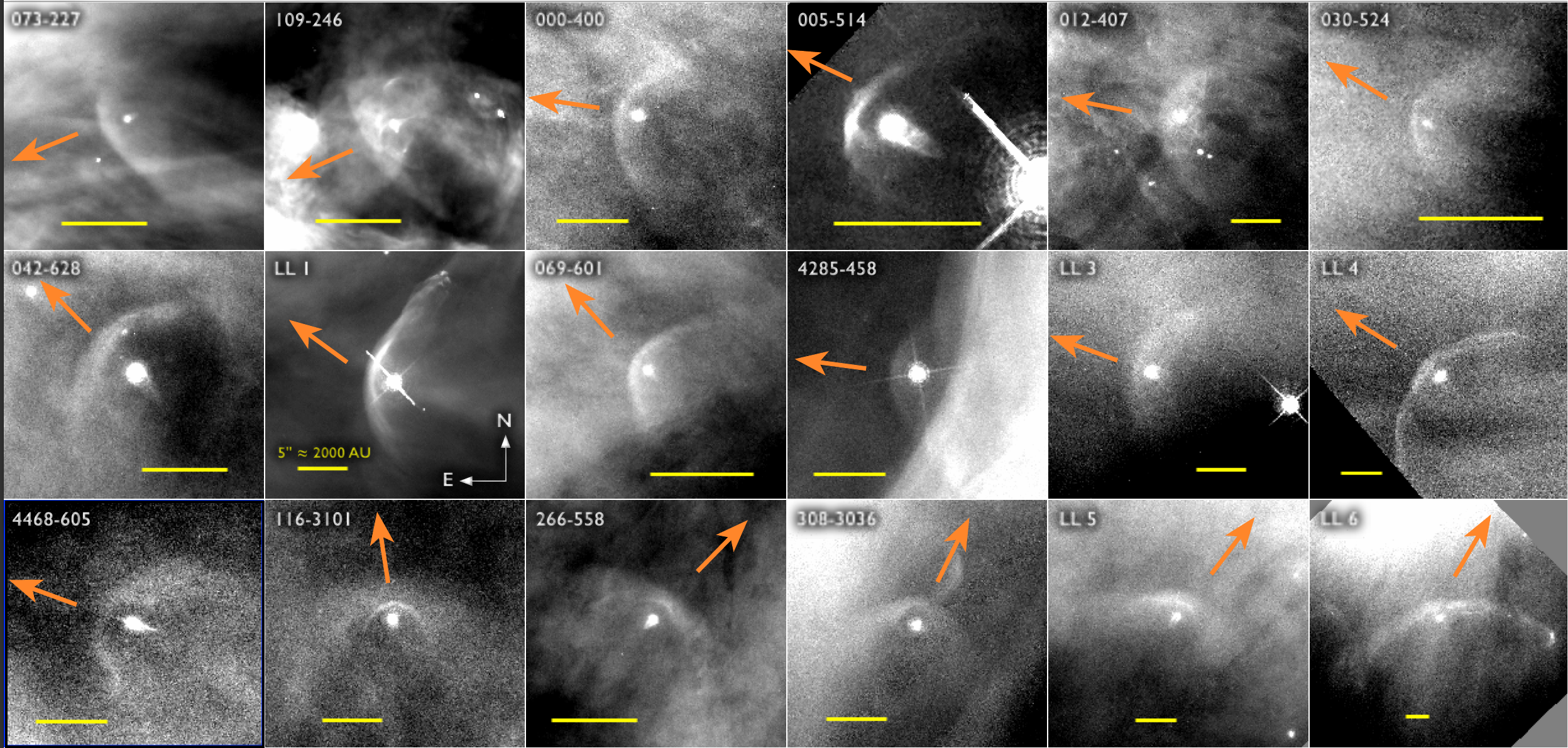}
  \vspace*{-\baselineskip}
  \caption[]{Stationary bow shock arcs in the Orion Nebula.  Images
    are of H\(\alpha\) plus [\ion{N}{ii}] emission through the HST ACS f658n
    filter. The angular scale of each image is different, with the
    yellow horizontal lines indicating \(5''\), corresponding to a
    physical scale of \(\approx\SI{0.01}{pc}\).  Orange arrows indicate the
    direction to the principal ionizing star, \thC. }
  \label{fig:ll-arcs}
\end{figure*}

\begin{figure}
  \centering
  \includegraphics[width=\linewidth]{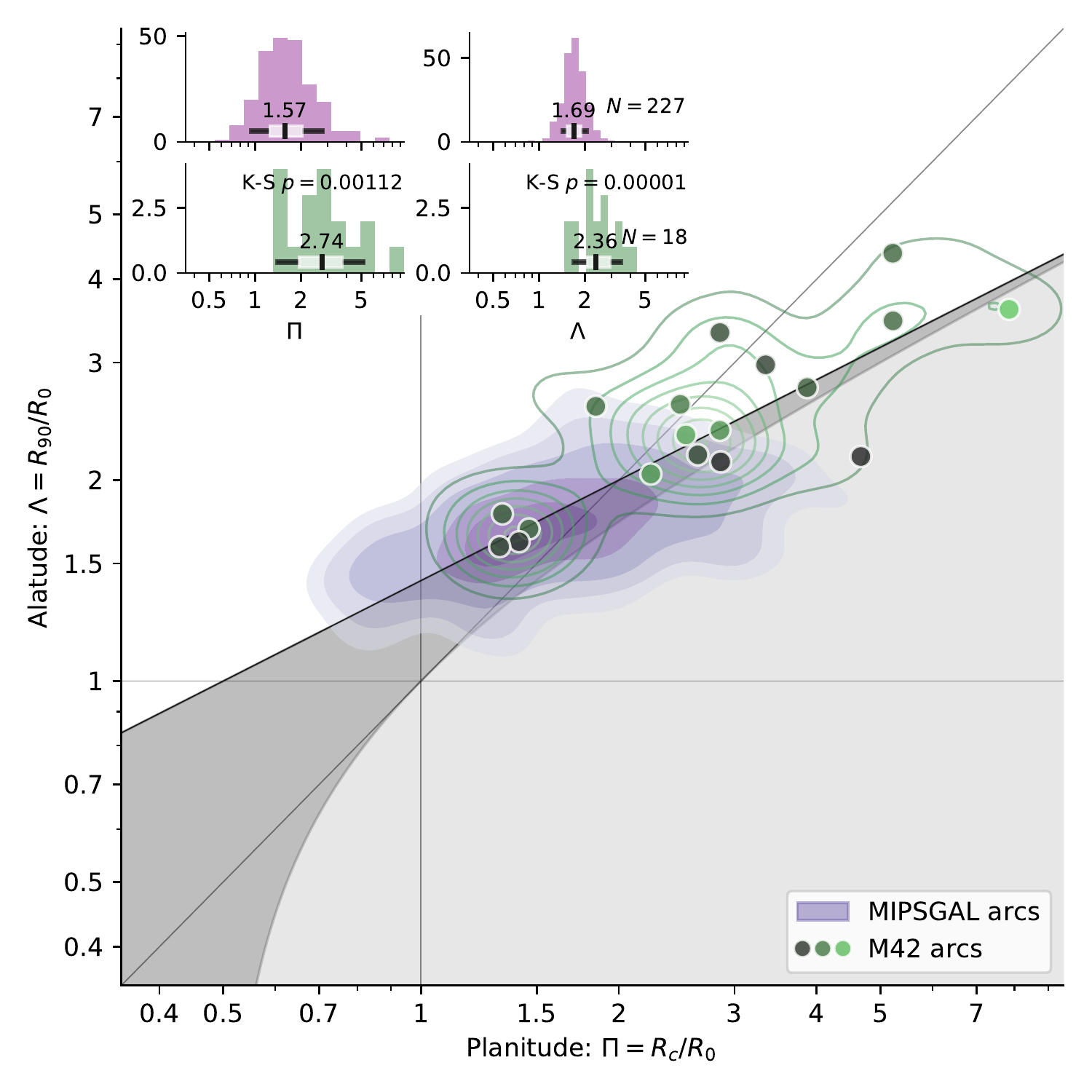}
  \vspace*{-\baselineskip}
  \caption[]{Shape comparison between Orion Nebula arcs (green) and OB
    star bow shocks (purple).  Other details of the plots are as in
    Figs.~\ref{fig:mipsgal-shapes} and~\ref{fig:mipsgal-uncorrelated},
    except that a larger KDE smoothing bandwidth of
    \(0.06 \times 0.045\) is used for the Orion Nebula sample.  The colors
    of the individual points for the Orion Nebula sample show the
    size-to-distance ratio, \(R_0/D\), on a logarithmic scale from
    \(R_0/D = 0.002\) (dark green) to \(R_0/D = 0.018\) (light
    green).}
  \label{fig:ll-compare-mipsgal}
\end{figure}

We obtain a third sample of bow shocks from a catalog of stationary
emission line arcs in the Orion Nebula \citep{Bally:2000a}, which are
detected via their H\(\alpha\) emission in HST~ACS surveys of the nebula
\citep{Bally:2006a, Robberto:2013a}.  Out of a total of 73 such
objects \citep{Gutierrez-Soto:2015a} we have selected 18 where the
observations are of sufficient quality to reliably measure the
planitude and alatude.  Images in the ACS f658n filter of these
sources are shown in Figure~\ref{fig:ll-arcs}.  Some of the sources
were identified as LL~Orionis-type objects by \citet{Bally:2001a} and
are identified as LL~1--6. The remaining sources are labeled by their
coordinate-based designation according to the system of
\citet{ODell:1994a}.  The determination of the bow shock shape for
each source was carried out as described in \S~7 of Paper~0, using the
``ridge'' method for manually tracing the arc (see Fig.~28 of
Paper~0).  In some sources, such as LL~1, LL~6, 109--246, and 4468--605,
a high-velocity jet flow is also seen to issue from the star
\citep{Bally:2006a, Henney:2013a}.  We took care when tracing the arcs
of such sources to avoid regions where jet knot emission is projected
on the bow shock shell.

The planitude and alatude are found by circle fitting using the
algorithm described in Paper~0's Appendix~E, as implemented in the
python program
\texttt{circle-fit.py}.\footnote{\url{https://github.com/div-B-equals-0/circle-fit}}
The only adjustable parameter of the algorithm is \(\Delta\theta\), the maximum
angle from the bow shock axis of points that are included in the
circle fit.  We tested all values of \(\Delta\theta\) in \ang{5} increments from
\ang{45} to \ang{80} and found that \(\Delta\theta = \ang{55}\) to \ang{70}
gives stable results for all sources.  The accuracy of the fitted
planitudes can be estimated from the dispersion in values of \(\Pi\) for
the same source with different \(\Delta\theta\), which we find to be
\(\approx 10\%\).  The fitted alatudes have a much smaller dispersion
\(\approx 1 \%\), but in this case the accuracy is limited by the asymmetry
of the wings, as characterized by \(\Delta\Lambda/\Lambda\) with an RMS value of
\(\approx 15\%\).

The resultant shape distribution for the Orion Nebula arcs is shown in
Figure~\ref{fig:ll-compare-mipsgal}, where it is compared with the OB
star shapes from \S~\ref{sec:ob-shapes}.  It can be seen that both the
planitude and alatude distributions are shifted towards higher values,
with median values of \(\Pi = 2.70\) and \(\Lambda = 2.55\).  The difference
from the MIPSGAL sources (\(\Pi = 1.57\) and \(\Lambda = 1.69\)) is highly
statistically significant, as indicated by extremely low \(p\)-values
for the Kuiper tests.  It is also at least 4 times larger
than the systematic fitting uncertainties discussed in the previous
paragraph.  On closer inspection, it appears that the shape
distribution is bimodal.  Four of the sources (005--514, 4468--605,
116--3101, 308--3036) are similar to typical OB stars, with
\(\Lambda \approx \Pi \approx 1.5\), whereas the remainder are concentrated in the range
\(\Pi = 2\) to \(4\), with an extended tail towards higher planitude.
The alatudes mainly follow the line for parabolas,
\(\Lambda = (2\Pi)^{1/2}\) (see App.~C of Paper~0), and the deviations from
this line can be readily understood in terms of peculiar morphologies
of the source (Fig.~\ref{fig:ll-arcs}).  For instance, LL~4 and
266--558 both show large alatudes for their respective planitudes and
these are both sources where the bow shock wings seem to bend outwards
instead of following the curve from the apex region.  In contrast,
030--524 has a small alatude for its planitude and this is a very
asymmetric source with an apparent corner to the shell on one side.

Uniquely among the three bow shock datasets, for the Orion Nebula arcs
we know the source of the external flow.  This is not the stellar wind
from the high-mass stars at the center of the cluster, which has
insufficient momentum at the position of the arcs \citep{Bally:2000a},
nor is it due to stars' own motions, since the space velocities of all
these sources are very low, conforming to the velocity dispersion
\(\sigma_{\text{1D}} \approx \SI{2}{km.s^{-1}}\) of the Orion Nebular Cluster
\citep{Dzib:2017a, Kim:2018b}.  Instead it is due to the transonic
champagne flow of photoionized gas away from the core of the nebula
\citep{Zuckerman:1973a, Henney:2005a}, which has its density peak in
the Orion~S region \citep{Weilbacher:2015a}, located roughly \(30''\)
SW of the dominant Trapezium star \thC.  The bow shock sources are
located at much larger distances of \(D = 90''\) to \(700''\) from
\thC, so that the ratio \(R_0/D\) can be used as a proxy for the local
divergence of the champagne flow on the scale of the bow shock (notice
that the symmetry axes of the bow shocks in Fig.~\ref{fig:ll-arcs} are
always roughly parallel to the direction to \thC{}).  This is
indicated by the color of the points in
Figure~\ref{fig:ll-compare-mipsgal}, with lighter colors corresponding
to a larger \(R_0/D\) and hence a more divergent external flow.  It
can be seen that there is some correlation between the bow shock
shapes and \(R_0/D\): the source with the largest \(\Pi\), 109--246, also
has the largest \(R_0/D\), while the four sources with low \(\Pi\) and
\(\Lambda\) all have low \(R_0/D\).  However, the correlation is far from
perfect, with a Pearson correlation coefficient of only \(r = 0.47\).

 % end input ./sec-quadrics-observations.tex
 % start input ./sec-standing-wave.tex

\section{Discussion}
\label{sec:discussion}

In this section, we discuss the physical implications of our empirical
findings regarding bow shock shapes.  Our most reliable result is the
average shape of the OB bow shocks from the 227 MIPSGAL sources with
quality rating of 3~stars or higher.  This yields mean values of
\(\Pi = 1.78 \pm 0.06\) and \(\Lambda = 1.72 \pm 0.02\), or median values of
\(\Pi = 1.57\) and \(\Lambda = 1.69\).  The uncertainty quoted on the mean
values is the ``standard error of the mean'':
\(\text{sem} = \sigma / \sqrt{n}\), where \(\sigma\) is the rms dispersion and
\(n\) is the number of sources.  Note that in the case of the
planitude \(\text{sem}(\Pi) = 0.06\) is considerably smaller than
\(\text{mean}(\Pi) - \text{median}(\Pi) = 0.21\), so the latter would be a
more conservative estimate of the uncertainty.\footnote{This is
  because the distribution of \(\Pi\) is approximately log-normal, which
  yields a significant tail towards high values when converted to
  linear space.}  These values can be compared with the predictions of
the thin-shell wilkinoid model \citep{Wilkin:1996a}, which are
\(\Pi = 1.67\), \(\Lambda = 1.73\) when the bow shock axis lies in the plane
of the sky (following Paper~0, this is defined as the zero point of
the inclination angle, \(i\)).  When the axis is inclined, both
planitude and alatude are predicted to decrease but not by very much,
tending to \(\Pi = 1.5\), \(\Lambda = 1.63\) as
\(\abs{i} \to \ang{90}\) (see \S~5.3 of Paper~0).  The median observed
value falls squarely inside this range for both the planitude and
alatude, which is a remarkable triumph for the \citet{Wilkin:1996a}
model.

\subsection{Diversity in bow shock shapes}
\label{sec:diversity-bow-shock}

On the other hand, turning now to the \emph{variety} of bow shock
shapes, we see that the wilkinoid can no longer explain our results.
The rms dispersions of the planitude and alatude distributions for the
MIPSGAL sources are \(\sigma(\Pi) = 0.87\) and
\(\sigma(\Lambda) = 0.30\) (Tab.~\ref{tab:big-p}), which are respectively 5 times
and 3 times larger than the total range of variation of \(\Pi\) and
\(\Lambda\) predicted for the wilkinoid surface.  Although some of the
dispersion is due to uncertainties in the observations and the fitting
algorithm, this contribution is expected to be small, especially for
the larger, well-resolved sources, for which systematic uncertainties
in the methods for determining \(\Pi\) and \(\Lambda\) will
dominate. Conservative upper limits to the relative size of these
uncertainties were estimated in \S~7 of Paper~0 to be \(< 20\%\) for
\(\Pi\) and \(< 10\%\) for \(\Lambda\), whereas the observed dispersions are
roughly twice as large: \(\sigma(\Pi)/\Pi = 55\%\) and
\(\sigma(\Lambda)/\Lambda = 18\%\).  Furthermore, the variations in planitude and
alatude are readily apparent by eye, as is demonstrated by the example
bow shock images shown in Figure~\ref{fig:mipsgal-shapes}b.  Sources
such as K123 have very typical shapes and fall near the center of the
\(\Pi\)--\(\Lambda\) distribution, whereas high-\(\Pi\) sources such as K447 have
a very flat apex region, while low-\(\Pi\) sources such as K566 have a
pointier, almost triangular apex.  High-\(\Lambda\) sources, such as K517,
have very open wings that bend away from the star, while
low-\(\Lambda\) sources such as K489 have closed wings and a semi-circular
appearance.

\subsubsection{The influence of projection effects}
\label{sec:infl-proj-effects}

In Paper~0 we found that certain bow shock shapes can show a much
greater variation in their projected appearance as a function of
inclination angle than is seen for the wilkinoid.  For example, the
cantoids and ancantoids, which have asymptotically hyperbolic far
wings, can shift towards higher apparent planitude and alatude as the
inclination increases, generally with \(\Lambda \ge \Pi\) (Fig.~20 of Paper~0).
This might plausibly explain the vertical spur towards higher
\(\Lambda\) seen in the empirical distribution (\S~\ref{sec:ob-shapes}). A
different behavior is shown by bow shocks with very flat apex regions,
such as the MHD simulation from \citet{Meyer:2017a} that is analyzed
in the \S~6 of Paper~0.  This shows a high planitude \(\Pi\) when the
orientation is exactly edge-on, but \(\Pi\) decreases sharply along a
roughly horizontal track as the inclination \(\abs{i}\) increases
(Fig.~25 of Paper~0).  This is similar to the principal axis of
variation of the observed shapes (e.g.,
Fig.~\ref{fig:mipsgal-shapes}a).

If such variations in orientation do make a significant contribution
to the observed distribution of bow shock shapes in the
\(\Pi\)-\(\Lambda\) plane, then various predictions follow, which might be
observationally tested.  High-planitude sources with \(\Pi > 3\) would
be expected to have low inclinations, \(\abs{i} < \ang{30}\), whereas
high-alatude sources with \(\Lambda > 2\) would be expected to have high
inclinations, \(\abs{i} > \ang{45}\).  Unfortunately, determination of
the inclination for individual sources requires high resolution
spectroscopy of emission lines in order to map the kinematics of the
flow in the bow shock shell \citep[e.g.,][]{Henney:2013a}.  This is
not currently available for the majority of the MIPSGAL sources, which
are detected only by their dust continuum emission.  A further
prediction for the high-alatude sources is that the environmental flow
should be divergent rather than plane-parallel, in order to give a
cantoid shape instead of a wilkinoid.  This would tend to favor
``weather-vane'' cases, where the interstellar medium is flowing past
the star, and disfavor ``runaway'' cases, where the star is moving
through a static medium.  However, in \S~\ref{sec:corr-shape} we found
no significant difference in the shape distributions as a function of
the bow shock environment.  Figure~\ref{fig:mipsgal-uncorrelated}a
shows that the alatudes of sources that are facing \hii{} regions or
\SI{8}{\um} bright-rimmed clouds (and therefore might be expected to
be immersed in a champagne flow) are no higher than sources that are
isolated.

\subsubsection{Perturbations to the bow shape}
\label{sec:pert-bow-shape}

\begin{figure}
  (a)\\
  \includegraphics[width=\linewidth]
  {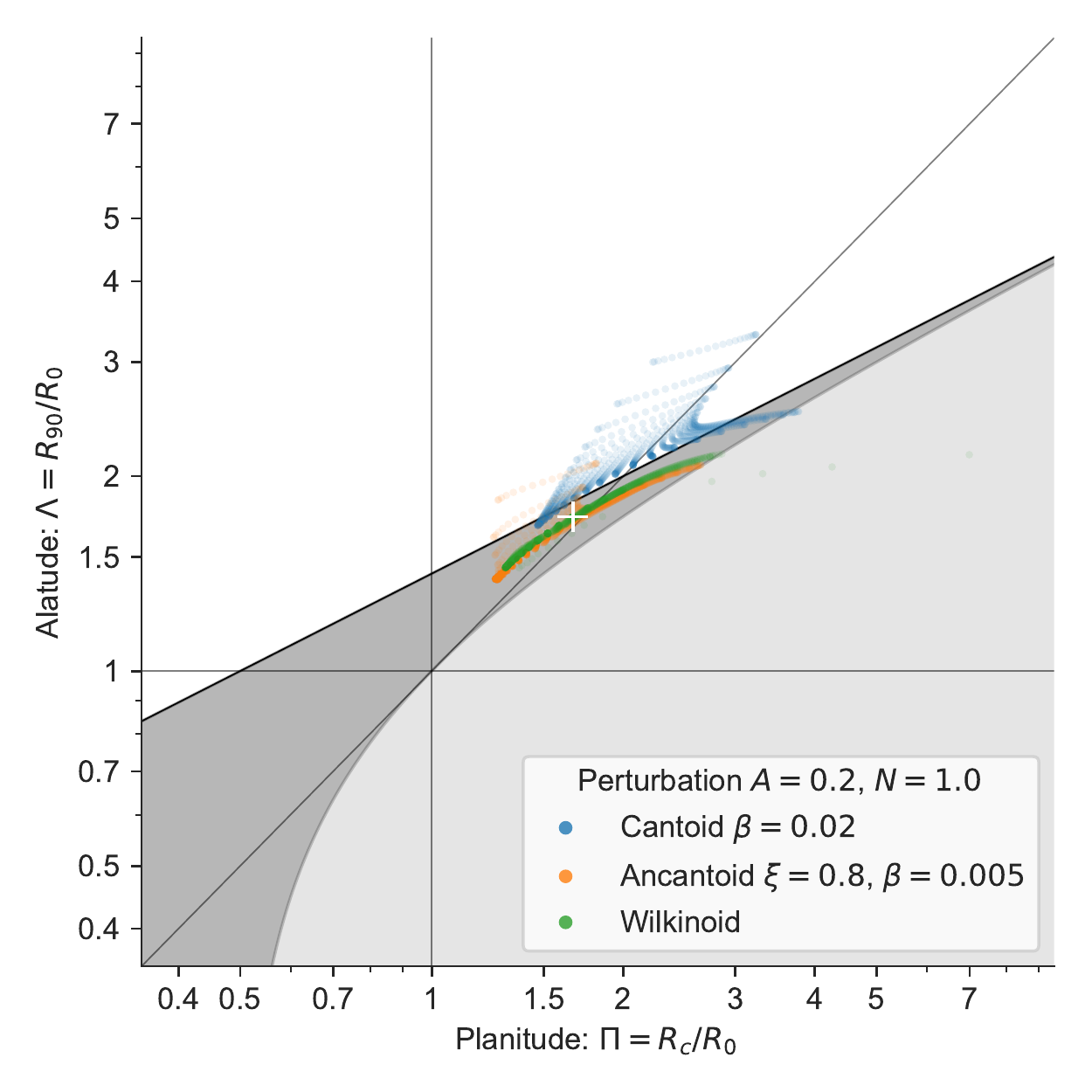}
  (b)\\
  \includegraphics[width=\linewidth]
  {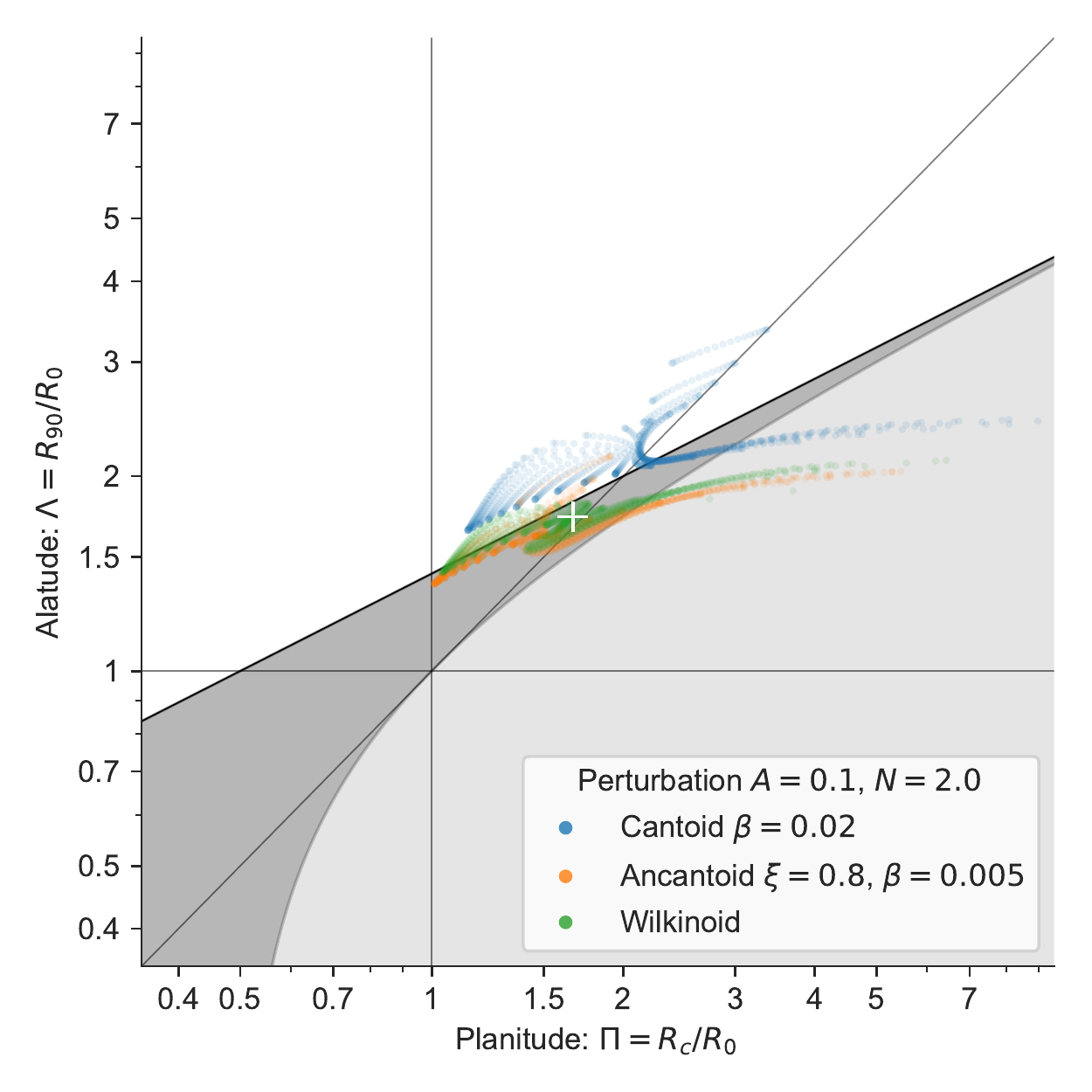}
  \vspace*{-\baselineskip}
  \caption{Diagnostic diagram for perturbed shapes from standing wave
    oscillations.  Each model is characterized by a base shape
    (colored symbols, as described in key) and an amplitude, \(A\),
    and wavenumber, \(N\), of the oscillation: (a)~breathing mode with
    \(N = 1\), \(A = 0.2\); (b)~curling mode with \(N = 2\),
    \(A = 0.1\) (see Fig.~\ref{fig:perturb-shapes} for the
    phase-dependent intrinsic shapes).  The plotted points show the
    varying planitude and alatude of the projected bow shock shapes
    with uniform sampling over an entire period of the oscillation and
    for varying inclinations (sampled according to an isotropic
    distribution of orientations).  Each individual point is plotted
    with a low opacity so that the crowding of points in certain
    regions of the plane can be appreciated.}
  \label{fig:perturb-Rc-R90}
\end{figure}

An alternative explanation for the variety of observed bow shock
shapes is that they are due to time-dependent perturbations to an
underlying base shape.  For instance, multiple studies have shown that
stellar bow shock shells can be unstable \citep{Dgani:1996a,
  Dgani:1996b, Blondin:1998a, Comeron:1998a, Meyer:2014a}, leading to
large amplitude oscillations in the shell shape.  The oscillations are
found to be most vigorous when the post shock cooling is highly
efficient, allowing the formation of a thin shell (see Paper~I).  Even
in cases where the shell is stable, oscillations may be driven by
periodic variations in the stellar wind mass-loss rate or velocity, or
by inhomogeneities in the ambient stream.  Rather than using a
particular dynamical model of these oscillations, we instead crudely
simulate their effect by assuming a constant amplitude standing wave
perturbation to the base shape, as described in
Appendix~\ref{sec:perturbed-bows}.  Example results are shown in
Figure~\ref{fig:perturb-Rc-R90} for an ensemble of bow shocks with
different orientations and phases of oscillation, considering three
different underlying base shapes.  It can be seen that modest
amplitudes of 10 to 20\% can give rise to a distribution in \(\Pi\) and
\(\Lambda\) similar to that observed for the MIPSGAL sources when the
oscillation wavelength is of the same order as the bow shock size.

An attractive feature of the oscillation hypothesis is that it
naturally explains why we find little correlation between the bow
shock shape and other source parameters (\S~\ref{sec:corr-shape}),
since the instantaneous shape at any instant is largely a matter of
chance rather than being due to any intrinsic property of the source.
The one significant correlation that we do find is that the alatude
distribution is broader for bow shocks with larger angular sizes.
This might be explained if the relative amplitude of oscillations were
higher for sources with more powerful winds. \citet{Meyer:2016a} in
their Fig.~2 plot the ``axis ratio'' (which is our \(\Lambda^{-1}\)) as a
function of apex distance for a large set of hydrodynamic bow shock
simulations.  They find the most unstable bow shocks (with the largest
spread in \(\Lambda\)) to be those associated with the highest mass stars in
a relatively dense medium, which have apex distances of
\SIrange{0.3}{1}{pc}.  This corresponds to \(R_0 > 15''\) for
distances less than \SI{4}{kpc}, which is larger than the median
angular size for the MIPSGAL sources, making this a plausible explanation for our statistical result.

\subsection{Variations between different source classes}
\label{sec:vari-betw-diff}

We now address the difference in shape distribution between the
different classes of source.  Compared with the OB star bow shocks,
the cool star sample from \citet{Cox:2012a}\footnote{Three of these
  sources are high-mass red supergiant (RSG) stars, while the
  remaining 13 are intermediate-mass asymptotic giant branch (AGB)
  stars.} shows a significantly smaller alatude of
\(\Lambda = 1.41 \pm 0.03\) (see
Fig.~\ref{fig:herschel-compare-mipsgal}).\footnote{Although the
  planitude also appears to be slightly smaller, this is not very
  statistically significant due to the small number of cool star
  sources and the large width of the OB star planitude distribution.}
Such a closed shape for the wings is inconsistent with the wilkinoid
value of \(\Lambda = \text{\numrange{1.63}{1.73}}\), which is surprising
given that the emission shells in these sources are relatively narrow
(Figs.~\ref{fig:herschel-arc-fits}
and~\ref{fig:herschel-arc-fits-poor}), so one might have thought that
the thin-shell approximation of \citet{Wilkin:1996a} would be
\emph{more} appropriate than for the OB stars, but this is clearly not
the case.  One possible explanation for this might be that the bow
shocks have not had time to reach a steady-state configuration, as was
suggested by \citet{Mohamed:2012a} for the case of \(\alpha\)~Ori.  The
dynamical timescales, \(R_0 / V\wind\), for the cool star bow shocks
are of order \SI{e4}{yr} and numerical simulations \citetext{e.g.,
  Fig.~11 of \citealp{Mohamed:2012a} and Fig.~2 of
  \citealp{van-Marle:2014a}} show that the bow shock wings take of
order \SIrange{3e4}{e5}{yr} to fully unfold.  However, this is still
short compared with typical RSG and AGB lifetimes, so while it might
apply to a single source it does not work as an explanation for an
entire class of sources, especially given that the alatude and
planitude distributions for these sources are so narrow
(Fig.~\ref{fig:herschel-compare-mipsgal}).  A more promising
explanation is that the shape difference reflects a different origin
for the infrared-emitting dust.  In hot stars, the stellar wind is
dust-free, so the only dust shell comes from the interstellar medium
and lies outside the astropause (contact discontinuity).  In cool
stars, the stellar wind is also dusty, which could give rise to
significant infrared emission from near the stellar wind's termination
shock, as has been found from hydrodynamic simulations
\citep{Meyer:2014b}.  This is certainly the case for at least one of
our cool star sources, the C-AGB star CW~Leo (IRC+10216), where
ultraviolet GALEX observations \citep{Sahai:2010a} clearly show
\emph{both} the outer shock and the wind termination shock, and
comparison with the Herschel images demonstrate that the infrared dust
emission is associated with the latter.

In the case of the Orion Nebula bow shocks
(\S~\ref{sec:stat-emiss-line}), we have the opposite situation, where
the alatude and planitude are both significantly larger on average
than for the OB star sources (Fig.~\ref{fig:ll-compare-mipsgal}).  On
the face of it, this is surprising because there are two differences
between the source classes that would tend to work in the other
direction.  First, many of the Orion sources are proplyds, or
externally illuminated photoevaporating disks \citep{ODell:2008b}, in
which the inner wind is not isotropic but instead is mildly
concentrated towards the symmetry axis \citep{Garcia-Arredondo:2001a},
so that simple models predict an \textit{ancantoid} shape
\citep[\S~5]{Tarango-Yong:2018a} that is less open than for an
isotropic wind.  Second, as was the case with the cool star sources,
both the shocked inner wind and the shocked outer stream are expected
to contribute to the emission arcs.  Indeed, in several of the sources
of Figure~\ref{fig:ll-arcs} (LL~3, 116--3101, 266-558, 308-3036, and
LL~5), a two-component emission structure is apparent.  In the case of
the cool star bow shocks, we invoked this as a possible explanation of
their \emph{low} alatude (see previous paragraph), so
some stronger countervailing factor is necessary in order to explain
why the opposite is seen in the Orion Nebula bow shocks.

Four possible origins for this countervailing factor suggest
themselves: (i)~the external flow may be more divergent in the Orion
sources; (ii)~the Mach number of the external flow may be
systematically lower; (iii)~the fact that the arcs are observed in
recombination line emission instead of dust continuum may cause
different observational biases; (iv)~the shapes may be influenced by
collimated jet outflows from the young stars.  We now address each of
these in turn.

In case~(i) one would expect \(\Pi\) and \(\Lambda\) to be positively
correlated with \(R_0/D\), where \(R_0\) is the bow shock size
(star--apex separation) and \(D\) is the distance from the center of
divergence of the external flow.  In \S~\ref{sec:stat-emiss-line} we
found that these are indeed correlated, but only weakly.  A more
serious objection to this idea is that according to hypersonic
thin-shell models \citep{Canto:1996} the momentum ratio parameter
\(\beta\) must be relatively large in order to give significantly open bow
shock shapes.  From Figure~20 of Paper~0 we see that
\(\beta > \num{e-3}\) is required in order to give
\(\Pi, \Lambda > 2\).  However, for small \(\beta\) one has
\(\beta \approx (R_0/D)^2\), which yields \(\beta = \num{4e-6}\) to \num{3e-4} for
our sources, so that divergence ought to have little effect on the
shapes.

Case~(ii) arises from the fact that, if the hypersonic assumption is
relaxed, then the opening angle between the outer bow shock and the
contact discontinuity in the wings becomes increasingly large as the
Mach number, \(\M\), drops towards unity.  This is a consequence of
the \textit{shock polar} relation for oblique shocks \citep[\S\S~92
and 113]{Landau:1987a} and applies to both the radiative and
non-radiative case.  It will tend to produce more open shapes, at
least in the case that the emission is dominated by the outer shell.
At the same time, the relative thickness of the shell, \(h/R_0\), is
predicted to increase as \(\M\) decreases \citetext{e.g., eq~[35] of
  Paper~I}.  We have therefore measured \(h/R_0\) for the Orion
sources, finding values ranging from \numrange{0.3}{0.8} with median
of \num{0.5}.  We have not measured \(h/R_0\) for the full MIPSGAL
sample, but Figure~11 of Paper~III gives the values for the sub-sample
of OB bow shocks studied by \citep{Kobulnicky:2018a}.  The median
value is 0.2, implying thinner shells than in the Orion Nebula bow
shocks, which lends support to the idea that the Mach number may be
lower in the latter.  However, among the Orion sources we find that
\(h/R_0\) is completely uncorrelated with either \(\Pi\) or
\(\Lambda\) (\(r = 0.02\) and \(-0.05\), respectively).

Case~(iii) attempts to explain the difference in shapes as a result of
an ``optical illusion'' whereby the OB star bow shocks are in reality
more open than they appear. The hydrogen recombination line surface
brightness from an isothermal shell is proportional to the emission
measure (line of sight integral of the product of proton and electron
densities), whereas the mid-infrared surface brightness is not simply
proportional to the dust column density, since it results from the
reprocessing of stellar radiation.  For typical bow shock grain
temperatures \citetext{\(T = 50\) to \SI{100}{K},
  \citealp{Kobulnicky:2017a}} the \SI{24}{\um} band used for the OB
star sample lies on the short wavelength Wien side of the dust
emission spectrum, which gives a very steep radial dependence of the
emissivity.  \citet{Acreman:2016a} calculate synthetic emission maps
from hydrodynamical simulations and show that, even for a relatively
thin bow shock shell, the emission arc seen in H\(\alpha\) tends to be more
open than the arc seen in the mid-infrared (see their Fig.~3).  The
effect should be even larger for thicker shells, as demonstrated by
\citet{Mackey:2016a}, who simulate the subsonic motion of an O star
through its \hii{} region and show that this can give rise to an
\emph{apparent} bow shock at \SI{24}{\um} even when there is no
corresponding dense shell at all.

Finally, case~(iv) arises from the observation that several of the
Orion sources possess high velocity (\(> \SI{100}{km.s^{-1}}\))
collimated jets \citep{Bally:2006a}, which produce strings of emission
knots that partially overlap the wings of the bow shock.  In the case
of 109--246, LL~1, LL~4, LL~5, and LL~6, the projected jet axis is
roughly perpendicular to the projected bow shock axis, and two of
these (109--246 and LL~6) have the most extreme high values of \(\Pi\)
and \(\Lambda\).  There are several other Orion Nebula bow shock sources
with perpendicular jets (LL~2, 203--3039, 261--3018, 344--3020, and
LL~7), which we omitted from our sample because of difficulty in
measuring \(\Lambda\), but the majority also have large planitudes
\(\Pi > 5\). Only one source, 4468--605, has a jet oriented parallel to
the bow shock axis, and this has the smallest \(\Pi\) and third-smallest
\(\Lambda\) of the sample.  The circumstantial evidence thus points towards
the jets playing some role in shaping the bow shocks when they are
oriented perpendicular to the axis, even though we took care when
tracing the bow shock ridges to avoid any region with superimposed jet
knots.  This would break the cylindrical symmetry of the bow shock,
which should have a kinematic signature.  However, in the only two
sources that have been studied kinematically \citep{Henney:2013a}, the
bow shock shell does not show any sign of the red/blue asymmetry that
is seen in the jet knots, which argues against any such association.

 % end input ./sec-standing-wave.tex
 % start input ./sec-obs-conclusions.tex

\section{Summary}
\label{sec:conclusion}

We have presented a statistical study of the shapes of three different
classes of stellar bow shocks, as characterized by their flatness of
apex (planitude, \(\Pi\)) and openness of wings (alatude,
\(\Lambda\)), following the terminology of
\citet[Paper~0]{Tarango-Yong:2018a}.  Our principal findings are as
follows:
\begin{enumerate}[1.]
\item Bow shocks driven by hot OB stars, from the mid-infrared catalog
  of \citet{Kobulnicky:2016a}, have an average shape
  \((\Pi, \Lambda) \approx (1.6, 1.7)\) that is consistent with predictions of the
  \citet{Wilkin:1996a} analytic model, but the dispersion in observed
  shapes \((\sigma_\Pi, \sigma_\Lambda) \approx (0.9, 0.3)\) is many times larger than
  predicted by that model.
\item The bow shock shapes show little correlation with other
  parameters of the source, such as stellar magnitude, Galactic
  latitude or longitude, extinction, or type of environment (cluster
  versus isolated).  The only exception is that the dispersion in
  alatude is higher for bow shocks of larger angular size.
\item A possible explanation for the previous results is that the
  variation in shapes is caused by time dependent oscillations in the
  bow shock surface, with relative amplitude of 10 to 20\% and
  wavelength of order the bow shock size.  The oscillations may either
  be due to dynamic instabilities in the bow shock shell or be driven
  by temporal variations in the stellar wind.  If the oscillations
  were more vigorous for stars with more powerful winds, it could
  explain the correlation with angular size.
\item Bow shocks driven by cool luminous stars (red supergiants and
  asymptotic giant branch stars), from the catalog of
  \citep{Cox:2012a}, have an average shape
  \((\Pi, \Lambda) \approx (1.5, 1.4)\), which has a significantly smaller alatude
  than the OB star sources, and which is not consistent with the
  \citet{Wilkin:1996a} model.  We suggest that this may be due to
  their dust emission being dominated by shocked stellar wind
  material, instead of shocked ambient material as is the case with
  the OB stars.
\item Bow shocks driven by proplyds and other young stars in the outer
  regions of the Orion Nebula, from the catalogs of
  \citet{Bally:2006a} and \citet{Gutierrez-Soto:2015a}, have an
  average shape of \((\Pi, \Lambda) \approx (2.7, 2.4)\), with a significant tail up
  to \((\Pi, \Lambda) \approx (7, 4)\).  A minority of these sources
  (\(\approx 20\%\)) have shapes similar to the OB star bow shocks, but the
  remainder have much flatter apexes and more open wings.  We suggest
  several possible mechanisms to explain this difference: divergent
  ambient flow; low Mach number; observational biases; influence of
  collimated jets, but the available evidence for and against each of
  these is mixed.
\end{enumerate}

 % end input ./sec-obs-conclusions.tex
 
\section*{Acknowledgements}
We are grateful for financial support provided by Dirección General de
Asuntos del Personal Académico, Universidad Nacional Autónoma de
México, through grants Programa de Apoyo a Proyectos de Investigación
e Inovación Tecnológica IN111215 and IN107019.  

\bibliographystyle{mnras}

\begin{thebibliography}{}
\makeatletter
\relax
\def\mn@urlcharsother{\let\do\@makeother \do\$\do\&\do\#\do\^\do\_\do\%\do\~}
\def\mn@doi{\begingroup\mn@urlcharsother \@ifnextchar [ {\mn@doi@}
  {\mn@doi@[]}}
\def\mn@doi@[#1]#2{\def\@tempa{#1}\ifx\@tempa\@empty \href
  {http://dx.doi.org/#2} {doi:#2}\else \href {http://dx.doi.org/#2} {#1}\fi
  \endgroup}
\def\mn@eprint#1#2{\mn@eprint@#1:#2::\@nil}
\def\mn@eprint@arXiv#1{\href {http://arxiv.org/abs/#1} {{\tt arXiv:#1}}}
\def\mn@eprint@dblp#1{\href {http://dblp.uni-trier.de/rec/bibtex/#1.xml}
  {dblp:#1}}
\def\mn@eprint@#1:#2:#3:#4\@nil{\def\@tempa {#1}\def\@tempb {#2}\def\@tempc
  {#3}\ifx \@tempc \@empty \let \@tempc \@tempb \let \@tempb \@tempa \fi \ifx
  \@tempb \@empty \def\@tempb {arXiv}\fi \@ifundefined
  {mn@eprint@\@tempb}{\@tempb:\@tempc}{\expandafter \expandafter \csname
  mn@eprint@\@tempb\endcsname \expandafter{\@tempc}}}

\bibitem[\protect\citeauthoryear{{Acreman}, {Stevens}  \& {Harries}}{{Acreman}
  et~al.}{2016}]{Acreman:2016a}
{Acreman} D.~M.,  {Stevens} I.~R.,   {Harries} T.~J.,  2016, \mnras, 456, 136

\bibitem[\protect\citeauthoryear{Anderson \& Darling}{Anderson \&
  Darling}{1952}]{Anderson:1952a}
Anderson T.~W.,  Darling D.~A.,  1952, Ann. Math. Statist., 23, 193

\bibitem[\protect\citeauthoryear{{Arthur} \& {Hoare}}{{Arthur} \&
  {Hoare}}{2006}]{Arthur:2006a}
{Arthur} S.~J.,  {Hoare} M.~G.,  2006, \apjs, 165, 283

\bibitem[\protect\citeauthoryear{{Astropy Collaboration} et~al.,}{{Astropy
  Collaboration} et~al.}{2018}]{Astropy-Collaboration:2018a}
{Astropy Collaboration} et~al., 2018, \aj, 156, 123

\bibitem[\protect\citeauthoryear{{Bally} \& {Reipurth}}{{Bally} \&
  {Reipurth}}{2001}]{Bally:2001a}
{Bally} J.,  {Reipurth} B.,  2001, \apj, 546, 299

\bibitem[\protect\citeauthoryear{{Bally}, {Sutherland}, {Devine}  \&
  {Johnstone}}{{Bally} et~al.}{1998}]{Bally:1998a}
{Bally} J.,  {Sutherland} R.~S.,  {Devine} D.,   {Johnstone} D.,  1998, \aj,
  116, 293

\bibitem[\protect\citeauthoryear{{Bally}, {O'Dell}  \& {McCaughrean}}{{Bally}
  et~al.}{2000}]{Bally:2000a}
{Bally} J.,  {O'Dell} C.~R.,   {McCaughrean} M.~J.,  2000, \aj, 119, 2919

\bibitem[\protect\citeauthoryear{{Bally}, {Licht}, {Smith}  \&
  {Walawender}}{{Bally} et~al.}{2006}]{Bally:2006a}
{Bally} J.,  {Licht} D.,  {Smith} N.,   {Walawender} J.,  2006, \aj, 131, 473

\bibitem[\protect\citeauthoryear{{Baranov}, {Krasnobaev}  \&
  {Kulikovskii}}{{Baranov} et~al.}{1970}]{Baranov:1970a}
{Baranov} V.~B.,  {Krasnobaev} K.~V.,   {Kulikovskii} A.~G.,  1970, Akademiia
  Nauk SSSR Doklady, 194, 41

\bibitem[\protect\citeauthoryear{{Baranov}, {Krasnobaev}  \&
  {Kulikovskii}}{{Baranov} et~al.}{1971}]{Baranov:1971a}
{Baranov} V.~B.,  {Krasnobaev} K.~V.,   {Kulikovskii} A.~G.,  1971, Soviet
  Physics Doklady, 15, 791

\bibitem[\protect\citeauthoryear{{Benaglia}, {Romero}, {Mart{\'{\i}}}, {Peri}
  \& {Araudo}}{{Benaglia} et~al.}{2010}]{Benaglia:2010a}
{Benaglia} P.,  {Romero} G.~E.,  {Mart{\'{\i}}} J.,  {Peri} C.~S.,   {Araudo}
  A.~T.,  2010, \aap, 517, L10

\bibitem[\protect\citeauthoryear{{Blondin} \& {Koerwer}}{{Blondin} \&
  {Koerwer}}{1998}]{Blondin:1998a}
{Blondin} J.~M.,  {Koerwer} J.~F.,  1998, \na, 3, 571

\bibitem[\protect\citeauthoryear{{Bodensteiner}, {Baade}, {Greiner}  \&
  {Langer}}{{Bodensteiner} et~al.}{2018}]{Bodensteiner:2018a}
{Bodensteiner} J.,  {Baade} D.,  {Greiner} J.,   {Langer} N.,  2018, \aap, 618,
  A110

\bibitem[\protect\citeauthoryear{{Brown} \& {Bomans}}{{Brown} \&
  {Bomans}}{2005}]{Brown:2005a}
{Brown} D.,  {Bomans} D.~J.,  2005, \aap, 439, 183

\bibitem[\protect\citeauthoryear{Brown \& Forsythe}{Brown \&
  Forsythe}{1974}]{Brown:1974a}
Brown M.~B.,  Forsythe A.~B.,  1974, Journal of the American Statistical
  Association, 69, 364

\bibitem[\protect\citeauthoryear{{Brownsberger} \& {Romani}}{{Brownsberger} \&
  {Romani}}{2014}]{Brownsberger:2014a}
{Brownsberger} S.,  {Romani} R.~W.,  2014, \apj, 784, 154

\bibitem[\protect\citeauthoryear{{Canto}, {Raga}  \& {Wilkin}}{{Canto}
  et~al.}{1996}]{Canto:1996}
{Canto} J.,  {Raga} A.~C.,   {Wilkin} F.~P.,  1996, \mn@doi [\apj]
  {10.1086/177820}, \href {http://adsabs.harvard.edu/abs/1996ApJ...469..729C}
  {469, 729}

\bibitem[\protect\citeauthoryear{{Carey} et~al.,}{{Carey}
  et~al.}{2009}]{Carey:2009a}
{Carey} S.~J.,  et~al., 2009, \pasp, 121, 76

\bibitem[\protect\citeauthoryear{{Comeron} \& {Kaper}}{{Comeron} \&
  {Kaper}}{1998}]{Comeron:1998a}
{Comeron} F.,  {Kaper} L.,  1998, \aap, 338, 273

\bibitem[\protect\citeauthoryear{{Cox} et~al.,}{{Cox} et~al.}{2012}]{Cox:2012a}
{Cox} N.~L.~J.,  et~al., 2012, \aap, 537, A35

\bibitem[\protect\citeauthoryear{Cureton}{Cureton}{1956}]{Cureton:1956a}
Cureton E.~E.,  1956, Psychometrika, 21, 287

\bibitem[\protect\citeauthoryear{{Cyganowski}, {Reid}, {Fish}  \&
  {Ho}}{{Cyganowski} et~al.}{2003}]{Cyganowski:2003a}
{Cyganowski} C.~J.,  {Reid} M.~J.,  {Fish} V.~L.,   {Ho} P.~T.~P.,  2003, \apj,
  596, 344

\bibitem[\protect\citeauthoryear{{Dgani}, {van Buren}  \&
  {Noriega-Crespo}}{{Dgani} et~al.}{1996a}]{Dgani:1996a}
{Dgani} R.,  {van Buren} D.,   {Noriega-Crespo} A.,  1996a, \apj, 461, 372

\bibitem[\protect\citeauthoryear{{Dgani}, {van Buren}  \&
  {Noriega-Crespo}}{{Dgani} et~al.}{1996b}]{Dgani:1996b}
{Dgani} R.,  {van Buren} D.,   {Noriega-Crespo} A.,  1996b, \apj, 461, 927

\bibitem[\protect\citeauthoryear{{Dyson} \& {de Vries}}{{Dyson} \& {de
  Vries}}{1972}]{Dyson:1972a}
{Dyson} J.~E.,  {de Vries} J.,  1972, \aap, 20, 223

\bibitem[\protect\citeauthoryear{{Dzib} et~al.,}{{Dzib}
  et~al.}{2017}]{Dzib:2017a}
{Dzib} S.~A.,  et~al., 2017, \apj, 834, 139

\bibitem[\protect\citeauthoryear{{Eker} et~al.,}{{Eker}
  et~al.}{2015}]{Eker:2015a}
{Eker} Z.,  et~al., 2015, \aj, 149, 131

\bibitem[\protect\citeauthoryear{{Garc{\'{\i}}a-Arredondo}, {Henney}  \&
  {Arthur}}{{Garc{\'{\i}}a-Arredondo} et~al.}{2001}]{Garcia-Arredondo:2001a}
{Garc{\'{\i}}a-Arredondo} F.,  {Henney} W.~J.,   {Arthur} S.~J.,  2001, \apj,
  561, 830

\bibitem[\protect\citeauthoryear{{Geballe}, {Rigaut}, {Roy}  \&
  {Draine}}{{Geballe} et~al.}{2004}]{Geballe:2004a}
{Geballe} T.~R.,  {Rigaut} F.,  {Roy} J.-R.,   {Draine} B.~T.,  2004, \apj,
  602, 770

\bibitem[\protect\citeauthoryear{{Groenewegen} et~al.,}{{Groenewegen}
  et~al.}{2011}]{Groenewegen:2011a}
{Groenewegen} M.~A.~T.,  et~al., 2011, \aap, 526, A162

\bibitem[\protect\citeauthoryear{{Gull} \& {Sofia}}{{Gull} \&
  {Sofia}}{1979}]{Gull:1979a}
{Gull} T.~R.,  {Sofia} S.,  1979, \apj, 230, 782

\bibitem[\protect\citeauthoryear{Guti{\'e}rrez-Soto}{Guti{\'e}rrez-Soto}{2015}]{Gutierrez-Soto:2015a}
Guti{\'e}rrez-Soto L.~{\'A}.,  2015, Master's thesis, Universidad Nacional
  Aut{\'o}noma de M{\'e}xico

\bibitem[\protect\citeauthoryear{{Gvaramadze} \& {Bomans}}{{Gvaramadze} \&
  {Bomans}}{2008}]{Gvaramadze:2008a}
{Gvaramadze} V.~V.,  {Bomans} D.~J.,  2008, \aap, 490, 1071

\bibitem[\protect\citeauthoryear{{Hayward}, {Houck}  \& {Miles}}{{Hayward}
  et~al.}{1994}]{Hayward:1994a}
{Hayward} T.~L.,  {Houck} J.~R.,   {Miles} J.~W.,  1994, \apj, 433, 157

\bibitem[\protect\citeauthoryear{Head, Holman, Lanfear, Kahn  \& Jennions}{Head
  et~al.}{2015}]{Head:2015a}
Head M.~L.,  Holman L.,  Lanfear R.,  Kahn A.~T.,   Jennions M.~D.,  2015, PLOS
  Biology, 13, 1

\bibitem[\protect\citeauthoryear{{Henney} \& {Arthur}}{{Henney} \&
  {Arthur}}{2019a}]{Henney:2019a}
{Henney} W.~J.,  {Arthur} S.~J.,  2019a, \mnras, \href
  {http://adsabs.harvard.edu/abs/2019arXiv190303737H} {486, 3423 (Paper I)}

\bibitem[\protect\citeauthoryear{{Henney} \& {Arthur}}{{Henney} \&
  {Arthur}}{2019b}]{Henney:2019b}
{Henney} W.~J.,  {Arthur} S.~J.,  2019b, \mnras, \href
  {http://adsabs.harvard.edu/abs/2019arXiv190307774H} {486, 4423 (Paper II)}

\bibitem[\protect\citeauthoryear{{Henney} \& {Arthur}}{{Henney} \&
  {Arthur}}{2019c}]{Henney:2019c}
{Henney} W.~J.,  {Arthur} S.~J.,  2019c, arXiv e-prints, \href
  {http://adsabs.harvard.edu/abs/2019arXiv190400343H} {1904.00343 MNRAS
  submitted (Paper III)}

\bibitem[\protect\citeauthoryear{{Henney}, {Arthur}  \&
  {Garc{\'{\i}}a-D{\'{\i}}az}}{{Henney} et~al.}{2005}]{Henney:2005a}
{Henney} W.~J.,  {Arthur} S.~J.,   {Garc{\'{\i}}a-D{\'{\i}}az} M.~T.,  2005,
  \apj, 627, 813

\bibitem[\protect\citeauthoryear{{Henney}, {Garc{\'{\i}}a-D{\'{\i}}az},
  {O'Dell}  \& {Rubin}}{{Henney} et~al.}{2013}]{Henney:2013a}
{Henney} W.~J.,  {Garc{\'{\i}}a-D{\'{\i}}az} M.~T.,  {O'Dell} C.~R.,   {Rubin}
  R.~H.,  2013, \mnras, 428, 691

\bibitem[\protect\citeauthoryear{{Immer}, {Cyganowski}, {Reid}  \&
  {Menten}}{{Immer} et~al.}{2014}]{Immer:2014a}
{Immer} K.,  {Cyganowski} C.,  {Reid} M.~J.,   {Menten} K.~M.,  2014, \aap,
  563, A39

\bibitem[\protect\citeauthoryear{{Indebetouw} et~al.,}{{Indebetouw}
  et~al.}{2005}]{Indebetouw:2005a}
{Indebetouw} R.,  et~al., 2005, \apj, 619, 931

\bibitem[\protect\citeauthoryear{{Johnstone}, {Hollenbach}  \&
  {Bally}}{{Johnstone} et~al.}{1998}]{Johnstone:1998a}
{Johnstone} D.,  {Hollenbach} D.,   {Bally} J.,  1998, \apj, 499, 758

\bibitem[\protect\citeauthoryear{{Joye} \& {Mandel}}{{Joye} \&
  {Mandel}}{2003}]{Joye:2003a}
{Joye} W.~A.,  {Mandel} E.,  2003, in {Payne} H.~E.,  {Jedrzejewski} R.~I.,
  {Hook} R.~N.,  eds,  Astronomical Society of the Pacific Conference Series
  Vol. 295, Astronomical Data Analysis Software and Systems XII. p.~489

\bibitem[\protect\citeauthoryear{{Kim}, {Lu}, {Konopacky}, {Chu}, {Toller},
  {Anderson}, {Theissen}  \& {Morris}}{{Kim} et~al.}{2018}]{Kim:2018b}
{Kim} D.,  {Lu} J.~R.,  {Konopacky} Q.,  {Chu} L.,  {Toller} E.,  {Anderson}
  J.,  {Theissen} C.~A.,   {Morris} M.~R.,  2018, arXiv e-prints

\bibitem[\protect\citeauthoryear{{Klaassen} et~al.,}{{Klaassen}
  et~al.}{2018}]{Klaassen:2018a}
{Klaassen} P.~D.,  et~al., 2018, \aap, 611, A99

\bibitem[\protect\citeauthoryear{{Kobulnicky}, {Gilbert}  \&
  {Kiminki}}{{Kobulnicky} et~al.}{2010}]{Kobulnicky:2010a}
{Kobulnicky} H.~A.,  {Gilbert} I.~J.,   {Kiminki} D.~C.,  2010, \apj, 710, 549

\bibitem[\protect\citeauthoryear{{Kobulnicky} et~al.,}{{Kobulnicky}
  et~al.}{2016}]{Kobulnicky:2016a}
{Kobulnicky} H.~A.,  et~al., 2016, \apjs, 227, 18

\bibitem[\protect\citeauthoryear{{Kobulnicky}, {Schurhammer}, {Baldwin},
  {Chick}, {Dixon}, {Lee}  \& {Povich}}{{Kobulnicky}
  et~al.}{2017}]{Kobulnicky:2017a}
{Kobulnicky} H.~A.,  {Schurhammer} D.~P.,  {Baldwin} D.~J.,  {Chick} W.~T.,
  {Dixon} D.~M.,  {Lee} D.,   {Povich} M.~S.,  2017, \aj, 154, 201

\bibitem[\protect\citeauthoryear{{Kobulnicky}, {Chick}  \&
  {Povich}}{{Kobulnicky} et~al.}{2018}]{Kobulnicky:2018a}
{Kobulnicky} H.~A.,  {Chick} W.~T.,   {Povich} M.~S.,  2018, \apj, 856, 74
  (K18)

\bibitem[\protect\citeauthoryear{{Kulkarni} \& {Hester}}{{Kulkarni} \&
  {Hester}}{1988}]{Kulkarni:1988a}
{Kulkarni} S.~R.,  {Hester} J.~J.,  1988, \nat, 335, 801

\bibitem[\protect\citeauthoryear{Landau \& Lifshitz}{Landau \&
  Lifshitz}{1987}]{Landau:1987a}
Landau L.~D.,  Lifshitz E.~M.,  1987, Fluid Mechanics, Second Edition: Volume 6
  (Course of Theoretical Physics), 2 edn.
Course of theoretical physics / by L. D. Landau and E. M. Lifshitz, Vol. 6,
  Butterworth-Heinemann

\bibitem[\protect\citeauthoryear{Leiva-Murillo \&
  Art{\'e}s-Rodr\'{\i}guez}{Leiva-Murillo \&
  Art{\'e}s-Rodr\'{\i}guez}{2012}]{Leiva-Murillo:2012a}
Leiva-Murillo J.~M.,  Art{\'e}s-Rodr\'{\i}guez A.,  2012, Pattern Recogn.
  Lett., 33, 1717

\bibitem[\protect\citeauthoryear{{Mac Low}, {van Buren}, {Wood}  \&
  {Churchwell}}{{Mac Low} et~al.}{1991}]{Mac-Low:1991a}
{Mac Low} M.-M.,  {van Buren} D.,  {Wood} D.~O.~S.,   {Churchwell} E.,  1991,
  \apj, 369, 395

\bibitem[\protect\citeauthoryear{{Mackey}, {Gvaramadze}, {Mohamed}  \&
  {Langer}}{{Mackey} et~al.}{2015}]{Mackey:2015a}
{Mackey} J.,  {Gvaramadze} V.~V.,  {Mohamed} S.,   {Langer} N.,  2015, \aap,
  573, A10

\bibitem[\protect\citeauthoryear{{Mackey}, {Haworth}, {Gvaramadze}, {Mohamed},
  {Langer}  \& {Harries}}{{Mackey} et~al.}{2016}]{Mackey:2016a}
{Mackey} J.,  {Haworth} T.~J.,  {Gvaramadze} V.~V.,  {Mohamed} S.,  {Langer}
  N.,   {Harries} T.~J.,  2016, \aap, 586, A114

\bibitem[\protect\citeauthoryear{{Majewski}, {Zasowski}  \&
  {Nidever}}{{Majewski} et~al.}{2011}]{Majewski:2011a}
{Majewski} S.~R.,  {Zasowski} G.,   {Nidever} D.~L.,  2011, \apj, 739, 25

\bibitem[\protect\citeauthoryear{Makarov \& Simonova}{Makarov \&
  Simonova}{2017}]{Makarov:2017a}
Makarov A.~A.,  Simonova G.~I.,  2017, Journal of Mathematical Sciences, 221,
  580

\bibitem[\protect\citeauthoryear{Mann \& Whitney}{Mann \&
  Whitney}{1947}]{Mann:1947a}
Mann H.~B.,  Whitney D.~R.,  1947, The Annals of Mathematical Statistics, 18,
  50

\bibitem[\protect\citeauthoryear{{Meyer}, {Gvaramadze}, {Langer}, {Mackey},
  {Boumis}  \& {Mohamed}}{{Meyer} et~al.}{2014a}]{Meyer:2014a}
{Meyer} D.~M.-A.,  {Gvaramadze} V.~V.,  {Langer} N.,  {Mackey} J.,  {Boumis}
  P.,   {Mohamed} S.,  2014a, \mnras, 439, L41

\bibitem[\protect\citeauthoryear{{Meyer}, {Mackey}, {Langer}, {Gvaramadze},
  {Mignone}, {Izzard}  \& {Kaper}}{{Meyer} et~al.}{2014b}]{Meyer:2014b}
{Meyer} D.~M.-A.,  {Mackey} J.,  {Langer} N.,  {Gvaramadze} V.~V.,  {Mignone}
  A.,  {Izzard} R.~G.,   {Kaper} L.,  2014b, \mnras, 444, 2754

\bibitem[\protect\citeauthoryear{{Meyer}, {van Marle}, {Kuiper}  \&
  {Kley}}{{Meyer} et~al.}{2016}]{Meyer:2016a}
{Meyer} D.~M.-A.,  {van Marle} A.-J.,  {Kuiper} R.,   {Kley} W.,  2016, \mnras,
  459, 1146

\bibitem[\protect\citeauthoryear{{Meyer}, {Mignone}, {Kuiper}, {Raga}  \&
  {Kley}}{{Meyer} et~al.}{2017}]{Meyer:2017a}
{Meyer} D.~M.-A.,  {Mignone} A.,  {Kuiper} R.,  {Raga} A.~C.,   {Kley} W.,
  2017, \mnras, 464, 3229

\bibitem[\protect\citeauthoryear{{Mohamed}, {Mackey}  \& {Langer}}{{Mohamed}
  et~al.}{2012}]{Mohamed:2012a}
{Mohamed} S.,  {Mackey} J.,   {Langer} N.,  2012, \aap, 541, A1

\bibitem[\protect\citeauthoryear{{Noriega-Crespo}, {van Buren}  \&
  {Dgani}}{{Noriega-Crespo} et~al.}{1997}]{Noriega-Crespo:1997b}
{Noriega-Crespo} A.,  {van Buren} D.,   {Dgani} R.,  1997, \aj, 113, 780

\bibitem[\protect\citeauthoryear{{O'Dell}}{{O'Dell}}{2001}]{ODell:2001c}
{O'Dell} C.~R.,  2001, \aj, 122, 2662

\bibitem[\protect\citeauthoryear{{O'Dell} \& {Wen}}{{O'Dell} \&
  {Wen}}{1994}]{ODell:1994a}
{O'Dell} C.~R.,  {Wen} Z.,  1994, \apj, 436, 194

\bibitem[\protect\citeauthoryear{{O'Dell}, {Wen}  \& {Hu}}{{O'Dell}
  et~al.}{1993}]{ODell:1993a}
{O'Dell} C.~R.,  {Wen} Z.,   {Hu} X.,  1993, \apj, 410, 696

\bibitem[\protect\citeauthoryear{{O'Dell}, {Muench}, {Smith}  \&
  {Zapata}}{{O'Dell} et~al.}{2008}]{ODell:2008b}
{O'Dell} C.~R.,  {Muench} A.,  {Smith} N.,   {Zapata} L.,  2008, {Star
  Formation in the Orion Nebula II: Gas, Dust, Proplyds and Outflows}.
p.~544

\bibitem[\protect\citeauthoryear{{Paltani}}{{Paltani}}{2004}]{Paltani:2004a}
{Paltani} S.,  2004, \aap, 420, 789

\bibitem[\protect\citeauthoryear{{Peri}, {Benaglia}, {Brookes}, {Stevens}  \&
  {Isequilla}}{{Peri} et~al.}{2012}]{Peri:2012a}
{Peri} C.~S.,  {Benaglia} P.,  {Brookes} D.~P.,  {Stevens} I.~R.,   {Isequilla}
  N.~L.,  2012, \aap, 538, A108

\bibitem[\protect\citeauthoryear{{Peri}, {Benaglia}  \& {Isequilla}}{{Peri}
  et~al.}{2015}]{Peri:2015a}
{Peri} C.~S.,  {Benaglia} P.,   {Isequilla} N.~L.,  2015, \aap, 578, A45

\bibitem[\protect\citeauthoryear{{Pikel'ner}}{{Pikel'ner}}{1968}]{Pikelner:1968a}
{Pikel'ner} S.~B.,  1968, \aplett, 2, 97

\bibitem[\protect\citeauthoryear{{Povich}, {Benjamin}, {Whitney}, {Babler},
  {Indebetouw}, {Meade}  \& {Churchwell}}{{Povich} et~al.}{2008}]{Povich:2008a}
{Povich} M.~S.,  {Benjamin} R.~A.,  {Whitney} B.~A.,  {Babler} B.~L.,
  {Indebetouw} R.,  {Meade} M.~R.,   {Churchwell} E.,  2008, \apj, 689, 242

\bibitem[\protect\citeauthoryear{{Puls} et~al.,}{{Puls}
  et~al.}{1996}]{Puls:1996a}
{Puls} J.,  et~al., 1996, \aap, 305, 171

\bibitem[\protect\citeauthoryear{{Reid} \& {Ho}}{{Reid} \&
  {Ho}}{1985}]{Reid:1985a}
{Reid} M.~J.,  {Ho} P.~T.~P.,  1985, \apjl, 288, L17

\bibitem[\protect\citeauthoryear{{Robberto} et~al.,}{{Robberto}
  et~al.}{2005}]{Robberto:2005a}
{Robberto} M.,  et~al., 2005, \aj, 129, 1534

\bibitem[\protect\citeauthoryear{{Robberto} et~al.,}{{Robberto}
  et~al.}{2013}]{Robberto:2013a}
{Robberto} M.,  et~al., 2013, \apjs, 207, 10

\bibitem[\protect\citeauthoryear{{Sahai} \& {Chronopoulos}}{{Sahai} \&
  {Chronopoulos}}{2010}]{Sahai:2010a}
{Sahai} R.,  {Chronopoulos} C.~K.,  2010, \apjl, 711, L53

\bibitem[\protect\citeauthoryear{{Sahai}, {Güsten}  \& {Morris}}{{Sahai}
  et~al.}{2012}]{Sahai:2012b}
{Sahai} R.,  {Güsten} R.,   {Morris} M.~R.,  2012, \apjl, 761, L21

\bibitem[\protect\citeauthoryear{{Sanchez-Bermudez}, {Schödel}, {Alberdi},
  {Muzić}, {Hummel}  \& {Pott}}{{Sanchez-Bermudez}
  et~al.}{2014}]{Sanchez-Bermudez:2014a}
{Sanchez-Bermudez} J.,  {Schödel} R.,  {Alberdi} A.,  {Muzić} K.,  {Hummel}
  C.~A.,   {Pott} J.-U.,  2014, \aap, 567, A21

\bibitem[\protect\citeauthoryear{Scholz \& Stephens}{Scholz \&
  Stephens}{1987}]{Scholz:1987a}
Scholz F.~W.,  Stephens M.~A.,  1987, Journal of the American Statistical
  Association, 82, 918

\bibitem[\protect\citeauthoryear{Scott}{Scott}{2015}]{Scott:2015a}
Scott D.~W.,  2015, Multivariate density estimation: theory, practice, and
  visualization, 2nd edn.
John Wiley \& Sons

\bibitem[\protect\citeauthoryear{{Sexton}, {Povich}, {Smith}, {Babler}, {Meade}
   \& {Rudolph}}{{Sexton} et~al.}{2015}]{Sexton:2015b}
{Sexton} R.~O.,  {Povich} M.~S.,  {Smith} N.,  {Babler} B.~L.,  {Meade} M.~R.,
   {Rudolph} A.~L.,  2015, \mnras, 446, 1047

\bibitem[\protect\citeauthoryear{{Smith}, {Bally}, {Shuping}, {Morris}  \&
  {Kassis}}{{Smith} et~al.}{2005}]{Smith:2005a}
{Smith} N.,  {Bally} J.,  {Shuping} R.~Y.,  {Morris} M.,   {Kassis} M.,  2005,
  \aj, 130, 1763

\bibitem[\protect\citeauthoryear{{Steggles}, {Hoare}  \& {Pittard}}{{Steggles}
  et~al.}{2017}]{Steggles:2017a}
{Steggles} H.~G.,  {Hoare} M.~G.,   {Pittard} J.~M.,  2017, \mnras, 466, 4573

\bibitem[\protect\citeauthoryear{Stephens}{Stephens}{1970}]{Stephens:1970a}
Stephens M.~A.,  1970, Journal of the Royal Statistical Society. Series B
  (Methodological), 32, 115

\bibitem[\protect\citeauthoryear{{Tanner}, {Ghez}, {Morris}  \&
  {Christou}}{{Tanner} et~al.}{2005}]{Tanner:2005a}
{Tanner} A.,  {Ghez} A.~M.,  {Morris} M.~R.,   {Christou} J.~C.,  2005, \apj,
  624, 742

\bibitem[\protect\citeauthoryear{{Tarango-Yong} \& {Henney}}{{Tarango-Yong} \&
  {Henney}}{2018}]{Tarango-Yong:2018a}
{Tarango-Yong} J.~A.,  {Henney} W.~J.,  2018, \mnras, 477, 2431 (Paper 0)

\bibitem[\protect\citeauthoryear{{Ueta} et~al.,}{{Ueta}
  et~al.}{2006}]{Ueta:2006a}
{Ueta} T.,  et~al., 2006, \apjl, 648, L39

\bibitem[\protect\citeauthoryear{{Ueta} et~al.,}{{Ueta}
  et~al.}{2008}]{Ueta:2008a}
{Ueta} T.,  et~al., 2008, \pasj, 60, S407

\bibitem[\protect\citeauthoryear{{Weilbacher} et~al.,}{{Weilbacher}
  et~al.}{2015}]{Weilbacher:2015a}
{Weilbacher} P.~M.,  et~al., 2015, \aap, 582, A114

\bibitem[\protect\citeauthoryear{{Werner} et~al.,}{{Werner}
  et~al.}{2004}]{Werner:2004a}
{Werner} M.~W.,  et~al., 2004, \apjs, 154, 1

\bibitem[\protect\citeauthoryear{{Wilkin}}{{Wilkin}}{1996}]{Wilkin:1996a}
{Wilkin} F.~P.,  1996, \apjl, 459, L31

\bibitem[\protect\citeauthoryear{{Wood} \& {Churchwell}}{{Wood} \&
  {Churchwell}}{1989}]{Wood:1989a}
{Wood} D.~O.~S.,  {Churchwell} E.,  1989, \apjs, 69, 831

\bibitem[\protect\citeauthoryear{{Wright} et~al.,}{{Wright}
  et~al.}{2010}]{Wright:2010a}
{Wright} E.~L.,  et~al., 2010, \aj, 140, 1868

\bibitem[\protect\citeauthoryear{{Zuckerman}}{{Zuckerman}}{1973}]{Zuckerman:1973a}
{Zuckerman} B.,  1973, \apj, 183, 863

\bibitem[\protect\citeauthoryear{{van Buren} \& {McCray}}{{van Buren} \&
  {McCray}}{1988}]{van-Buren:1988a}
{van Buren} D.,  {McCray} R.,  1988, \apjl, 329, L93

\bibitem[\protect\citeauthoryear{{van Buren}, {Mac Low}, {Wood}  \&
  {Churchwell}}{{van Buren} et~al.}{1990}]{van-Buren:1990a}
{van Buren} D.,  {Mac Low} M.-M.,  {Wood} D.~O.~S.,   {Churchwell} E.,  1990,
  \apj, 353, 570

\bibitem[\protect\citeauthoryear{{van Buren}, {Noriega-Crespo}  \&
  {Dgani}}{{van Buren} et~al.}{1995}]{van-Buren:1995a}
{van Buren} D.,  {Noriega-Crespo} A.,   {Dgani} R.,  1995, \aj, 110, 2914

\bibitem[\protect\citeauthoryear{{van Marle}, {Decin}  \& {Meliani}}{{van
  Marle} et~al.}{2014}]{van-Marle:2014a}
{van Marle} A.~J.,  {Decin} L.,   {Meliani} Z.,  2014, \aap, 561, A152

\makeatother
\end{thebibliography}

\appendix
% start input ./app-p-values.tex

\section{Distribution of p-values for all correlations tested}
\label{sec:distr-p-values}

\begin{figure}
  (a)\\
  \includegraphics[width=\linewidth]{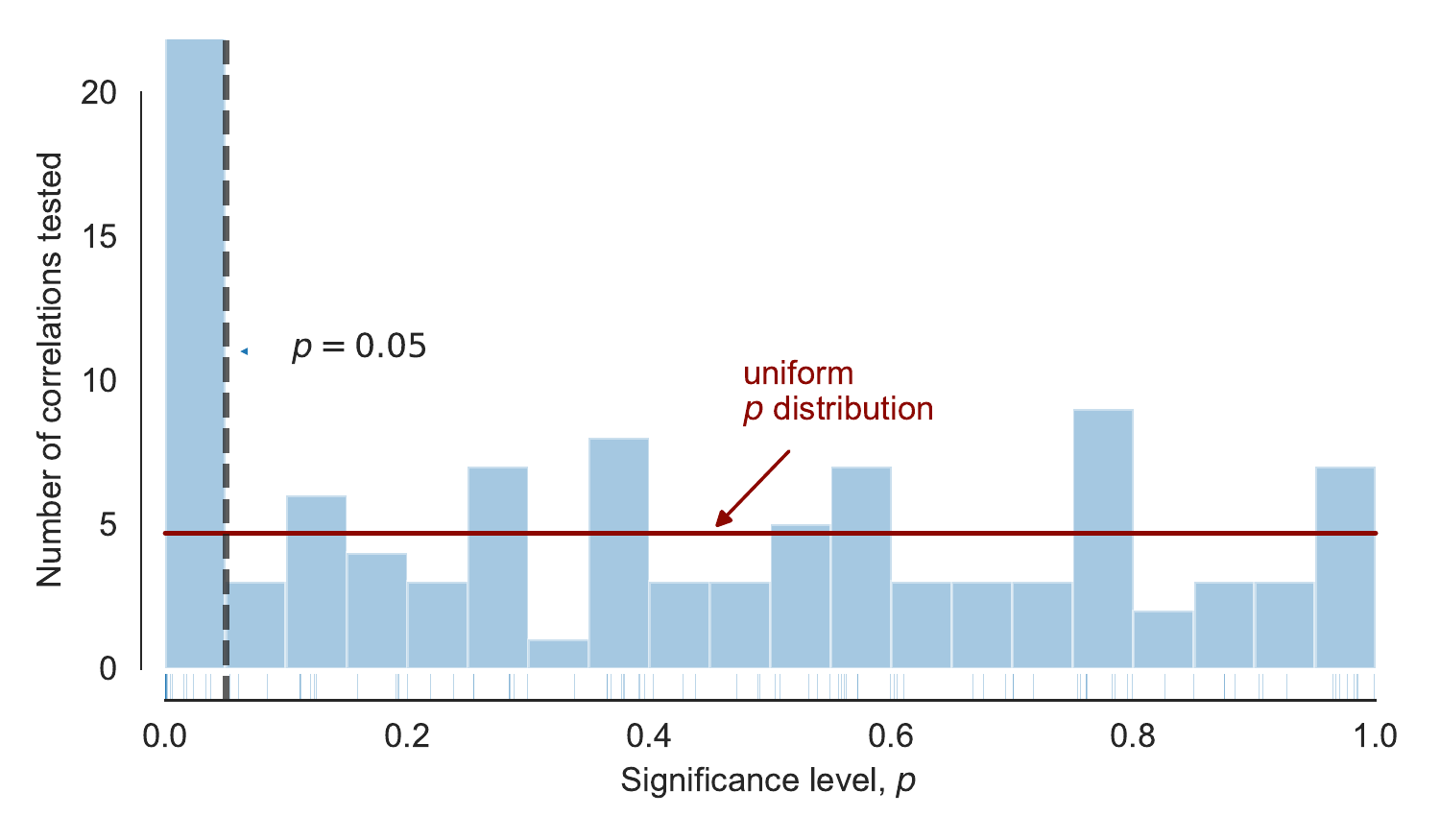}\\
  (b)\\
  \includegraphics[width=\linewidth]{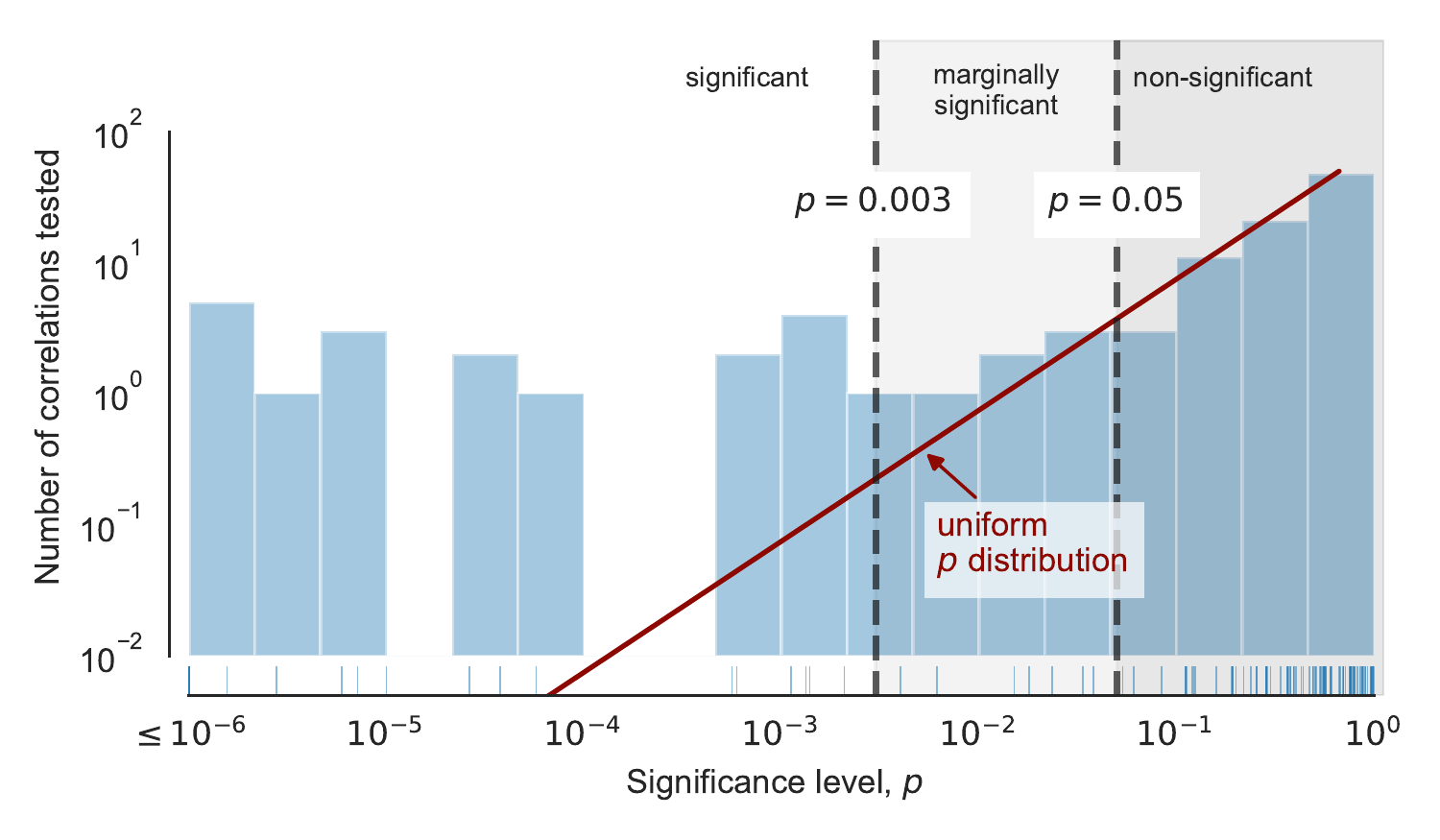}
  \caption{Histogram of \(p\)-values for all non-parametric 2-sample
    tests listed in Table~\ref{tab:big-p}. (a)~Uniformly spaced linear
    bins and linear vertical axis. (b)~Uniformly spaced logarithmic
    bins and logarithmic vertical axis, with all values
    \(p \le 10^{-6}\) included in the leftmost bin.  Short thin vertical
    lines above the horizontal axis show the individual values.  The
    thick vertical dashed lines show the traditional threshold values
    for significance: \(p = 0.003\) (\(\approx 3 \sigma\)) and
    \(p = 0.05\) (\(\approx 2 \sigma\)). The red solid line shows the uniform
    distribution of \(p\)-values that would be expected if the null
    hypothesis were always true, that is, if no significant
    correlations existed.}
  \label{fig:histo-p-values}
\end{figure}

\begin{table*}
  \sisetup{detect-all=true, detect-inline-weight=math}
  \sisetup{round-mode=figures, round-precision=2}
  \sisetup{table-align-exponent = false}
  \setlength\tabcolsep{2pt}
  \caption{Results of all statistical tests performed on observed bow
    shock shape parameters. Significant correlations are shown in
    \textbf{bold}, marginally significant correlations in
    \textit{italic}}
  \label{tab:big-p}
  % start input ./figs/mipsgal-summary-stats-table-body.tex
	\newlength\Width\settowidth\Width{Comparison}
	\begin{tabular}{
	  @{} ll @{\quad }
	  S[round-mode=places]S[round-mode=places]
	  S[round-mode=places]S[round-mode=places]
	  SS
	  @{\quad} SSS[round-mode=places]
	  @{\quad}
	  S[table-format = +1.2e1]
	  S[table-format = +1.2e1]
	  S[table-format = +1.2e1] @{}
	  }\toprule
	  & {Dependent}
	  & \multicolumn{2}{c}{Mean}
	  & \multicolumn{2}{c}{Std.\ Dev.}
	  & \multicolumn{2}{c}{Obs.\ Disp.}
	  & \multicolumn{3}{c @{} }{\dotfill Effect sizes\dotfill }
	  & \multicolumn{3}{c @{}}{Non-parametric test \(p\)-values} \\ 
	  {Comparison} & {Variable}
	  & {\(\langle \text{A} \rangle\)} & {\(\langle \text{B} \rangle\)}
	  & {\(\sigma_{\text{A}}\)} & {\(\sigma_{\text{B}}\)}
	  & {\(\langle \epsilon_{\text{A}} \rangle\)} & {\(\langle \epsilon_{\text{B}} \rangle\)}
	  & {\(r_b\)} & {Cohen \(d\)} & {\(\sigma_{\text{A}}/\sigma_{\text{B}}\)}
	  & {K--S} & {Rank} &  {B--F}\\
	  {\makebox[\Width]{(1)}} & \multicolumn{1}{c@{\quad}}{(2)}
	  & {(3)} & {(4)}
	  & {(5)} & {(6)}
	  & {(7)} & {(8)}
	  & {(9)} & {(10)} & {(11)}
	  & {(12)} & {(13)} & {(14)}  \\  
	  \midrule\multicolumn{10}{@{} l @{}}{\itshape Median split of continuous independent variables}\\
\addlinespace
Faint/bright & \(\Pi\) & 1.655 & 1.917 & 0.631 & 1.045 & 0.097 & 0.078 & 0.123 & 0.303 & \itshape 1.654 & 0.572 & 0.111 & \itshape 0.0335\\
\(H\) magnitude & \(\Lambda\) & 1.677 & 1.768 & 0.269 & 0.316 & 0.231 & 0.236 & \itshape 0.174 & \itshape 0.308 & 1.175 & \itshape 0.018 & \itshape 0.0235 & 0.125\\
\(n_{\text{A}} =  n_{\text{B}} = 113\) & \(\Delta \Lambda\) & 0.183 & 0.198 & 0.161 & 0.163 &   &   & 0.07 & 0.093 & 1.013 & 0.472 & 0.365 & 0.761\\
\addlinespace
Low/high & \(\Pi\) & 1.766 & 1.803 & 0.975 & 0.755 & 0.114 & 0.062 & 0.1 & 0.043 & 0.774 & 0.508 & 0.192 & 0.599\\
bow shock size, \(R_0\) & \(\Lambda\) & 1.707 & 1.739 & 0.25 & 0.336 & 0.256 & 0.212 & 0.061 & 0.11 & \bfseries 1.342 & \bfseries 0.00208 & 0.428 & \bfseries 0.00139\\
\(n_{\text{A}} =  n_{\text{B}} = 113\) & \(\Delta \Lambda\) & 0.175 & 0.204 & 0.157 & 0.165 &   &   & 0.091 & 0.176 & 1.054 & \itshape 0.00612 & 0.238 & 0.193\\
\addlinespace
Low/high & \(\Pi\) & 1.725 & 1.846 & 0.822 & 0.917 & 0.091 & 0.085 & 0.082 & 0.139 & 1.116 & 0.379 & 0.285 & 0.982\\
extinction, \(A_K\) & \(\Lambda\) & 1.703 & 1.742 & 0.267 & 0.323 & 0.233 & 0.235 & 0.04 & 0.132 & 1.213 & 0.572 & 0.602 & 0.123\\
\(n_{\text{A}} =  n_{\text{B}} = 113\) & \(\Delta \Lambda\) & 0.186 & 0.195 & 0.138 & 0.183 &   &   & -0.039 & 0.057 & 1.326 & 0.0609 & 0.61 & 0.112\\
\addlinespace
Low/high & \(\Pi\) & 1.706 & 1.862 & 0.727 & 0.988 & 0.085 & 0.091 & 0.069 & 0.181 & 1.358 & \itshape 0.0379 & 0.368 & 0.0842\\
\(\vert{}b\vert\) & \(\Lambda\) & 1.722 & 1.724 & 0.328 & 0.261 & 0.234 & 0.234 & 0.02 & 0.008 & 0.796 & 0.255 & 0.795 & 0.0534\\
\(n_{\text{A}} =  n_{\text{B}} = 113\) & \(\Delta \Lambda\) & 0.188 & 0.191 & 0.161 & 0.162 &   &   & 0.009 & 0.021 & 1.005 & 0.884 & 0.907 & 0.694\\
\addlinespace
High/low & \(\Pi\) & 1.807 & 1.751 & 0.946 & 0.742 & 0.09 & 0.084 & -0.0 & -0.064 & 0.785 & 0.985 & 0.999 & 0.549\\
\(\cos \ell\) & \(\Lambda\) & 1.734 & 1.707 & 0.279 & 0.321 & 0.241 & 0.223 & -0.049 & -0.093 & 1.152 & 0.504 & 0.532 & 0.159\\
\(n_{\text{A}}, n_{\text{B}} = 137, 90\) & \(\Delta \Lambda\) & 0.182 & 0.201 & 0.155 & 0.171 &   &   & 0.054 & 0.122 & 1.1 & 0.761 & 0.491 & 0.365\\
\midrule
\multicolumn{10}{@{} l @{}}{\itshape Categorical independent variables}\\
\addlinespace
Environment: & \(\Pi\) & 1.757 & 1.852 & 0.854 & 0.899 & 0.087 & 0.083 & 0.042 & 0.11 & 1.053 & 0.717 & 0.676 & 0.377\\
Isolated vs Facing & \(\Lambda\) & 1.735 & 1.693 & 0.283 & 0.338 & 0.238 & 0.218 & -0.07 & -0.142 & 1.194 & 0.965 & 0.49 & 0.392\\
\(n_{\text{A}}, n_{\text{B}} = 170, 41\) & \(\Delta \Lambda\) & 0.19 & 0.195 & 0.161 & 0.172 &   &   & -0.019 & 0.034 & 1.066 & 0.556 & 0.85 & 0.438\\
\addlinespace
Environment: & \(\Pi\) & 1.757 & 1.907 & 0.854 & 0.955 & 0.087 & 0.105 & 0.024 & 0.174 & 1.118 & 0.284 & 0.875 & 0.255\\
Isolated vs \hii & \(\Lambda\) & 1.735 & 1.68 & 0.283 & 0.309 & 0.238 & 0.233 & -0.13 & -0.193 & 1.092 & 0.701 & 0.391 & 0.782\\
\(n_{\text{A}}, n_{\text{B}} = 170, 16\) & \(\Delta \Lambda\) & 0.19 & 0.175 & 0.161 & 0.138 &   &   & -0.048 & -0.095 & 0.855 & 0.785 & 0.754 & 0.799\\
\addlinespace
Single/multiple & \(\Pi\) & 1.767 & 1.833 & 0.825 & 0.988 & 0.09 & 0.08 & 0.027 & 0.076 & 1.198 & 0.977 & 0.756 & 0.605\\
source candidate & \(\Lambda\) & 1.709 & 1.762 & 0.289 & 0.315 & 0.23 & 0.243 & 0.074 & 0.177 & 1.09 & 0.111 & 0.396 & 0.338\\
\(n_{\text{A}}, n_{\text{B}} = 167, 60\) & \(\Delta \Lambda\) & 0.184 & 0.206 & 0.162 & 0.16 &   &   & 0.093 & 0.136 & 0.987 & 0.559 & 0.284 & 0.97\\
\addlinespace
With/without & \(\Pi\) & 1.714 & 1.802 & 0.598 & 0.926 & 0.091 & 0.087 & -0.012 & 0.101 & 1.547 & 0.875 & 0.904 & 0.2\\
\SI{8}{\um} emission & \(\Lambda\) & 1.734 & 1.721 & 0.289 & 0.298 & 0.223 & 0.237 & -0.042 & -0.044 & 1.031 & 0.299 & 0.667 & 0.563\\
\(n_{\text{A}}, n_{\text{B}} = 45, 182\) & \(\Delta \Lambda\) & 0.203 & 0.186 & 0.205 & 0.149 &   &   & 0.021 & -0.106 & 0.726 & 0.967 & 0.826 & 0.219\\
\addlinespace
3-star vs (4+5)-star & \(\Pi\) & 1.632 & 2.0 & 0.91 & 0.764 & 0.106 & 0.061 & \bfseries 0.386 & \bfseries 0.431 & 0.84 & \bfseries 3.75e-05 & \bfseries 7.66e-07 & 0.762\\
 & \(\Lambda\) & 1.661 & 1.812 & 0.288 & 0.286 & 0.247 & 0.216 & \bfseries 0.328 & \bfseries 0.525 & 0.99 & \bfseries 0.000562 & \bfseries 2.63e-05 & 0.403\\
\(n_{\text{A}}, n_{\text{B}} = 133, 94\) & \(\Delta \Lambda\) & 0.189 & 0.19 & 0.162 & 0.161 &   &   & -0.007 & 0.009 & 0.994 & 0.539 & 0.927 & 0.561\\
\midrule
\multicolumn{10}{@{} l @{}}{\itshape Intercomparison with other datasets}\\
\addlinespace
MIPS vs Orion & \(\Pi\) & 1.784 & 3.09 & 0.871 & 1.666 &   &   & \bfseries 0.57 & \bfseries 1.37 & \bfseries 1.911 & \bfseries 0.00112 & \bfseries 5.73e-05 & \bfseries 0.00133\\
 & \(\Lambda\) & 1.723 & 2.532 & 0.297 & 0.749 &   &   & \bfseries 0.71 & \bfseries 2.308 & \bfseries 2.526 & \bfseries 9.99e-06 & \bfseries 5.49e-07 & \bfseries 1.49e-09\\
\(n_{\text{A}}, n_{\text{B}} = 227, 18\) & \(\Delta \Lambda\) & 0.19 & 0.658 & 0.162 & 0.513 &   &   & \bfseries 0.642 & \bfseries 2.243 & \bfseries 3.173 & \bfseries 7.14e-06 & \bfseries 5.95e-06 & \bfseries 1.07e-10\\
\addlinespace
MIPS vs RSG & \(\Pi\) & 1.784 & 1.479 & 0.871 & 0.227 &   &   & -0.159 & -0.362 & \itshape 0.26 & \bfseries 0.000594 & 0.288 & \itshape 0.0151\\
 & \(\Lambda\) & 1.723 & 1.41 & 0.297 & 0.101 &   &   & \bfseries -0.719 & \bfseries -1.089 & \itshape 0.339 & \bfseries 2.78e-06 & \bfseries 1.56e-06 & \itshape 0.004\\
\(n_{\text{A}}, n_{\text{B}} = 227, 16\) & \(\Delta \Lambda\) & 0.19 & 0.147 & 0.162 & 0.076 &   &   & -0.072 & -0.267 & 0.469 & 0.191 & 0.701 & 0.12
\\
\bottomrule
\multicolumn{14}{@{}p{\linewidth}@{}}{
  \textit{Description of columns:}
  (Col.~1)~How the two A/B source sub-samples are defined, also giving the size of each sub-sample, \(n_{\text{A}}\) and~\(n_{\text{B}}\).
  (Col.~2)~Dependent variable whose distribution is compared between the two sub-samples.
  (Cols.~3--6)~Mean and standard deviation, \(\sigma\), of the dependent variable for each of the two sub-samples.
  (Cols.~7--8)~Mean over each sub-sample of the observational dispersion (\(\epsilon\), standard deviation) of radii that contribute to the dependent variable for each individual source, as in steps~\ref{step:R0} and \ref{step:R90} of \S~\ref{sec:autom-trac-fitt}.  Note that in the case of \(\Pi\), this is \(\epsilon(R_0)\), and so is not a direct measure of the observational uncertainty in \(\Pi\). 
  (Cols.~9--11)~Standardized ``effect sizes'', which are dimensionless measures of the difference in the distribution of the dependent variable between the two sub-samples.
  (Col.~9)~Rank biserial correlation coefficient \citep{Cureton:1956a}, which is obtained by considering all \(n_{\text{A}} n_{\text{B}}\) pair-wise comparisons of the dependent variable between a source in sub-sample~A and a source in sub-sample~B.  It is the difference between the fraction of such comparisons ``won'' by sub-sample~A and those ``won'' by sub-sample~B, and thus may vary between \(-1\) and \(+1\). 
  (Col.~10)~Cohen's \(d\), which is a dimensionless mean difference: \(d = (\langle \text{A} \rangle - \langle \text{B} \rangle) / \sigma_{\text{pool}} \), where \(\sigma_{\text{pool}} = (n_{\text{A}} \sigma_{\text{A}}^2 + n_{\text{B}} \sigma_{\text{B}}^2)^{1/2} / \sqrt{n_{\text{A}} + n_{\text{B}}}\) is the pooled standard deviation.
  (Col.~11)~Ratio of standard deviations between the two sub-samples.
  (Cols.~12--14)~Probabilities (\(p\)-values) of the two sub-samples being as different as observed if they were to be drawn from the same population, according to three different non-parametric tests.
  (Col.~12)~Kuiper 2-sample test, which is a general test of similarity between two distributions that is designed to retain sensitivity to differences in the tails of the distributions.
  (Col.~13)~Mann--Whitney--Wilcoxon \(U\) test \citep{Mann:1947a}, which is sensitive to differences in the central value of the distributions.
  (Col.~14)~Brown--Forsythe test for equality of variance \citep{Brown:1974a}
}
\end{tabular}
 % end input ./figs/mipsgal-summary-stats-table-body.tex
 \end{table*}

Results from all the statistical tests of the shape distributions
discussed in \S~\ref{sec:corr-shape} are given in
Table~\ref{tab:big-p}.  The \(p\)-values are the probability of
finding a difference between two populations at least as large as what
is observed \emph{given} that there is no difference in the underlying
distribution from which the two populations are drawn (that is, given
that the null hypothesis is true).  Conventionally, the null
hypothesis is rejected at a certain significance threshold \(\alpha\) when
\(p < \alpha\).  Since we are blindly testing many different hypotheses at
once, the commonly used \(\alpha = 0.05\) threshold is too lenient.  In
Figure~\ref{fig:histo-p-values} we analyse the frequency distribution
of \(p\) from all our tests \citep[see][]{Head:2015a}, finding a
systematic excess over a uniform distribution only for \(p < 0.01\).
We therefore take \(\alpha = 0.003\) as the optimum threshold in order to
balance the risks of false positives and false negatives. A false
positive is the erroneous rejection of the null hypothesis (the
spurious detection of a correlation that is not really there), while a
false negative is failing to detect a true correlation.

\section{Perturbed bow shocks}
\label{sec:perturbed-bows}

\begin{figure*}
  \centering
  \includegraphics[width=\linewidth]{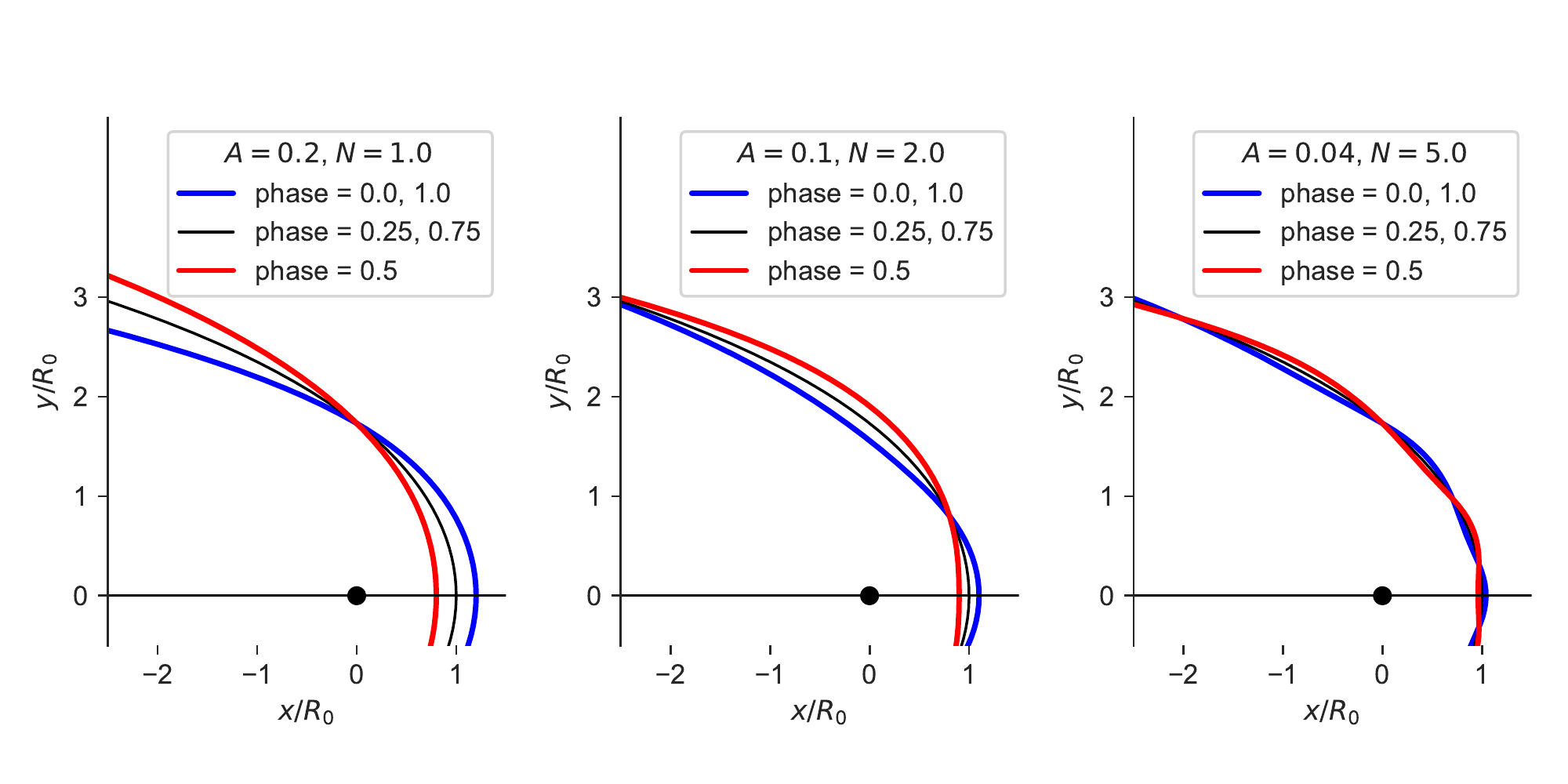}
  \caption{Small-amplitude standing wave perturbations to wilkinoid
    bow shapes.  The maximum deviations from the base shape are seen
    at phases \(\phi = 0\) (blue line) and \(\phi = 0.5\) (red line), while
    the perturbation is zero at \(\phi = 0.25\) and \(0.75\) (black
    line).  Results are shown left to right for increasing wave
    numbers \(N\) and decreasing amplitudes \(A\): (a)~\(A = 0.2\),
    \(N = 1.0\), (b)~\(A = 0.1\), \(N = 2.0\), (a)~\(A = 0.04\),
    \(N = 5.0\).  The maximum curvature, proportional to \(A N\) is
    the same in all three cases.}
  \label{fig:perturb-shapes}
\end{figure*}
\begin{figure*}
  \centering
  \includegraphics[width=\linewidth]
  {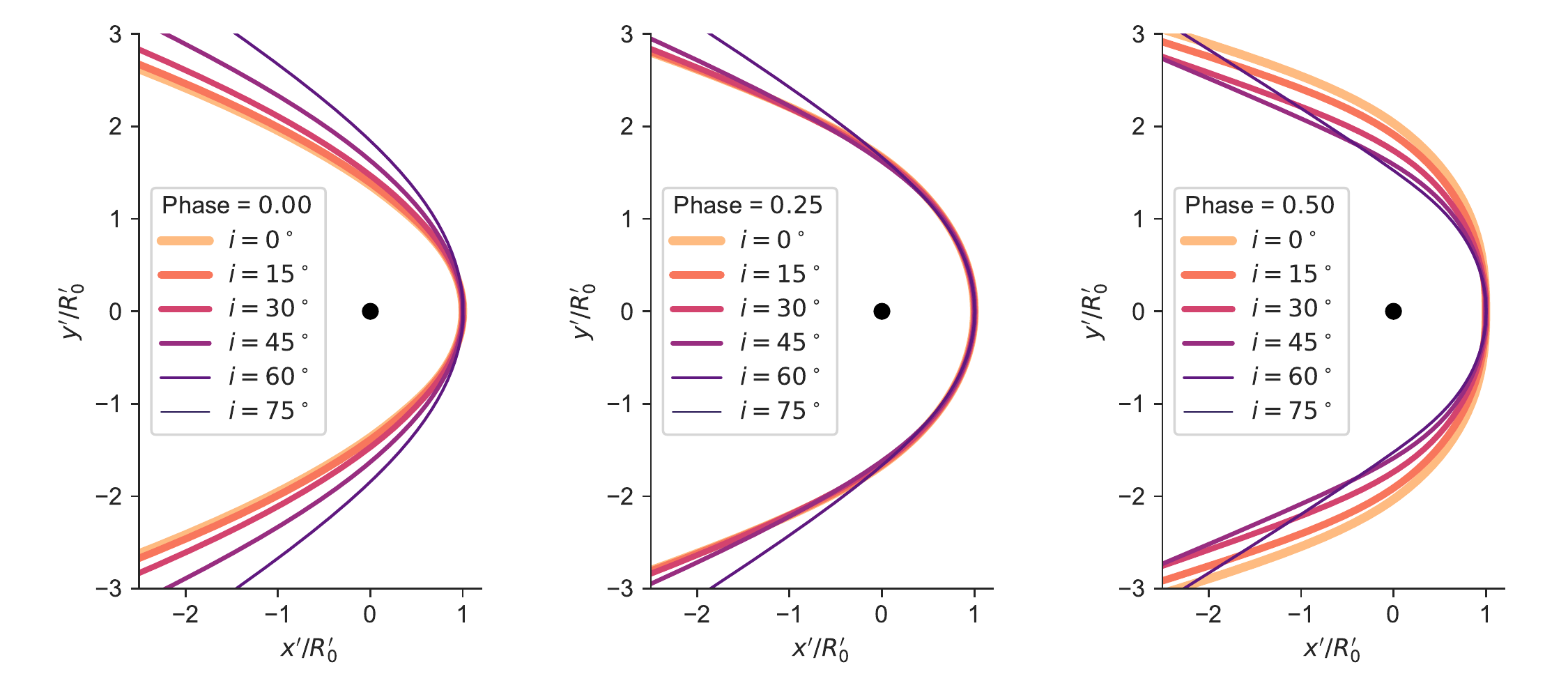}
  \caption{Plane-of-sky projections of perturbed bow shapes.  In all
    cases, the base bow shape is ancantoid with \(\xi = 0.8\),
    \(\beta = 0.005\) and the perturbation is the curling mode shown in
    the central panel of Fig.~\ref{fig:perturb-shapes}, with amplitude
    \(A = 0.1\) and wave number \(N = 2.0\). Results are shown for
    inclination angles \(i = 0\) to \(i = 75^\circ\) (indicated by line
    color and thickness, see key) and for different fractional phases
    of the oscillation: (a)~\(\varphi = 0.0\), (b)~\(\varphi = 0.25\),
    (c)~\(\varphi = 0.50\). Unlike in Fig.~\ref{fig:perturb-shapes}, the
    spatial coordinates are normalized to the instantaneous projected
    apex radius \(R_0'\) at each phase, so the apex does not appear to
    move.}.
  \label{fig:perturb-xy-prime}
\end{figure*}

In this appendix, we present a highly idealized model for small,
time-varying perturbations to a steady-state bow shock shape, such as
those discussed in \S~5 of Paper~0.  These perturbations may be due to
periodic variations in the momentum-loss rate of one of the winds, or
due to dynamical instabilities in the shocked shell.

We consider fractional perturbations \(\Delta(\theta, t)\) of a base shape
\(R(\theta)\), such that
\(R(\theta) \to [1 + \Delta(\theta, t)] R(\theta)\).  For simplicity,
\(\Delta(\theta, t)\) is a standing wave of constant amplitude \(A\), which is
periodic in \(\theta\), with wave number \(N\).  We assume that the
oscillation occurs simultaneously and coherently at all azimuths, so
that cylindrical symmetry is maintained.  This implies that
\(\Delta(\theta, t)\) must be even in \(\theta\), so can be expressed as
\begin{equation}
  \label{eq:standing-wave}
  \Delta(\theta, t) = A \cos(N \theta) \cos(2\pi \varphi) . 
\end{equation}
For waves with period \(P\), the fractional phase \(\varphi\) will
vary with time \(t\) as
\begin{equation}
  \label{eq:fractional-phase}
  \varphi(t) = (\varphi_0 + t/P) \bmod 1.0\ ,
\end{equation}
where \(\varphi_0\) is an arbitrary reference phase.

Example oscillations with wave numbers \(N = 1.0\), \(2.0\), and
\(5.0\) superimposed on a wilkinoid base shape are shown in
Figure~\ref{fig:perturb-shapes}.  There are \(N\) nodes of the
oscillation between \(\theta = [0, \pi]\), always with an antinode at the apex
(\(\theta = 0\)), as required by symmetry.  So, with \(N = 1.0\) there is a
node (fixed point) in the near wing at \(\theta = \pi/2\), but an antinode in
the far wing at \(\theta = \pi\), which is in antiphase with the oscillation
of the apex, giving rise to a large-scale ``breathing'' mode of
oscillation.  With \(N = 2.0\), there are nodes at \(\theta = \pi/4\) and
\(3\pi/4\), while the antiphase antinode has moved to the near wing at
\(\theta = \pi/2\).  There is still an antinode in the far wing at
\(\theta = \pi\) but it is now in phase with the apex, giving rise to a
``curling-up/straightening-out'' mode of oscillation.  With
\(N = 5.0\), there are many more nodes and antinodes, giving a
``ringing'' mode of oscillation.  Note that all our examples have
\(A \propto 1/N\) in order to keep the local curvature relatively low.  If
the product \(A N\) is not small compared to unity, then the local
curvature can be so extreme as to reverse the concave shape of the
base bow shape, producing locally convex regions.

If the bow shape is viewed at different inclinations, then the effect
of the oscillations on the projected shape will vary.  In particular,
the apex-to-wing interval in body-frame angle changes from
\(\theta = [0, \pi/2]\) at \(i = 0\) to
\(\theta = [\theta_0, \theta_{90}]\) for general \(i\), see equations~(18) and (21)
of Paper~0.  The difference \(\theta_{90} - \theta_0\) is always a decreasing
function of \(|i|\), so the oscillations of the tangent line become
increasingly stretched out as the inclination increases.  This effect
can be seen in Figure~\ref{fig:perturb-xy-prime}, which shows an
example of the variation in projected perturbed shape with inclination
angle for 3 different phases, this time for an ancantoid base shape
and the \(N = 2.0\) perturbation shown in
Figure~\ref{fig:perturb-shapes}b.  The most marked changes with phase
are seen for low inclinations (light colored lines), whereas the
changes are smaller, although still noticeable, for
\(|i| \ge 45^\circ\). If \(A N\) exceeds about 0.5, then the local curvature
of the perturbations is so extreme that multiple tangent lines exist
at intermediate inclinations, which produces the appearance of
additional incomplete bright arcs inside the main arc of the bow.

 % end input ./app-p-values.tex

\bsp	
\label{lastpage}

\end{document}